\documentclass[twocolumn]{aastex61}

\usepackage{graphicx}
\usepackage{txfonts}
\usepackage[toc,page]{appendix}
\usepackage{amsfonts}
\usepackage{amssymb}
\usepackage{mathrsfs}
\usepackage{booktabs}

\newcommand{\grad}{\hspace{-2mm}$\phantom{a}^{\circ}$}
\received{?}
\revised{?}
\accepted{2018 April 04}
\submitjournal{ApJ}

\shorttitle{Circumnuclear star-formation and AGN activity}
\shortauthors{Esparza-Arredondo et al.}

\begin{document}

\title{Circumnuclear star-formation and AGN activity: \\ Clues from surface brightness radial profile of PAHs and [SIV]}

\correspondingauthor{Donaji Esparza-Arredondo}
\email{d.esparza@crya.unam.mx}

\author{Donaji Esparza-Arredondo}
\affiliation{Instituto de Radioastronom\'ia y Astrof\'isica, Universidad Nacional Aut\'onoma de M\'exico, Campus Morelia, \\ Apartado Postal 3-72, 58090, Morelia, Michoac\'an, M\'exico}

\author{Omaira Gonz\'alez-Mart\'in}
\affiliation{Instituto de Radioastronom\'ia y Astrof\'isica, Universidad Nacional Aut\'onoma de M\'exico, Campus Morelia, \\ Apartado Postal 3-72, 58090, Morelia, Michoac\'an, M\'exico}

\author{Deborah Dultzin}
\affiliation{Instituto de Astronom\'ia, Universidad Nacional Aut\'onoma de M\'exico, Apartado Postal 70-264, 04510, CDMX, M\'exico}

\author{Almudena Alonso-Herrero}
\affiliation{Centro de Astrobiolog\'ia (CAB, CSIC-INTA), ESAC Campus,  E-28692 Villanueva de la Ca\~nada, Madrid, Spain}
\affiliation{Department of Physics, University of Oxford, Oxford OX1 3RH, UK}
\affiliation{Department of Physics and Astronomy, University of Texas at San Antonio, San Antonio, TX 78249, USA}

\author{Cristina Ramos Almeida}
\affiliation{Instituto de Astrof\'isica de Canarias (IAC), C/V\'ia L\'actea, s/n, E-38205 La Laguna, Spain}
\affiliation{Departamento de Astrof\'isica, Universidad de La Laguna, E-38206 La Laguna, Tenerife, Spain}

\author{Tanio D\'iaz-Santos}
\affiliation{N\'ucleo de Astronom\'ia de la Facultad de Ingenier\'ia, Universidad Diego Portales, Av. Ejercito Libertador 441, Santiago, Chile}

\author{Ismael Garc\'ia-Bernete}
\affiliation{Instituto de Astrof\'isica de Canarias (IAC), C/V\'ia L\'actea, s/n, E-38205 La Laguna, Spain}

\author{Mariela Martinez-Paredes}
\affiliation{Instituto de Radioastronom\'ia y Astrof\'isica, Universidad Nacional Aut\'onoma de M\'exico, Campus Morelia, \\ Apartado Postal 3-72, 58090, Morelia, Michoac\'an, M\'exico}

\author{Jose Miguel Rodr\'iguez-Espinosa}
\affiliation{Instituto de Astrof\'isica de Canarias (IAC), C/V\'ia L\'actea, s/n, E-38205 La Laguna, Spain}
\affiliation{Departamento de Astrof\'isica, Universidad de La Laguna, E-38206 La Laguna, Tenerife, Spain}

\begin{abstract}
We studied the circumnuclear MIR emission in a sample of 19 local active galactic nuclei (AGN) with high spatial resolution spectra using T-ReCS (Gemini) and CanariCam (GTC), together with IRS/\emph{Spitzer} observations. We measured the flux and the equivalent width for the $\rm{11.3~\mu m}$ PAH feature and the [SIV] line emission as a function of galactocentric distance. This allowed to study the star formation (SF) at sub-kpc scales from the nucleus for a large sample of nearby AGN.
The [SIV] line emission could be tracing the AGN radiation field within a few thousand times the sublimation radius ($\rm{R_{sub}}$), but it often peaks at distances greater than 1000 $\rm{R_{sub}}$. One possibility is that the SF is contributing to the [SIV] total flux. We found an $\rm{11.3~\mu m}$ PAH emission deficit within the inner few tens of parsecs from the AGN. This deficit might be due to the destruction of the molecules responsible for this feature or the lack of SF at these distances. We found a sensible agreement in the expected shift of the relation of the AGN bolometric luminosity and the SF rate. This indicates that numerical models attributing the link between AGN activity and host galaxy growth to mergers are in agreement with our data, for most inner galaxy parts.
\end{abstract}

\keywords{ - - - -}

\section{Introduction}
The understanding of the coevolution of active galactic nuclei (AGN) and the host galaxy has been one of the greatest challenges in astronomy in the last decades. Several studies have discovered correlations between the mass of the super-massive black hole (SMBH), the mass of the bulge \citep{Magorrian98, Tremaine02, Marconi03, McConnell13}, and the bulge velocity dispersion \citep{Kormendy95, Ferrarese00}. However, the physical connection between these observational properties is still unclear. The study of SMBH accretion and circumnuclear\footnote{We considered circumnuclear scales at distances less than 1\,kpc.} star-formation (SF) can be the key. Some authors propose that the gas that moves toward the center is responsible for both the growth of the SMBH and the enhancement of SF \citep{Sanders88, Barnes91, Storchi01}. Other works suggest that quenching of SF is due to AGN feedback \citep[][and references therein]{Silk98, Vollmer13}.

Numerical simulations propose a scenario where large-scale processes can be related to small-scales phenomena close to the nucleus \citep[e.g.][]{Kawakatu08,Hopkins10, Neistein14, Volonteri15, Gutcke15}. According to these studies, major-mergers and even tidal interactions produce perturbations that can be correlated with the accretion of the SMBH and SF \citep[][]{Krongold02}. Other authors propose a scenario in which the radiation field of the SMBH is able to stop the SF, imposing a balance between the two \citep[e.g.][]{Wu09}.

The study of the neighborhood of AGN is very complex because the classic indicators of SF such as the ultraviolet (UV) continuum, Pa${\rm{\alpha}}$, and H${\rm{\alpha}}$ emission line are easily contaminated by the powerful AGN emission \citep[][and references therein]{Alonso-Herrero14}. However, the mid-infrared (MIR) wavebands are a powerful tool to disentangle SF and AGN contributions \citep[e.g.][]{Dultzin89, Gonzalez13,Alonso-Herrero14}.
Recently, new MIR spectroscopic data have provided opportunities to quantify the SF close ($\rm{<1\,kpc}$) to the AGN \citep[e.g.][]{Esquej14,Ruschel16}. The polycyclic aromatic hydrocarbons (PAHs) emission features at 3.3, 6.2, 7.7, 8.6, and $\rm{11.3~\mu m}$ contribute to MIR flux. The PAHs are composed of 20-100 atoms of carbon and hydrogen \citep{Millar93}. These features are powerful tools to study SF on the vicinity of AGN. These molecules have been studied in different objects associated with dust and gas including evolved stars, reflection nebulae, Orion bars, and star-forming regions \citep{Gillett73, Cohen86, Aitken84}.
It is known that the PAHs emissions are good tracers of young and massive stars (i.e. recent circumnuclear SF activity).
In particular, Starburst galaxies show a good correlation between the strength of the PAH and the IR luminosity, indicating they are good tracers of SF \citep[][]{Brandl06}.

\begin{table*}[ht]
\def\arraystretch{1.1}
\caption{General properties of sample}
\begin{center}
\begin{footnotesize}
\resizebox{18cm}{!} {
\begin{tabular}{cccccccccccccc}
\hline \hline
Object  & Type &  D    & L$\rm{_x}$\tablenotemark{A} & M$\rm{_{BH}}$\tablenotemark{B} & Instrument    & P.A. & Scale factor & Slit width (Nuclear) & \emph{Spitzer} slit width & Radius (in/out) & $\rm{R_{subl}}$ & $\rm{logd_{25}}$ & Ref. \\ 
        &      & (Mpc) & log(L(2-10 keV)) & log(M/M$_{\odot})$ &    & (degrees) &      & (arcsec/pc) & (arcsec/pc) &  (pc/pc) & (pc) & log(0.1 arcmin) &   \\
(1)     & (2)  & (3)   &    (4) & (5)      &    (6)        &  (7) &  (8)      &    (9)               & (10)                      & (11)            & (12) & (13) & (14)   \\ 
\hline
\decimals
NGC\,931     & Sy1   & $49.4$ & $43.3^{\rm{c}}$ & $8.3^{\rm{I}}$    & CanariCam & $80$ &  --   & $0.52/124.5$  &      --       & $93.4/242.8$  & $0.24$ & $1.39$ & 2  \\
Mrk\,1066    & Sy2   & $51.7$ & $42.9^{\rm{a}}$ & $7.0^{\rm{I}}$    & CanariCam & $315$  & 1.23  & $0.52/130.3$  & $3.7/927.3$   & $87.9/440.3$  & $0.09$ & $1.08$ & 2,4 \\
NGC\,1320    & Sy2   & $37.7$ & $42.5^{\rm{c}}$ & $7.2^{\rm{I}}$    & CanariCam & $315$   &  --   & $0.52/95.0$  &      --       & $149.7/406.4$  & $0.03$ & $1.27$ & 2   \\
NGC\,1386    & Sy2   & $16.2$ & $41.6^{\rm{a}}$ & $7.4^{\rm{I}}$    & T-ReCS    & $0$   & 1.17  & $0.31/24.4$  & $3.7/291.1$   & $31.8/88 .3$   & $0.02$ & $1.55$ & 1 \\
NGC\,1808    & Sy2   & $11.5$ & $39.7^{\rm{a}}$ & $6.7^{\rm{II}}$   & T-ReCS    & $45$  & 1.35  & $0.35/19.6$  & $3.7/207.2$   & $27.6/62.8$  & $0.002$ & $1.73$ & 1\\
NGC\,2992    & Sy1.8 & $31.6$ & $41.9^{\rm{c}}$ & $7.7^{\rm{I}}$    & CanariCam & $30$  & 0.4  & $0.52/79.7$  & $3.7/566.8$   & $59.7/741.2$   & $0.03$ & $1.47$ & 2  \\
NGC\,3081    & Sy2   & $32.5$ & $42.5^{\rm{b}}$ & $7.1^{\rm{II}}$  & T-ReCS    & $0$   & 0.96  & $0.65/102.4$ & $3.7/582.9$   & $63.6/205.0$ & 0.06 & $1.43$ & 1 \\
"    & ''    &   ''    &      ''         &      ''            &  ''       & $350$ & 0.92  &    ''         &    ''         & $91.9/205.1$  & " & '' \\
NGC\,3227    & Sy1.5 & $21.8$ & $42.1^{\rm{c}}$ & $7.6^{\rm{I}}$    & CanariCam & $0$  & 0.64  & $0.52/54.9$  & $3.7/391.0$  & $74.2/205.6$   & 0.04 & $1.60$ & 2 \\
NGC\,3281    & Sy2   & $21.8$ & $43.2^{\rm{a}}$ & $7.9^{\rm{II}}$  & T-ReCS    & $315$ & 0.61  & $0.35/77.6$  & $3.7/820.0$  & $109.4/248.6$ & $0.14$ & $1.49$ &1,4 \\
NGC\,4253*    & Sy1   & $55.4$ & $42.5^{\rm{c}}$ & $6.8^{\rm{III}}$   & CanariCam & $285$  & 0.75 & $0.52/139.6$ &  --  & $94.2/408.3$  & $0.06$ & $0.95$ & 2 \\
NGC\,4569    & Sy2   & $12.6$ & $39.4^{\rm{e}}$ & $7.8^{\rm{IV}}$    & CanariCam & $30$ &  --   & $0.52/31.7$  &      --       & $19.1/285.9$  & $0.004$ & $1.96$ & 2 \\
NGC\,5135*    & Sy2   & $58.6$ & $43.1^{\rm{b}}$ & $7.3^{\rm{II}}$  & T-ReCS    & $30$  & 1.33 & $0.70/199.0$ & --  & $89.3/1033.5$ & $0.13$ & $1.38$ & 1,3  \\
NGC\,5643    & Sy2   & $16.9$ & $42.6^{\rm{b}}$ & $7.4^{\rm{II}}$   & T-ReCS    & $80$  & 0.68  & $0.35/28.7$  & $3.7/303.1$  & $55.2/106.6$ & $0.07$ & $1.72$  & 1  \\
IC\,4518W*    & Sy2   & $69.6$ & $42.6^{\rm{b}}$ & $7.5^{\rm{V}}$   & T-ReCS    &  $5$  & 1.22 & $0.70/236.3$ & --  & $166.7/560.6$ & $0.07$ & $1.10$ & 1,3 \\
IC\,5063*    & Sy2   & $48.6$ & $42.9^{\rm{c}}$ & $7.7^{\rm{I}}$    & T-ReCS    & $303$ & 1.33  & $0.65/153.1$ & --  & $95.2/306.7$  & $0.09$ & $1.43$ & 1,5 \\
NGC\,7130    & Sy2   & $69.2$ & $42.9^{\rm{a}}$ & $7.6^{\rm{II}}$  & T-ReCS    & $348$ & 1.35  & $0.70/234.7$ & $3.7/1240.7$ & $135.4/496.6$ & $0.09$ & $1.19$ & 1,3\\
NGC\,7172    & Sy2   & $33.9$ & $42.7^{\rm{a}}$ & $7.7^{\rm{II}}$  & T-ReCS    & $60$  & 0.57  & $0.35/57.5$  & $3.7/608.0$  & $95.8/228.6$  & $0.08$ & $1.44$ & 1,6 \\
NGC\,7465    & Sy2   & $27.2$ & $41.4^{\rm{d}}$ & $7.6^{\rm{VI}}$ & CanariCam & $330$  &  --   & $0.52/68.6$  &      --       & $46.3/200.6$   & $0.03$ & $1.03$ & 2 \\
NGC\,7582    & Sy2   & $22.5$ & $42.6^{\rm{b}}$ & $7.1^{\rm{II}}$   & T-ReCS    & $0$   & 0.39  & $0.70/76.4$  & $3.7/403.6$  & $53.8/396.6$  & $0.07$ & $1.84$ & 1 \\
\hline 
\end{tabular}
}
\label{tab:observations}
\end{footnotesize}
\end{center}
\begin{footnotesize}
\tablecomments{ (1) Source name; (2) Type of sources according to \citet{Gonzalez13} or \citet{Alonso15}; (3) Distances calculated from redshift obtained from observations for $\rm{\Omega_{\Lambda} = 0.73}$, $\rm{\Omega_M = 0.27}$, and $\rm{H_0 = 70~km s^{-1} Mpc^{-1}}$ ;(4) X-ray luminosity;(5) Black hole mass; (6) Instrument used by each object; (7) Position angle; (8) Scale factor between T-ReCS/CanariCam and \emph{Spitzer} spectra. The mark ``*'' is used to identify the sources where we do not use the \emph{Spitzer} spectra; (9) Slit width for nuclear spectra; (10) Slit width for \emph{Spitzer} spectra; (11) The minimum and maximum radius used for the extended profiles (T-ReCS or CanariCam); (12) The sublimation radius; (13) The isophotal diameter; (14) Reference  where the observations were originally published. (1) \citet{Gonzalez13}, (2) \citet{Alonso15}, (3) \citet{Diaz10}, (4) \citet{Sales11}, (5) \citet{Young07}, and (6) \citet{Roche07}.}
\tablenotetext{A}{ The references for the X-ray results are: a) Gonz\'alez-Mart\'in et al. in preparation; b) \citet{Gonzalez13}; c) \citet{Liu14}; d) \citet{OSullivan01}; e) \citet{Ho01}.}
\tablenotetext{B}{ The black hole mass is calculated using the relation with the stellar velocity dispersion. References: I) \citet{Woo02}, II) \citet{Esquej14}, III) \citet{Woo15}, IV) \citet{Mason15}, V) \citet{Alonso13}, VI) \citet{Dudik05}.}
\end{footnotesize}
\end{table*}

Among these PAH features, the $\rm{11.3~\mu m}$ PAH feature has the advantage of being isolated (i.e. not blended) from others and is observable with ground-based telescopes (i.e. with enough spatial resolution to disentangle the contribution of the few tenths of pc from the nucleus in nearby galaxies). Indeed, the $\rm{11.3~\mu m}$ PAH emission feature has been used in several works to study the SF in the vicinity of AGN \citep[e.g.][]{Diaz10}. Recently, \citet{Esquej14} computed the star forming rate (SFR) from this feature and compared it with the AGN accretion rate. They confronted this relation with coevolution models elaborated by \citet{Hopkins10}. They found a good agreement between observations and theoretical models for physical scales of $\rm{\sim}$100\,pc. Recently, \citet{Ruschel16} have analyzed the circumnuclear SF in a sample of 15 AGN in order to investigate the validity of the same relation. They found that SF luminosities are correlated with the bolometric luminosity of the AGN (for objects with $\rm{L_{bol,AGN}\geq10^{42}~erg~s^{-1}}$).

The PAH features have been studied in the vicinity of the AGN of many galaxies. Some authors claim that these molecules are destroyed by the strong  AGN radiation field  \citep[][]{Voit92,Wu09,Diaz10}. \citet{Siebenmorgen04} and \citet{Ruschel14} have found evidence in favour of this destruction of PAHs in AGNs. Supporting this, the correlation between the strength of the PAH features and the IR luminosity appear to be absent or weak in AGN \citep{Weedman05}.  If this were the case, the PAH emission feature could not be used as a tracer of SF in AGN. In a more recent paper, it has been suggested that PAH emission might not be a good tracer of the SF within 1\,kpc around the AGN \citep{Jensen17}.

Against it, \citet{Alonso-Herrero14} concluded that at least those molecules responsible for the $\rm{11.3~\mu m}$ PAH feature survive in the nuclear environment as close as 10 pc from the nucleus for their sample of six local AGN \citep[see also][]{Ramos-Almeida14, Esquej14}. They propose that material in the dusty tori, nuclear gas disk, and/or host galaxies of AGN is likely providing the column density necessary to protect the PAH molecules from the AGN radiation field.

Here we investigate if the $\rm{11.3~\mu m}$ PAH can be used (and at which scales) as a tracer of SF, and we use it to get some a clues about the coevolution between the AGN and its host galaxy. For that purpose we have compiled a sample of high spatial resolution spectra ($\rm{8-13~\mu m}$) of local AGN observed with T-ReCS in the Gemini South observatory and CanariCam on the 10.4\,m Gran Telescopio CANARIAS (GTC). This allowed us to study the SF at different scales from the nucleus for a large sample of sources.
The coverage of these spectra will also allows us to analyze the origin of the [SIV] line emission at $\rm{10.5~\mu m}$. The [SIV] line arises from ions with an ionization potential of 35 eV. It has been proposed as an indicator of the AGN isotropic luminosity since it might come from the narrow line region \citep[NLR,][]{Dasyra11}. However, high spatial resolution MIR spectra indicate that this emission is not resolved at 100\,pc scales, against its NLR origin \citep{Honig08}. The [SIV] line emission at $\rm{10.5~\mu m}$ could also be related to star-forming regions \citep{Pereira10}. Our high spatial resolution spectra are very well suited to understand the origin of the [SIV] line emission. 

The main goal of this work is to address three questions: 1) the origin of [SIV] line emission, 2) the goodness of the $\rm{11.3~\mu m}$ PAH feature as tracer of SF in the vicinity of AGN, and 3) the connection between SF and AGN activity. 
The paper is organized as follows: Section \ref{sec:sample} presents our sample and the data reduction. Section \ref{sec:SpecAnalysis} presents the analysis of the spectra. Sections \ref{subsec:OSIV} and \ref{subsec:DiscPAHs}, a discussion of the main results in the framework of our goals. Finally, a brief summary is given in Section \ref{sec:SumCon}. Throughout this work, we assumed a $\Lambda$CDM cosmology with $\rm{\Omega_{\Lambda} = 0.73}$, $\rm{\Omega_M = 0.27}$, and $\rm{H_0 = 70~km s^{-1} Mpc^{-1}}$.

\section{Sample selection and data reduction}
\label{sec:sample}
\subsection{Sample}
Our sample consists of 19 local AGN with ground-based N band (i.e. $8 - 13~\mu$m) spectra available. All spectra have been observed with ground-based telescopes. These sources are taken from the samples of \citet{Gonzalez13} and \citet{Alonso15} which contain 22 and 45 local AGN, respectively. We have only included AGN showing extended emission. We considered the source as extended if we can detect emission of $\rm{11.3~\mu m}$ PAH feature or [SIV] line in more than three circumnuclear apertures (see Sect.3, for a detailed explanation on the aperture extraction procedure).

This sample is the largest reported where high resolution studies that has been done in the vicinity of AGN. However, note that this sample is not complete in any sense. Table \ref{tab:observations} shows the main observational details of the sample. Fifteen objects are type-2 Seyferts (Sy2) and four are type-1 Seyferts (Sy1). Our sample covers a range of X-ray luminosity (absorption corrected) of L$\rm{(2-10~keV)\sim 5 \times 10^{39} - 4 \times 10^{43}}$\,erg s$^{-1}$. The range of X-ray luminosity covers classical Seyfert galaxies and low luminosity AGN ($\rm{< 10^{42}}$\,erg s$^{-1}$). Appendix \ref{sec:Cat} has a short review of the published information on star-forming regions around these objects, when available. 

Eleven objects were observed with the Thermal-Region Camera Spectrograph \citep[T-ReCS,][]{Telesco98, DeBuizer05} located in the 8.1 m Gemini-South Telescope and published by \citet{Gonzalez13} (and references therein). The slit width used for the spectroscopy results in a spatial resolution between $\sim 20 - 250$\,pc (see column 9 in Table \ref{tab:observations}). 
The rest of the sources in our sample were obtained with CanariCam \citep{Telesco03} in the 10.4 m Gran Telescopio CANARIAS (GTC) and were published by \citet{Alonso15}. For these eight sources the slit width used for the spectroscopy results in a spatial resolution between $\sim 50-160$\,pc. The angular and spectral resolutions for both instruments (T-ReCS and CanariCam) is within an average of $\rm{FWHM\, \sim 0.3}$\,arcsec and $\rm{R \sim 100}$, respectively. Note that these spectral resolutions are not high enough to examined the width of [SIV] line. Indeed all the [SIV] emission lines reported here have a width compatible with the instrumental spectral resolution.

We have included the \emph{Spitzer}/IRS spectral data downloaded from the CASSIS\footnote{\url{http://cassis.astro.cornell.edu/atlas/}} catalog \citep[the Cornell AtlaS of \emph{Spitzer}/IRS Sources,][]{Lebouteiller11} to study larger regions in each galaxy. Note that the spectral resolution of \emph{Spitzer}/IRS ($\rm{R \sim 60-130}$) is similar to that obtained by our ground based observations. CASSIS provides flux-calibrated nuclear spectra associated with each observation. The \emph{Spitzer} spectra are not available for four of the sources in our sample (NGC\,931, NGC\,1320, NGC\,4569, and NGC\,7465). In four sources additional we did not use the \emph{Spitzer} data because the emission of the ground-based data extend up to the spatial resolution of \emph{Spitzer} data. Therefore this observations does not add extra information to our ground-based data.
Thus, we included \emph{Spitzer}/IRS spectra for eleven of the objects, column 10 in Table \ref{tab:observations} shows the \emph{Spitzer} radius spectra when we used them in the analysis. 

\citet{Gonzalez13} and \citet{Alonso15} focused their analyses on the nuclear emission. Also focusing on the central region, four sources have been observed with VISIR/VLT and reported by \citet{Honig10}. 
Furthermore, the MIR extended emission of some of our sources has been studied individually before. Three of our sources (NGC\,5135, IC\,4518W, and NGC\,7130) were studied by \citet{Diaz10}. They studied the extended emission of different features, including the $\rm{11.3~\mu m}$ and the [SIV] line, and they compared it with the \emph{Spitzer} spectra. Mrk\,1066 was analyzed by \citet{Ramos-Almeida14} and \citet{Alonso-Herrero14} to study the survival of the responsible molecules for the $\rm{11.3~\mu m}$ PAH feature in the close vicinity of an AGN. \citet{GarciaB15} studied the extended emission of NGC\,2992 up to $\rm{\sim}$3\,kpc, finding that PAH features might indicate that the bulk of this extended emission is dust heated by SF.
\citet{Esquej14} compared nuclear with larger apertures (using \emph{Spitzer} spectra) in 12 of our sources to study the correlation between SFR through the $\rm{11.3~\mu m}$ PAH feature and AGN accretion. These works will be compared with our results along this paper. 

\begin{table*}[ht]
\def\arraystretch{1.1}
\caption{Integrated fluxes and EWs for the nuclear and \emph{Spitzer} spectra}
\begin{center}
\begin{footnotesize}
\resizebox{16cm}{!} {
\begin{tabular}{ccccccccc}
\hline \hline 
Object     & \multicolumn{4}{c}{\textit{$\rm{Fluxes}$} ($\rm{10^{-13}\, erg \, s^{-1}\, cm^{-2}}$)} &   \multicolumn{4}{c}{EW ($10^{-3}\, \mu$m)}   \\
\cmidrule(lr){2-5} \cmidrule(r){6-9}
     & \multicolumn{2}{c}{PAH$_{\rm{11.3~ \mu m}}$} &  \multicolumn{2}{c}{[SIV]$_{\rm{10.5~ \mu m}}$} &  \multicolumn{2}{c}{PAH$_{\rm{11.3~ \mu m}}$}  &  \multicolumn{2}{c}{[SIV]$_{\rm{10.5~ \mu m}}$}   \\
\cmidrule(lr){2-3} \cmidrule(r){4-5} \cmidrule(lr){6-7} \cmidrule(lr){8-9}
     & Nuclear & \emph{Spitzer} & Nuclear & \emph{Spitzer} & Nuclear & \emph{Spitzer} & Nuclear  &    \emph{Spitzer}  \\
\hline  
NGC\,931     & $<~2$ & -- & $13~\pm~2$ & -- & $<~1$ & -- & $10~\pm~2$ & -- \\
Mrk\,1066   & $82~\pm~12$ & $264~\pm~40$ &  $<~0.1$ & $11~\pm~2$ & $118~\pm~18$ & $158~\pm~25$ & $<~0.2$ & $19~\pm~3$ \\
NGC\,1320   & $7~\pm~2$ & -- & $6~\pm~1$ & -- & $9~\pm~3$ & -- &  $7~\pm~1$ & -- \\
NGC\,1386   & $<~0.3$ & $16~\pm~2$ & $17~\pm~3$ & $23~\pm~4$ & $<~1$ & $14~\pm~2$ & $28~\pm~5$ & $26~\pm~4$ \\
NGC\,1808    & $154~\pm~24$ & $1176~\pm~178$ & $<~0.1$ & $<~7$ & $107~\pm~16$ & $167~\pm~26$ & $<~0.3$ & $<~2$ \\
NGC\,2992    & $<~6$ & $160~\pm~25$ & $4~\pm~1$ & $10~\pm~2$ & $<~22$ & $149~\pm~25$ & $17~\pm~3$ & $19~\pm~4$  \\
NGC\,3081    & $<~0.2$ & $6~\pm~2$ & $10~\pm~4$ & $29~\pm~12$ & $<~1$ & $9~\pm~2$ & $23~\pm~12$ & $39~\pm~6$  \\ 
NGC\,3081    & $<~0.1$ & $6~\pm~1$ & $12~\pm~4$ & $29~\pm~10$ & $<~0.1$ & $9~\pm~3$ & $26~\pm~8$ & $38~\pm~14$ \\
NGC\,3227    & $32~\pm~5$ & $176~\pm~29$ & $10~\pm~2$ & $9~\pm~3$ & $41~\pm~6$ & $119~\pm~21$ & $13~\pm~2$ & $11~\pm~3$ \\
NGC\,3281   & $<~0.3$ & $<~12$ & $10~\pm~2$ & $18~\pm~5$ & $<~0.4$ & $<~13$ & $15~\pm~2$ & $30~\pm~6$ \\
NGC\,4253    & $22~\pm~3$ & -- & $5~\pm~1$ & -- & $37~\pm~6$ & -- & $9~\pm~1$ & -- \\
NGC\,4569    & $32~\pm~5$ & -- & $<~0.02$ & -- & $115~\pm~18$ & -- & $<~0.1$ & --  \\
NGC\,5135   & $14~\pm~2$ & -- & $9\pm~1$ & -- & $40~\pm~6$ & -- & $28\pm~5$ & -- \\
NGC\,5643    & $5~\pm~1$ & $74~\pm~11$ & $10~\pm~2$ & $15~\pm~2$ & $12~\pm~2$ & $110~\pm~17$ & $29~\pm~5$ & $37~\pm~6$ \\
IC\,4518W   & $<~0.1$ & -- & $<~0.1$ & -- & $<~0.3$ & -- & $<~0.2$ & -- \\
IC\,5063     & $<~7$ & -- & $9~\pm~1$ & -- & $<~3$ & -- & $5~\pm~1$ & -- \\
NGC\,7130   & $39~\pm~6$ & $145~\pm~22$ & $4~\pm~1$ & $6~\pm~1$ & $74~\pm~11$ & $132~\pm~20$ & $12~\pm~2$ & $14~\pm~2$ \\
NGC\,7172    & $<~0.5$ & $33~\pm~5$ & $2.1~\pm~0.4$ & $5~\pm~1$ & $<~1$ & $70~\pm~11$ & $25~\pm~4$ & $24~\pm~4$ \\
NGC\,7465    & $14~\pm~2$ & -- & $<~0.1$ & -- & $54~\pm~9$ & -- & $<~0.3$ & --\\
NGC\,7582    & $8~\pm~1$ & $182~\pm~28$ & $<~0.1$ & $6~\pm~1$ & $27~\pm~4$ & $177~\pm~28$ & $<~0.1$ & $23~\pm~4$ \\   
\hline 
\end{tabular}
}
\label{tab:measure}
\end{footnotesize}
\tablecomments{ The symbol ``--'' indicates that \emph{Spitzer} spectra were not available.}
\end{center}
\end{table*}

\subsection{Data Reduction}

The data have been reduced using the RedCan pipeline \citep{Gonzalez13}. RedCan is a fully automated pipeline that was designed to efficiently exploit CanariCam data. Due to the similarities between CanariCam and T-ReCS low spectral resolution data, this pipeline can analyze successfully both sets of observations considered in this paper. RedCan is able to produce flux-calibrated images and 1D spectra. The main input is an ASCII file, which contains an observation list. The reduction process basically consists in eight steps: 1) identification of files, 2) flat fielding, 3) stacking, 4) image flux calibration, 5) wavelength calibration, 6) trace determination, 7) spectral extraction, and 8) spectral flux calibration and the combination of spectra. Within these steps, the subtraction of the sky background and rejection of bad images are also included. Flux-calibration is performed by observing standard stars taken immediately before or after the target.

\emph{Spitzer}/IRS spectra provided by CASSIS are already reduced. However, observations using data from both short-low and long-low spectra modules suffer from mismatches due to telescope pointing inaccuracy or due to a different spatial resolution of the IRS orders. This is not corrected in the final products given by CASSIS. Still, in this work it is not necessary to correct these mismatches, because we only considered one spectrum (SL1). This spectrum only covers a range between $\rm{7.5 - 15\, \mu m}$. Finally, the spectra are shifted to the rest-frame according to the distances of the objects (see Col. 3 in Table \ref{tab:observations}).

\section{Spectral analysis}
\label{sec:SpecAnalysis}
Nuclear spectra were first extracted as point-like sources using RedCan pipeline. These spectra show photometric errors typically of 11$\%$ in flux for all objects \citep{Alonso15}. We used these spectra as the nuclear component of our radial profile.

\begin{figure*}
\begin{center}
\includegraphics[width=1.9\columnwidth]{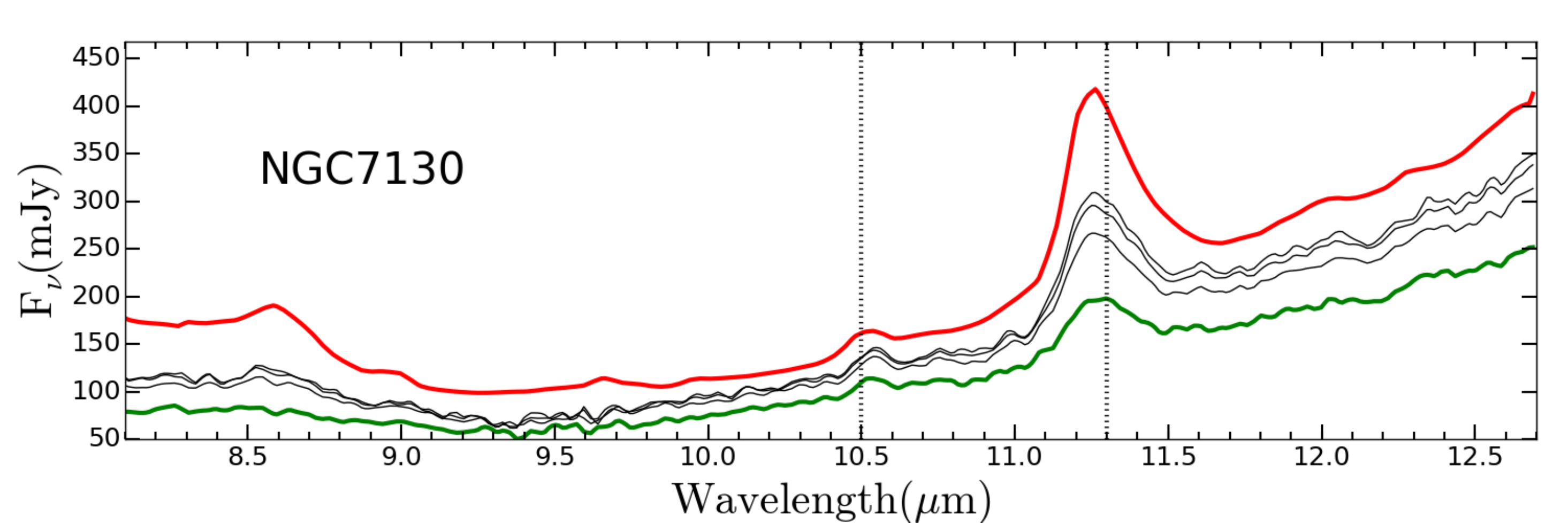}
\includegraphics[width=2.\columnwidth]{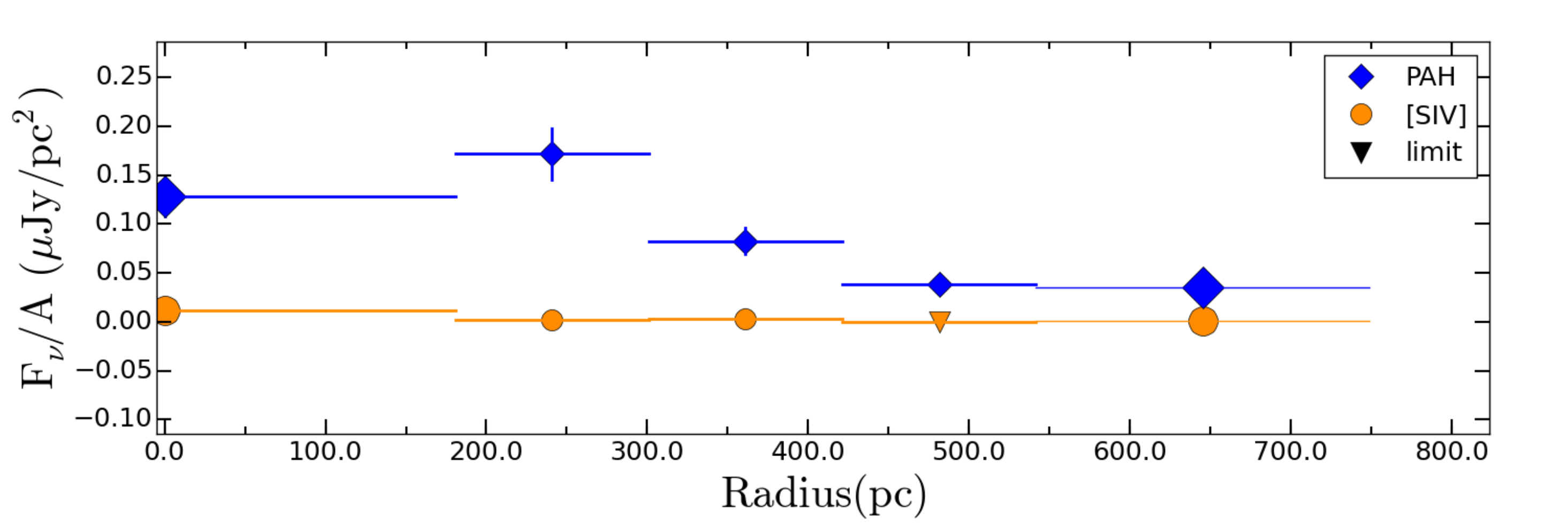}
\caption{(Top): Extracted spectra at different scales for NGC\,7130. The red line is the spectrum obtained by Spitzer, the green and black lines correspond to the nuclear spectrum and extended apertures spectra, respectively. The dotted lines show the PAH feature and [SIV] line emissions. (Bottom): Surface brightness radial profiles in units of $\rm{\mu Jy/pc^2}$ for NGC\,7130. We first extracted the flux at the radius of each aperture, and then we subtracted that of all inner apertures to get the flux of a ring.
The radial profile for $\rm{11.3~\mu m}$ PAH emission is presented with blue diamonds, while the radial profile for [SIV] line emission is shown with orange circles (the triangles are limits values). The larger symbols (diamonds or circles) corresponds to the nuclear and \emph{Spitzer} spectra, respectively. The rest of symbols represent the measurements for the extended apertures.}
\label{fig.2}
\end{center}
\end{figure*} 

\noindent
Then, in order to analyze the change in the spectrum at different distances from the nucleus, and study the circumnuclear emission, we have divided the spatial axis of the spectra into apertures at different radii. Thus, each aperture gives the spectrum of the extended emission within this radius together with the nuclear component.
The maximum radius is determined as the largest one where extended emission can be seen in the 2D spectra. The aperture increments are fixed to 4 pixels because it matches the FWHM of the average PSF in our observations. The extraction has been done using the extended source mode provided by RedCan. The minimum radius of the apertures is calculated as the first aperture where the $\rm{12~\mu m}$ continuum flux is greater (or equal) than the nuclear continuum flux\footnote{The nuclear continuum is extracted as a point source using the trace of the standard star \citep[see][for more details]{Gonzalez13}.}. Table \ref{tab:observations} (Col.11) reports the minimum (i.e. nuclear component extracted as point-like source) and maximum radius used for the extended profiles in units of pc. We are tracing minimum (maximum) extensions of $\rm{\sim}$20\,pc ($\rm{\sim}$ 1kpc) in the spatial direction of the 2D spectra.

Since we use T-ReCS/CanariCam and IRS/\emph{Spitzer} spectra together, we have studied the flux cross-calibration from both instruments, for which we have measured the $\rm{12~\mu m}$ continuum flux. We computed a $\rm{12~\mu m}$ radial continuum profile including both T-ReCS/CanariCam and \emph{Spitzer} fluxes (see Section 3.2 for more details about radial profiles). Then, we investigated whether the \emph{Spitzer} flux follows the extrapolation of the T-ReCS/CanariCam radial profile. We found that \emph{Spitzer} spectra (in the sources where it was used) do not extrapolate naturally from the radial distribution observed in high spatial resolution spectra. In five cases, the $\rm{12~\mu m}$ continuum flux for \emph{Spitzer} data is higher and in other six cases, it is lower than that of ground-based spectra. A larger integrated \emph{Spitzer} flux than that predicted by the extrapolation of the ground-based spectra is expected because they cover a different area although mapping the same aperture (3.7x 3.7 $\rm{arcsec^2}$ and slit-width x 3.7 $\rm{arcsec^2}$, for \emph{Spitzer} and ground-based data, respectively). Lower integrated \emph{Spitzer} flux is most certainly due to flux calibration uses in the ground-based spectra
due to the highly variable sky.
In order to correct this cross-calibration problem, we implemented a re-calibration of the T-ReCS/CanariCam data for each source.
This calibration was implemented as a scaled value for each source as the difference between the extrapolation of the fluxes given in the T-ReCS/CanariCam radial profiles and the \emph{Spitzer} flux at $\rm{12~\mu m}$. We then multiply the T-ReCS/CanariCam fluxes by this value (see Col. 8 in Table \ref{tab:observations}). Note that the scaled value is not within the reported error range for the \emph{Spitzer} or T-ReCS/CanariCam data. However, the correction applied is in general very small ($\rm{|F_{12~\mu m} (T-ReCS/CanariCam)/ F_{12~\mu m} (\emph{Spitzer})|\sim 1.3}$).

In Figure \ref{fig.2} (bottom) we show NGC\,7130 as an example of the data presented in this paper. This example clearly shows the PAH feature at $\rm{11.3~\mu m}$ and the [SIV] line in $\rm{10.5~\mu m}$. A similar figure for each object in our sample is included in Appendix~\ref{sec:Cat}.

\begin{table*}[ht]
\def\arraystretch{1.1}
\caption{PAH and [SIV] fluxes.}
\begin{center}
\begin{footnotesize}
\resizebox{14cm}{!} {
\begin{tabular}{ccccccccc}
\hline \hline 
Object     & \multicolumn{8}{c}{\textit{$\rm{Fluxes}$} ($\rm{10^{-13}\, erg \, s^{-1}\, cm^{-2}}$)} \\
\cmidrule(lr){2-9}
     & \multicolumn{4}{c}{PAH$_{\rm{11.3~ \mu m}}$} &  \multicolumn{4}{c}{[SIV]$_{\rm{10.5~ \mu m}}$} \\
\cmidrule(lr){2-5} \cmidrule(r){6-9}
     & 100\,pc & 200\,pc & 500\,pc &700\,pc & 100\,pc & 200\,pc & 500\,pc & 700\,pc \\
\hline 
NGC\,931    & -- & -- & -- & -- & $21 \pm 3$ & $35 \pm 6$ & $40 \pm 6$ & -- \\
Mrk\,1066   & -- & $160 \pm 25$ & $267 \pm 41$ & $273 \pm 42$ & -- & $9 \pm 1$ &  $12 \pm 2$ & $13 \pm 2$ \\
NGC\,1320  & -- & -- & -- & -- & $7 \pm 1$ & $8 \pm 1$ &  $11 \pm 2$ & -- \\
NGC\,1386  & -- & -- & -- & -- & $24 \pm 4$ & $24 \pm 4$ & -- & -- \\
NGC\,1808  & $486 \pm 74$ & $1065 \pm 161$ & -- & -- & -- & -- &  -- & -- \\
NGC\,2992  & -- & -- & -- & -- & $11 \pm 2$ & $13 \pm 2$ & $20 \pm 3$ & -- \\
NGC\,3081  & -- & -- & -- & -- & $17 \pm 3$ & $28 \pm 6$ &  -- & -- \\
NGC\,3081  & -- & -- & -- & -- & $16 \pm 3$ & $27 \pm 5$ &  -- & -- \\
NGC\,3227  & $80 \pm 12$ & $121 \pm 18$ & -- & -- & $9 \pm 2$ & $12 \pm 2$ &  -- & -- \\
NGC\,3281  & -- & -- & -- & -- & -- & $14 \pm 2$ &  -- & -- \\
NGC\,4253  & $43 \pm 7$ & $60 \pm 9$ & $64 \pm 10$ & -- & $6 \pm 1$ & $9 \pm 1$ &  $13 \pm 2$ & -- \\
NGC\,4569  & $307 \pm 47$ & $396 \pm 61$ & -- & -- & -- & -- &  -- & -- \\
NGC\,5135  & $4 \pm 1$ & $17 \pm 3$ & $60 \pm 9$ & $88 \pm 13$ & $13 \pm 2$ & $21 \pm 3$ & $32 \pm 5$ & $34 \pm 5$ \\
NGC\,5643  & $26 \pm 4$ & $74 \pm 11$ & -- & -- & $13 \pm 2$ & $15 \pm 2$ &  -- & -- \\
IC\,4518W   & -- & -- & -- & -- & -- & $7 \pm 1$ &  $10 \pm 2$ & $10 \pm 2$ \\
IC\,5063      & -- & -- & -- & -- & -- & $15 \pm 2$ &  -- & -- \\
NGC\,7130  & -- & $65 \pm 10$ & $90 \pm 14$ & $145 \pm 22$ & -- & $5 \pm 1$ &  $6 \pm 1$ & $6 \pm 1$ \\
NGC\,7172  & $5 \pm 1$ & $11 \pm 2$ & $33 \pm 5$ & -- & $5 \pm 1$ & $5 \pm 1$ &  $5 \pm 1$ & -- \\
NGC\,7465  & $35 \pm 5$ & $44 \pm 7$ & -- & -- & -- & -- &  -- & -- \\
NGC\,7582  & $26 \pm 4$ & $56 \pm 9$ & $179 \pm 27$ & $181 \pm 28$ & -- & -- &  -- & -- \\
\hline 
\end{tabular} 
}
\label{tab:measurenw}
\end{footnotesize}
\end{center}
\tablecomments{ These measurements have been obtained from interpolation at different distances from the nucleus (see text). Note that the symbol ``--''  indicates that we do not consider the measurement because the interpolated value is within nuclear radii or at larger radii than our outer radius for the sources.}
\end{table*}

\subsection{PAH feature and [SIV] line measurements}
\label{sec:measurementslines}

There are several methods to measure the fluxes of the PAH features. The best approach depends on the spectrum characteristics. For instance, PAHFIT \citep[][]{Smith07} or DecompIR \citep[][]{Mullaney11} are able to measure the PAH features and are very useful when the spectra are highly contaminated by its host galaxy emission. However, they require a wide spectral coverage in order to produce satisfactory results -- larger than that of the T-ReCS or CanariCam spectra presented here \citep[see][]{Esquej14}.
Instead, we followed the procedure described by \citet{Alonso-Herrero14} and \citet{Esquej14} to measure the flux and the equivalent width (EW). They use the method described by \citet{Hernan-Caballero11}, which is well suited for limited wavelengths (case of [SIV] line) or weak PAHs. Their method sets a local continuum by interpolating from two narrow bands (i.e. $\rm{10.7 - 10.9~\mu m}$ and $\rm{11.7-11.9~\mu m}$) at both sides of the PAH feature or at both sides of the [SIV] line emission (i.e. $\rm{10.35-10.40~\mu m}$ and $\rm{10.65-10.75~\mu m}$). Note that we selected these continuum ranges individually according to the particularities of each spectrum. This was done to optimize the measurement of the bands according to the natural width of the PAH feature.
After subtracting the underlying continuum, residual data were fitted using a Gaussian profile. We compared the fluxes obtained from the Gaussian fit and the direct integration in the case of the nuclear spectra. The discrepancy in the flux between the two methods for the nuclear spectrum is in average less than $3\%$ and $7\%$ for the PAH feature and the [SIV] line, respectively.
 
Then, the EW of the lines is measured by dividing the integrated flux by the interpolated continuum flux at the center. The uncertainties are obtained by Monte Carlo simulation using the calculated dispersion around the flux measurements. We have applied a smoothing to the high spatial resolution spectra to improve the signal-to-noise of the features. This smoothing was applied to the data using the average of three near spectral bins. The smoothing causes a peak dilution, which could dilute the emission lines if they are less than three points. Nevertheless, the lines that we studied are broad, therefore, we do not expect to have any significant effect on the results \citep[see][for more detail on the smoothing technique]{Alonso-Herrero14}.
Table \ref{tab:measure} shows integrated fluxes and EW measurements from each emission obtained with the nuclear and IRS/\emph{Spitzer} spectra.

\subsection{Surface brightness radial profiles}

We create surface brightness and EW radial profiles\footnote{We use the term radial profile for referring to the surface brightness radial profiles} for each object. We first extracted the flux at the radius of each aperture, and then we subtracted that of all inner apertures to get the flux of a ring. When the subtracted measurement was lower than 3$\sigma$ we considered it as a limit.
We have then divided each value by their respective area to correct for different aperture radii. In the case of the nucleus, the area is computed with the radius of the unresolved emission times the slit width. For the rest of the apertures, the area is calculated as the slit width times increment radius for the aperture (ie. 2 pixels, see Table\ref{tab:observations}, Col.7).

Figure \ref{fig.2} (bottom) shows the radial profile for the PAH feature at $\rm{11.3~\mu m}$ (blue diamonds) and [SIV] line emission at $\rm{10.5~\mu m}$ (orange circles).
Appendix \ref{sec:Cat} includes the radial profiles for the full sample.

In order to analyze the behaviour of the two emission features across the full sample, we calculated the integrated flux at fixed physical scales: 50\,pc, 100\,pc, 200\,pc, 500\,pc, and 1kpc. The measurements were calculated from a linear interpolation between the nearest points. Notice that we do not take into account the nuclear measurement to compute  these values at a fixed distance. These measurements are reported in Table \ref{tab:measurenw}.  We report measurements only when our radial profile includes these distances.

\begin{figure*}
\includegraphics[scale=0.30]{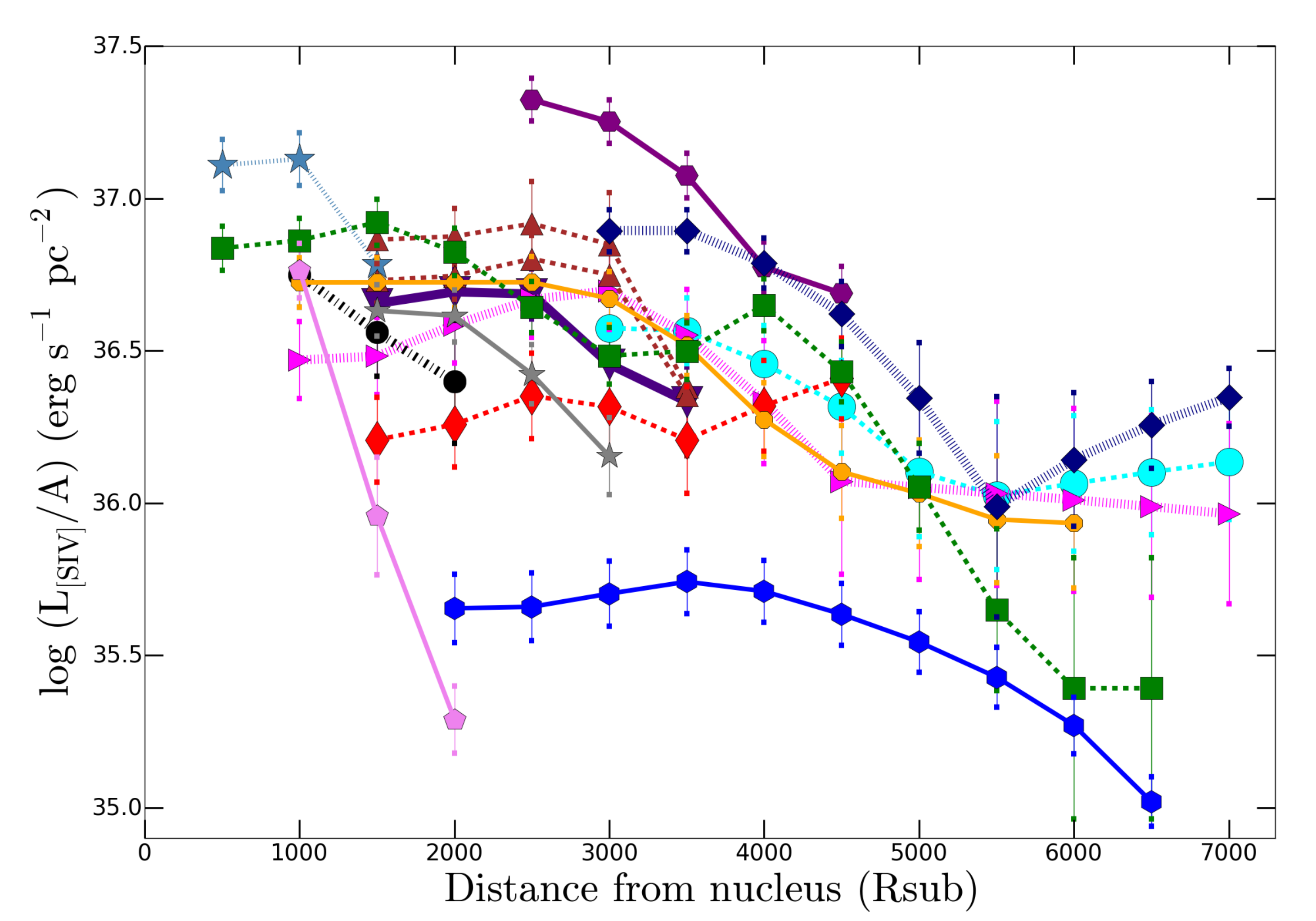}
\includegraphics[scale=0.30]{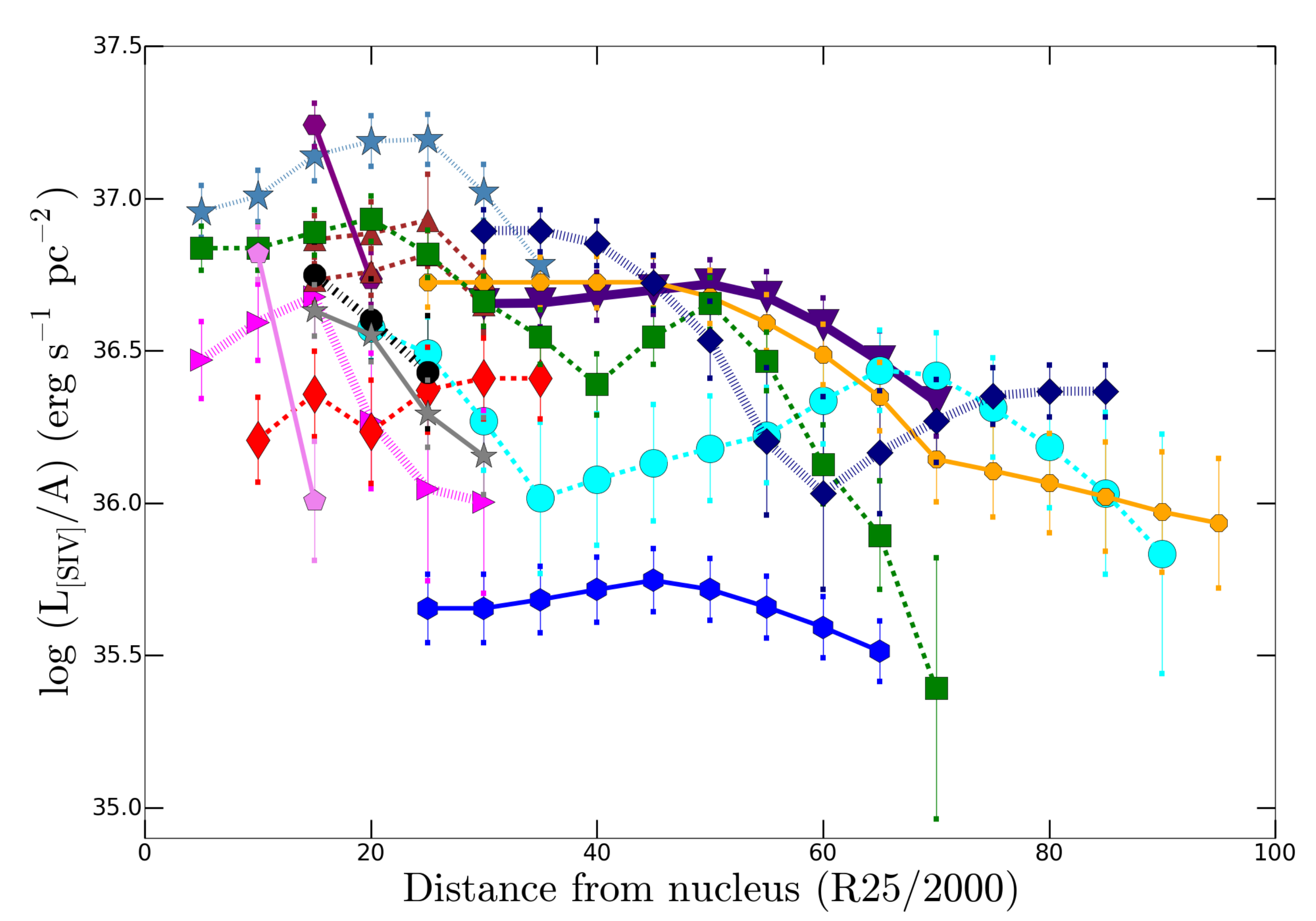}
\caption{Left: Luminosity of the [SIV] line as a function of distance from the nucleus in units of the sublimation radius. Right: Luminosity of the [SIV] line as a function of the distance from the nucleus in units of the isophotal radius divided by 2000 for each galaxy. In both figures, the symbols are measurements at fixed distances. The different lines link all the measurements for each object: 1)NGC\,931 (steel blue stars), 2) Mrk\,1066 (indigo triangles down), 3)NGC\,1320 (cyan circles), 4) NGC\,1386 (purple hexagons), 5) NGC\,2992 (magenta triangles right), 6) NGC\,3081 (brown triangles up), 7) NGC\,3227 (red thin diamonds), 8)NGC\,3281 (black circles), 9) NGC\,4253 (orange octagons), 10) NGC\,5135 (green squares), 11) NGC\,5643 (violet pentagons), 12) IC4518W (navy diamonds), 13) IC\,5963 (grey stars), and 14) NGC\,7130 (blue hexagons).}
\label{fig.3}
\end{figure*}

\begin{figure*}
\includegraphics[scale=0.25]{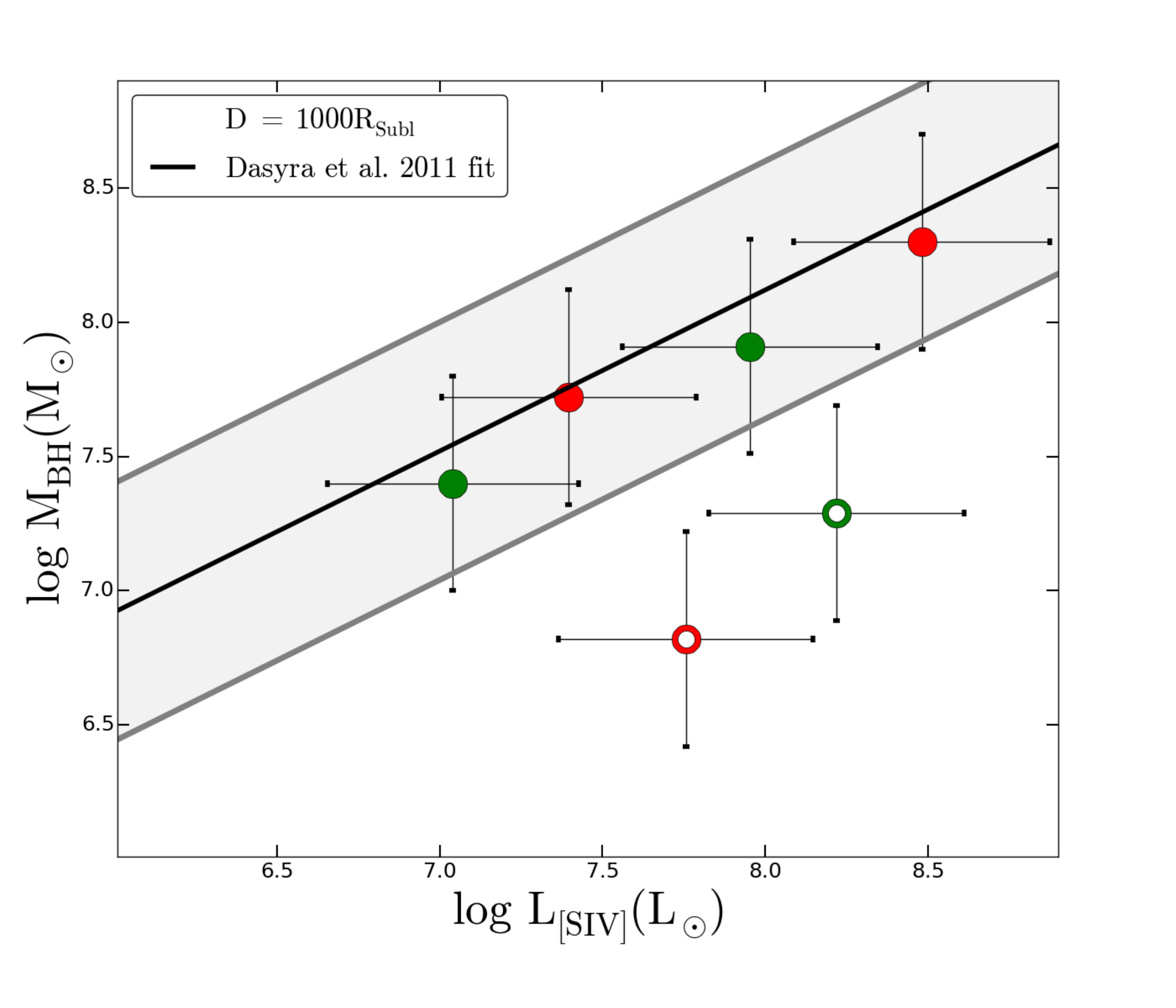}
\includegraphics[scale=0.25]{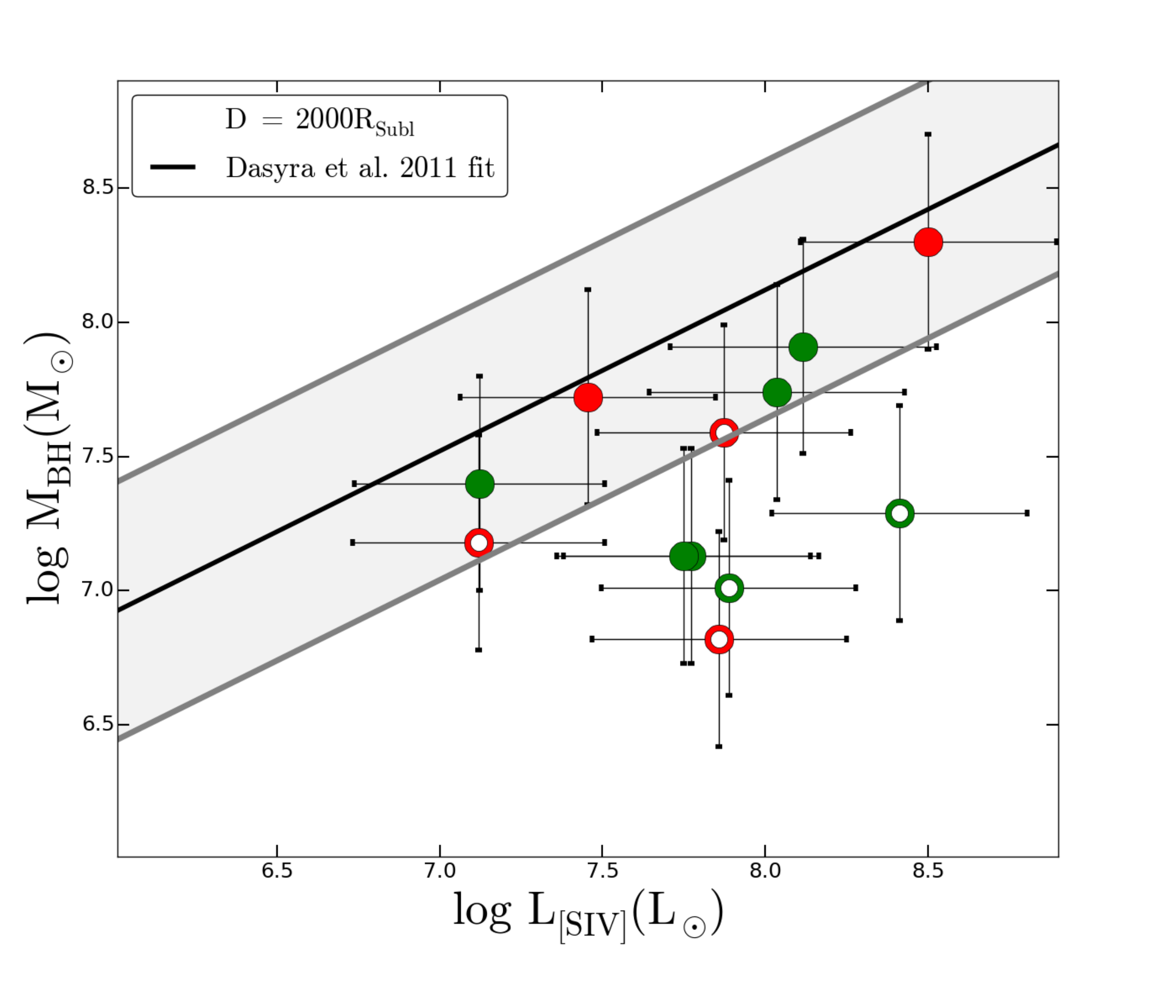}
\includegraphics[scale=0.25]{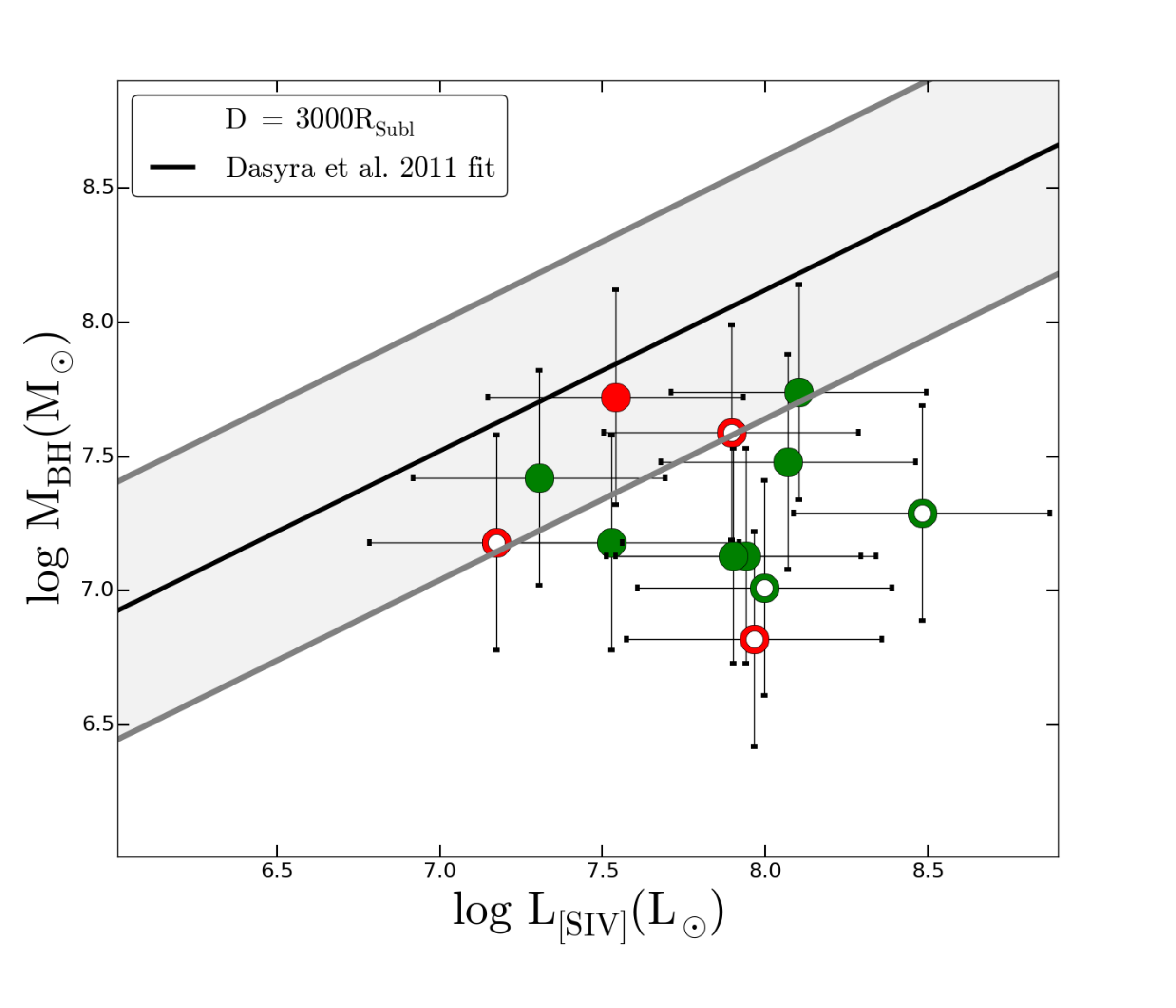}
\caption{The [SIV] line emission luminosity versus M$\rm{_{BH}}$ at 1000, 2000, and 3000 $\rm{R_{sub}}$. The white dots are sources where SF regions were previously reported at these spatial scales. The Sy1 and Sy2 are shown as red and green dots, respectively. The relation found by \citet{Dasyra11} is shown as a black-solid line in all panels.}
\label{fig.3.1}
\end{figure*}

\section{The origin of the [SIV] line emission}
\label{subsec:OSIV}
The [SIV] is an emission line typically observed in the planetary nebula, HII regions, and ULIRGs \citep[][]{Rank70,Holtz71,Gillett72} as well as AGN. The origin of the nuclear [SIV] line emission is controversial in the case of AGN. It can be produced in the NLR, and therefore be a good tracer of gas ionized by the AGN \citep[][]{Dasyra11}. However, it can also be related to star-forming regions due to its relatively low excitation potential \citep{Pereira10, Diaz10}.

\citet{Diaz10} studied four LIRG-type objects finding that half of the [SIV] line emission flux comes from the nucleus. Our sample has three objects in common with theirs (NGC\,5135, IC\,4518W, and NGC\,7130).
For NGC\,5135 they found that $\sim 40\%$ of [SIV] line emission integrated flux comes from the nuclear spectrum. Fairly consistent with that, we find that the nuclear spectrum contributes $\sim 35\%$ to the integrated flux of this emission line. They found that the [SIV] nuclear flux in IC\,4518W is smaller than the emission in the integrated spectrum by $\sim 22\%$.
We also agree that there is an excess of [SIV] emission at $\rm{0.5~arcsec ~(\sim 200}$\,pc), which is unrelated to the excess of $\rm{11.3~\mu m}$ PAH emission. \citet{Diaz10} suggested that this emission is associated with the central AGN.
 In NGC\,7130 we found that $70\%$ of the [SIV] flux comes from the nuclear spectrum, while \citet{Diaz10} found that the [SIV] nuclear emission corresponds to $50\%$ of the total flux.
In both IC\,4518W and NGC\,7130, star-forming regions near the nucleus have been found (see Appendix \ref{sec:Cat}). Based on the \emph{Spitzer} observations, \citet{Pereira10} could not conclude whether the [SIV] line emission is related with star-forming regions for this object due to poor data quality.
 However, they found that extended emission of the [SIV] line can be attributed to star-forming regions, using P$\alpha$ and H$\alpha$ images for the other three objects of their sample.

We explored the luminosity of the [SIV] emission line ($\rm{L_{[SIV]}}$) in the AGN environment by studying the radial profiles of $\rm{L_{[SIV]}/L_{Edd}}$ as a function of the sublimation radius \citep[$\rm{R_{sub}}$,][]{Nenkova08}. The latter was computed as:

\begin{equation}
R_{sub} \simeq 0.4 \left( \frac{L}{10^{45}\, erg^{-1}}\right)^{1/2} \left( \frac{1500K}{T_{sub}}\right)^{2.6}\,pc
\end{equation}

\noindent
where is assumed as $\rm{T_{sub} = 1400\,K}$.

We interpolated the given values of $\rm{L_{[SIV]}}$ to obtain measurements at the following distances from nucleus: 1000, 2000, 3000, 4000, 5000, 6000, and 7000 $\rm{R_{sub}}$.
This allowed us to compare the $\rm{L_{[SIV]}}$ at the same spatial scales. Figure \ref{fig.3} (left) shows the radial profiles of the $\rm{L_{[SIV]}}$ as a function of $\rm{R_{sub}}$ using the new measurements for the 13 AGN in our sample where we detected the [SIV] emission line in more than two apertures. The number of values included in the radial profiles vary due to the minimum and maximum distances from the nucleus that we can trace.
In general, the radial profiles in Figure \ref{fig.3} (left) show a chaotic behaviour.
It might be plausible that these profiles strongly depend on the gas suppliers around each AGN, adding scatter to the expected behaviour. The proper comparison between available gas around AGN and the [SIV] emission needs to be studied prior any further conclusions. Figure \ref{fig.3} (right) shows the radial profiles of the [SIV] line as a function of isophotal radius ($\rm{R_{25}}$) of the galaxy\footnote{The isophotal diameter of galaxy is the decimal logarithm of the length the projected major axis of a galaxy at isophotal level 25 mag/arcsec$^2$ in the B-band. See http://leda.univ-lyon1.fr/leda/param/logd25.html}. We found a similar chaotic behaviour.

\citet{Dasyra11} used the [NeV], [OIV], [NeIII], and [SIV] line emissions to study the kinematics of the NLR. They concluded that the M$\rm{_{BH}}$ and the gas velocity dispersion are related to the luminosity of these emission lines originating in the NLR. We studied the relationship between $\rm{M_{BH}}$ and the luminosity of the [SIV] line emission ($\rm{L_{[SIV]}}$) to interpret the origin of the line. Figure \ref{fig.3.1} shows this relation at 1000, 2000, and 3000 $\rm{R_{sub}}$ for the sources where we detected the [SIV] line emission at these scales. The solid line corresponds to the \citet{Dasyra11} relation:

\begin{equation}
\rm{log~(M_{BH}) = 0.6\times log~(L_{[SIV]}) + 3.32}
\label{eq:SIV}
\end{equation}
\noindent
This relation is based on the best-fit for their AGN sample using IRS/\emph{Spitzer} spectra and considering that the [SIV] line emission only comes from the NLR. The rms scatter computed for this relation is 0.48 dex (shaded area in Figure \ref{fig.3.1}).
For a few sources without significant star formation, the [SIV] line fluxes follow the \citet{Dasyra11} relation at scales of $\rm{\sim 1000\,R_{sub}}$. However, the sources move away from the relation with increasing distance from the nucleus. This result could be interpreted as HII regions, planetary nebulae or blue compact dwarfs contributing to the sulphur excitation, alongside with the AGN \citep[][]{Groves08}.
 
Even if the nuclear [SIV] line emission could arise from photoionization by the AGN in some of our sources, it could be strongly suppressed by dust because it is inside the broad $9.7 \mu$m silicates absorption feature \citep{Pereira10}. Therefore, it could not be an isotropic measurement of the AGN luminosity.
Moreover, the obscuration of the internal parts of the AGN by the dusty torus could also play a major role in the [SIV] line emission attenuation. This could be the case for NGC\,7172, showing a large value of the 9.7$\rm{\mu m}$ optical depth \citep[$\rm{\tau_{9.7\mu m}=1.9}$,][]{Gonzalez13}. Indeed a very weak detection of the [SIV] line emissions has been found for this object. We have considered the possibility that attenuation is affecting the [SIV] line emission in the inner parts. We found a deficit between nuclear and the first apertures in five sources (see Appendix \ref{sec:Cat}). We have compiled the nuclear $\rm{\tau_{9.7\mu m}}$ from \citet{Gonzalez13} and \citet{Alonso15}, but we did not find any relation between the $\rm{\tau_{9.7\mu m}}$ and the deficit on the [SIV] line emission flux. Also, \citet{Dasyra11} found that this obscuration does not significantly affect the relative flux of MIR lines. In summary, in six of the 13 sources we did not observe a common decrease in the radial profile, as we would expect if this line was caused by AGN photoionization.

\subsection{The [SIV] emission line versus the $\rm{11.3~\mu m}$ PAH feature.}

The [SIV] emission line could be produced by star-forming regions. If this is the case, we would expect a close resemblance between the [SIV] and PAH radial profiles at these radii.

We compared nine sources where the radial profiles of both the PAH feature and the [SIV] line show more than one measurement at different distances from the nucleus. In all sources, the radial profile for both emissions shows a complex behavior. In six\footnote{NGC\,2992, NGC\,3227, NGC\,5135, NGC\,5643, NGC\,7172, NGC\,7465.} of these nine sources it is clear that the behaviour of the radial profiles of both emissions are not related to each other at any distance.
Even with that, it could be the case that the star-forming regions traced by the $\rm{11.3~\mu m}$ PAH feature are not the same as those that give origin to the [SIV] emission in the majority of the sources.
A plausible explanation is that both emissions are tracing different stages of SF, and thus different degrees of ionizing fluxes. Ideally, to distinguish the type of stars that contribute to the emission of [SIV] line emission, high spatial resolution images of [SIV] line emission together with other tracers of SF related to different stages of the SF activity would be needed.

\begin{figure*}
\includegraphics[scale=0.30]{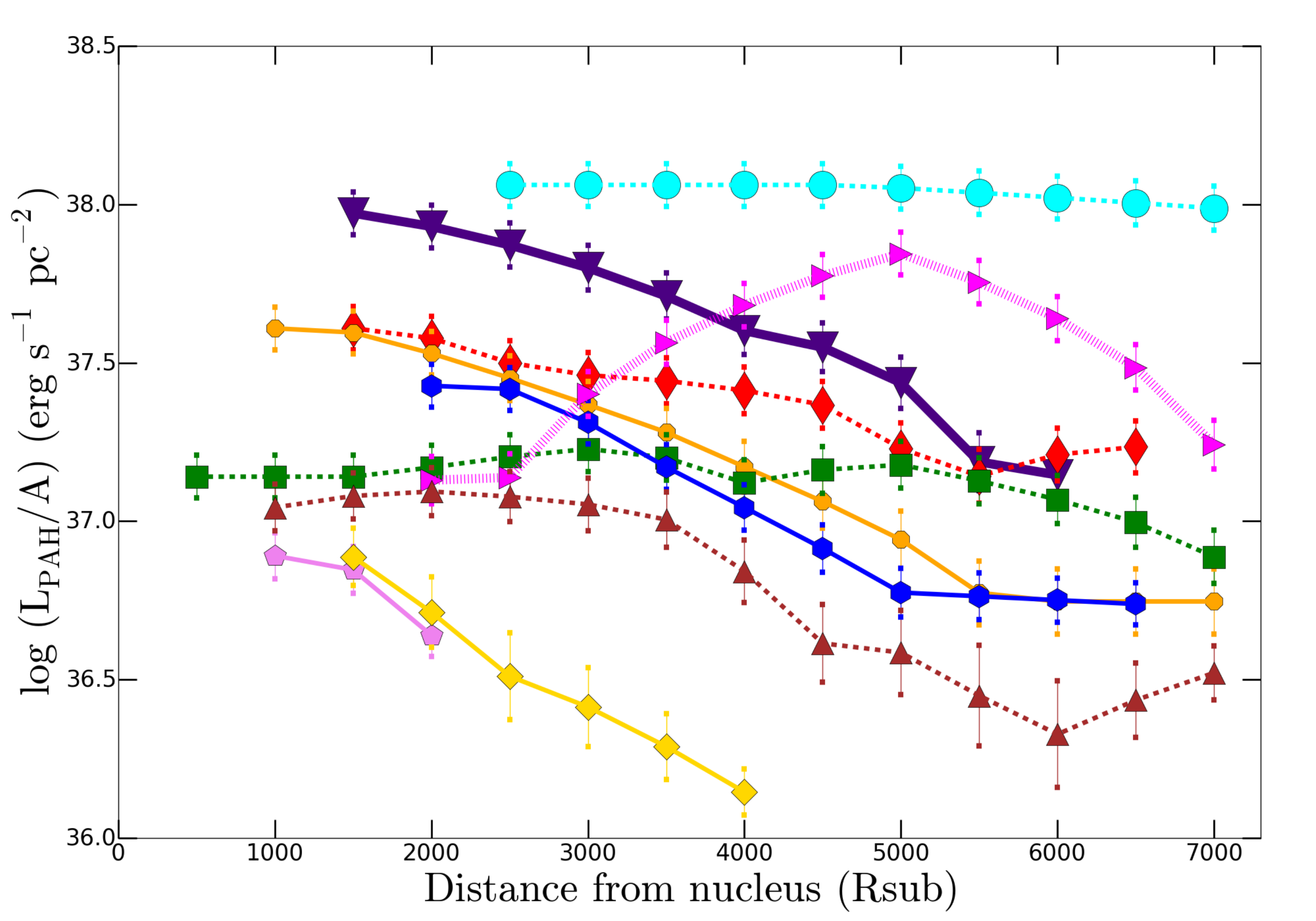}
\includegraphics[scale=0.30]{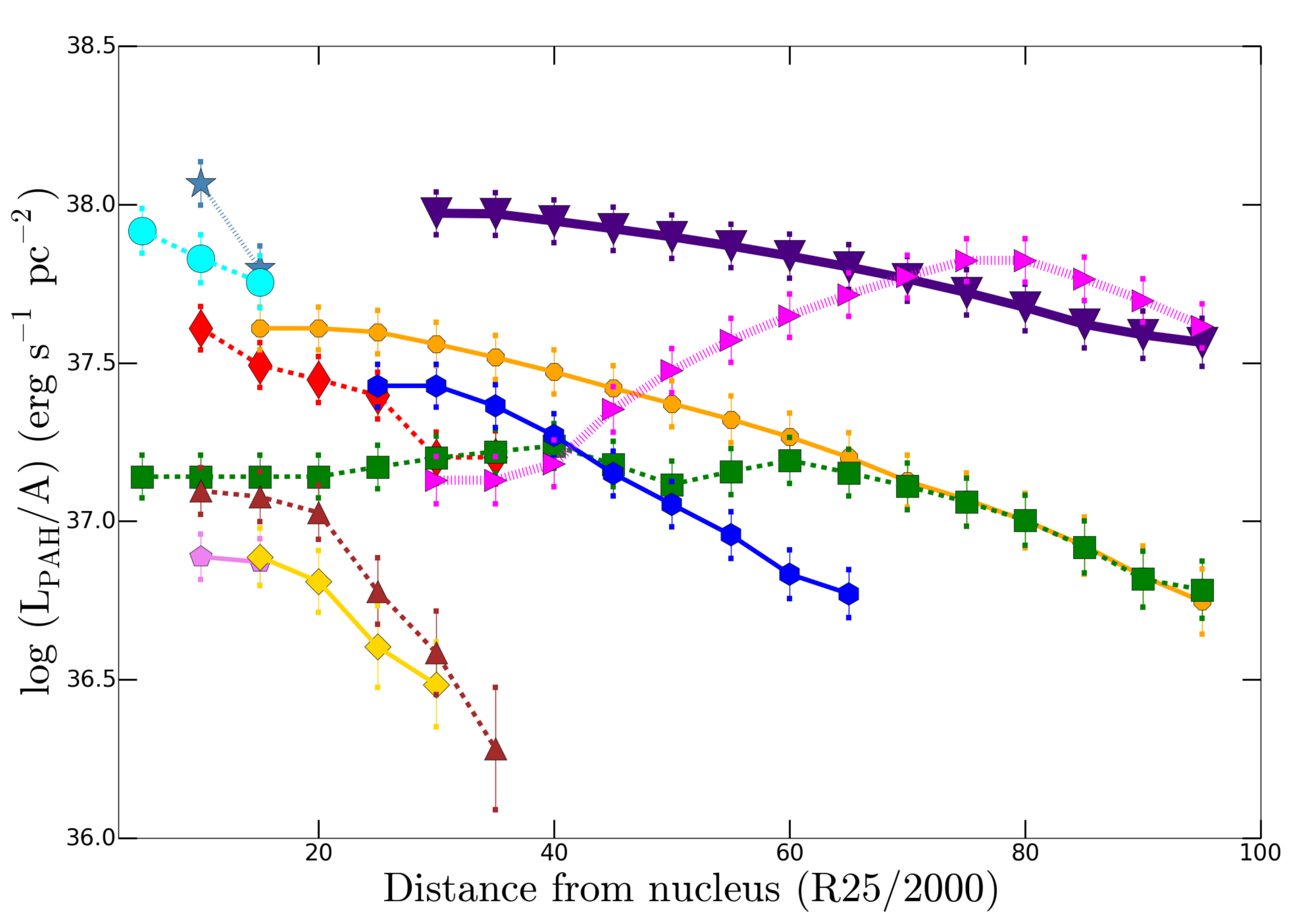}
\caption{Left: Luminosity of the $\rm{11.3\,\mu m}$ PAH as a function of distance from the nucleus in units of the sublimation radius. Right: Luminosity of the $\rm{11.3\,\mu m}$ PAH as a function of the distance from the nucleus in units of the isophotal radius divided by 2000 for each galaxy. In both figures, the symbols are measurements at fixed distances. The different lines link all the measurements for each object: 1) Mrk\,1066 (indigo triangles down), 2)NGC\,1808 (steel blue stars), 3) NGC\,3227 (red thin diamonds), 4) NGC\,4253 (orange octagons), 5) NGC\,4569 (cyan circles), 6) NGC\,5135 (green squares), 7) NGC\,5643 (violet pentagons), 8) NGC\,7130 (blue hexagons), 9) NGC\,7172 (gold diamonds), 10) NGC\,7465 (magenta triangles right), and 11) NGC\,7582 (brown triangles up)}.
\label{fig.3PAHs}
\end{figure*}

\section{The behaviour of the PAH emission feature}
\label{subsec:DiscPAHs}
In this section, we review the plausible dilution/destruction of PAHs in the innermost parts of the AGN (Sec. \ref{subsec:DPAHs}) and we use PAHs as tracers of SF to study the coevolution of the AGN and its host galaxy (Sec. \ref{subsec:Coev}).

\begin{figure}
\begin{center}
\includegraphics[scale=0.4]{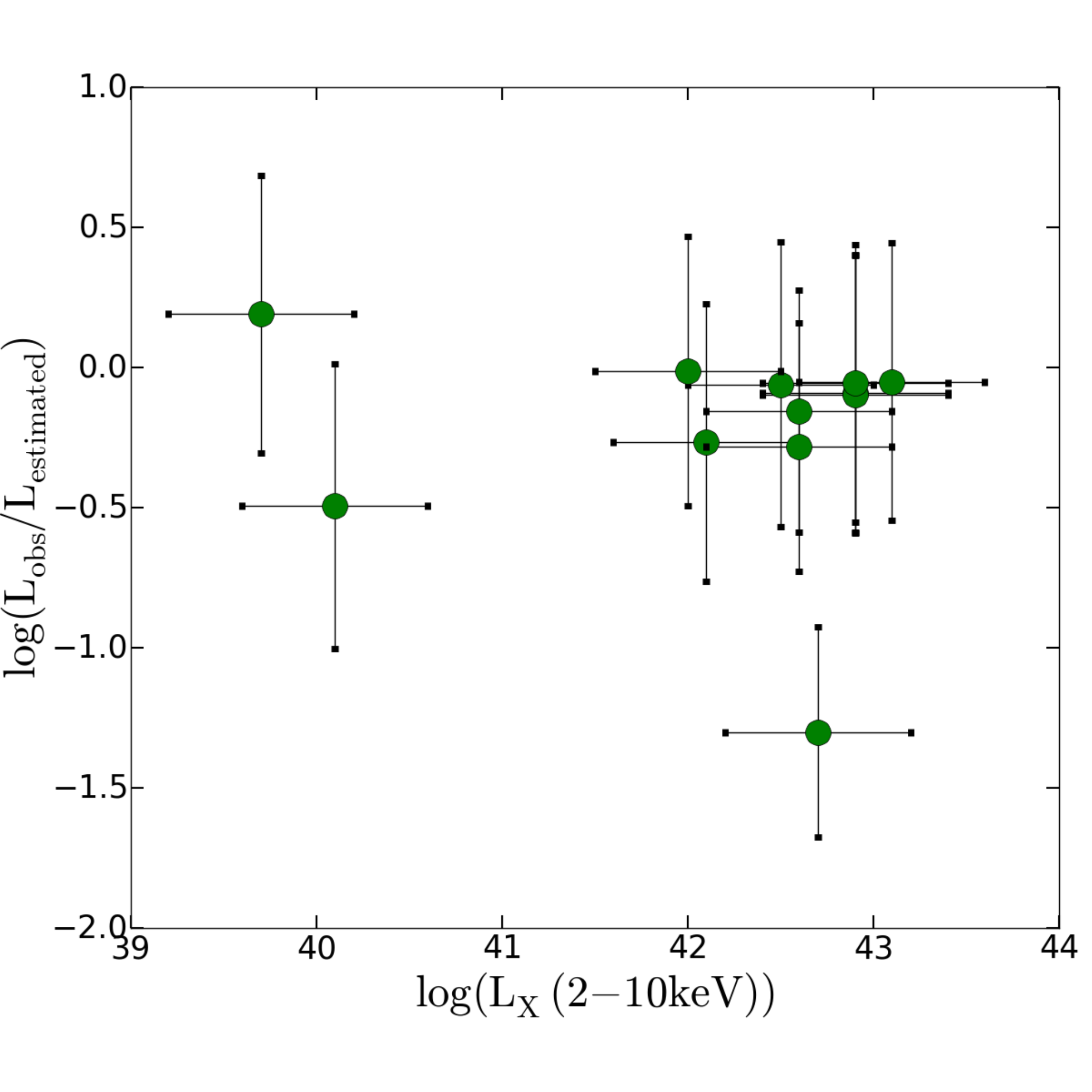}
\caption{X-ray luminosity versus PAH luminosity deficit (see text). This deficit is measurement as the ratio between the observed and the expected one. The expected PAH luminosity expected is estimated as the linear extrapolation to the center of the radial profile within 200\,pc.}
\label{Fig.LumXvsPAH}
\end{center} 
\end{figure}

\subsection{On the dilution/destruction of the nuclear PAHs}
\label{subsec:DPAHs}

The relation between the strength of the PAHs and IR luminosity is weak or absent in galaxies with AGN \citep{Weedman05, Siebenmorgen04}. An important implication of this is that PAHs might not be used as star-forming tracers in the surroundings of the nucleus because they can be destroyed by the AGN radiation field \citep[][]{Siebenmorgen04}. The AGN can directly modify PAH grain size distribution and even serve as the excitation source for some PAH emission \citep{Genzel98, Laurent00, Smith07}. On the other hand, PAHs could survive because they are shielded from the AGN radiation \citep{Goulding12}. Even more extremely, PAH could be induced by the AGN radiation field \citep[][]{Jensen17}.
\citet{Diamond10} found that the 6.2, 7.7 and $\rm{8.6~\mu m}$ PAH features are suppressed respect to the $\rm{11.3~\mu m}$ PAH feature in local Seyferts. They speculate that destruction of these features might be related to the fact that they are produced by the smallest aromatic molecules and, therefore, more easily destroyed. Following this argument, the molecules responsible of $\rm{11.3~\mu m}$ PAH emission could survive because they are more difficult to destroy. Already from IRAS data, it was pointed out that the emission at the $\rm{12~\mu m}$ band fits very well the predictions that follow from the emission modeling of transiently heated PAH molecules \citep{Dultzin89}. More recently, \citet{Diamond12} found a correlation between the nuclear SF ($\rm{<~1\,kpc}$) and SMBH accretion rate, where the nuclear SF is traced by the PAH at $\rm{11.3~\mu m}$ aromatic feature.

We detected the PAH feature at $\rm{11.3~\mu m}$ in 15 out of the 19 objects in our sample ($\sim 90\%$ of our sources) and 10 of these sources show nuclear PAHs ($\sim 58\%$ of our sample).  The $\rm{11.3~\mu m}$ PAH feature was measured in more than one aperture in 11 objects along the radial profiles\footnote{Another four objects of the sample show emission only in one aperture (NGC\,1386, NGC\,2992, NGC\,3081, and IC\,4518W). The other four sources do not show detection of the $\rm{11.3~\mu m}$ PAH feature.}. We found that in eight sources (except by NGC\,5643, NGC\,7172, and NGC\,7582) the nuclear EW of the PAH is larger than the one found in the first aperture\footnote{Note that we do not take into account IC\,4518W because the measurements at distances $< 400$\,pc are only upper limits.}.

In Figure \ref{fig.3PAHs}, we show the radial profiles of the $\rm{L_{11.3\,\mu m\, PAH}}$ as a function of $\rm{R_{sub}}$ (left) and $\rm{R_{25}/2000}$ (right). This figure is similar to Figure \ref{fig.3} from the previous section. We observed a complex behaviour. Increments and decrements at different distances were found.

Regarding the $\rm{11.3~\mu m}$ PAH nuclear flux are larger than those of the first aperture only in NGC\,1808 and NGC\,5135. The unresolved nucleus show lower than the first aperture PAH flux in most of our cases (seven out of 12, i.e 60\%). When observed, this decrement is seen within $\rm{\sim 100}$\,pc. Note that in many cases we do not see a decrease in the radial profile (as for example in NGC\,7582), but a drop between the nuclear and the first aperture (e.g. NGC\,7172).
Therefore, this decrement could be affecting even lower spatial scales. The explanation of this decrement are: 1) PAH dilution by AGN continuum\footnote{We refer to dilution as a decrease in equivalent width from the PAH feature due to the strength of the AGN continuum.}, 2) PAH destruction by the radiation field, 3) lack of the inner star formation, and 4) the existence of a nuclear ring. In the following we discuss these four possibilities.

\citet{Alonso-Herrero14} suggested that the apparent decrease in the EW of the PAH feature is an effect of the dilution of the PAH feature by the strong continuum of the AGN in the nuclear apertures. They indeed recovered an increase on the nuclear PAH flux toward the center in their sample of six local AGN. Meanwhile, the EW of the PAH feature showed an apparent decrease. We have not found a similar behaviour in any of our objects, but we only have one object in common with their analysis (MRK\,1066).
They computed the radial profile in isolated apertures at different distances from the nucleus. In our analysis, we have extracted spectra centred at the nucleus with different radii. Thus, each of our apertures includes the nuclear emission. In order to study the radial profile, we subtracted the previous inner aperture scaled to the area (see Section \ref{sec:SpecAnalysis}). This way, we avoided the dilution due to this effect. Thus, dilution cannot play a role in the lack of nuclear PAHs in the sources analysed here.

\citet{Siebenmorgen04} suggested that the suppression of PAH emission near the AGN may be due to the destruction of PAHs by the strong radiation field of the AGN. If this is the case, we would expect a relation between the PAH luminosity deficit and the X-ray luminosity as a tracer of the AGN bolometric luminosity. The stronger the AGN radiation field, the larger the nuclear PAH deficit. We have measured the PAH luminosity deficit from our radial profiles as the ratio between the expected and the observed one. We have estimated the expected nuclear PAH luminosity into two ways: (1) as the linear extrapolation of the radial profile within 200 pc; and (2) as the maximum of PAH emission within 200 pc. Figure \ref{Fig.LumXvsPAH}, shows the deficit obtained by extrapolation versus the X-ray luminosity. We do not find a relation between the PAH deficit and the AGN X-ray luminosity. Thus, from our data we have not found observational support for the destruction of the PAH features due to the AGN radiation field. However, we cannot rule out this hypothesis since more sensitive and better resolution observations are needed. For instance higher spatial resolution spectra could help to pinpoint the distance from the nucleus at which the PAH emission starts to show this deficit. In this sense it might be possible that the relation is missing due to a poor estimate of the PAH luminosity deficit.

Of course, a natural explanation of this inner deficit in the PAH feature is that there is a lack in SF toward the center. This is supported by the scenario in which the high-velocity winds or AGN-driven massive molecular outflows could be able to quench the surrounding SF \citep[][]{Cicone14,McAlpine15,Wylezalek16}. Another possible explanation for this deficit in PAHs in internal parts can be related to the dust/gas distribution which is ring-like rather than disk-like at the center \citep[e.g.][]{Ohsuga99,Yankulova99}.
In order to corroborate this, other measurements of the nuclear tracers of the SF must be compared with our PAH nuclear fluxes, isolating nuclear and circumnuclear emission.

\subsection{Hints on the coevolution of the AGN and its host galaxy}
\label{subsec:Coev}

\citet{Hopkins10} and \citet{Neistein14} have explored the correlation between BH accretion rate and the SFR through hydrodynamic simulations and semi-analytic models, respectively. 
\citet{Hopkins10} predicted the relation between BH accretion rate and SFR at different galactic scales. Their simulations start with a major galaxy merger of isolated bar-(un)stable disc galaxies. They found that nuclear SF is more coupled to AGN than the global SFR of the galaxy.
\citet{Neistein14} developed similar simulations including advection dominated accreting flow to account for the accretion processing low-luminosity AGN.
They observed a lack of correlation between SFR and AGN luminosity (related with BH accretion rate) at $z<1$ and $\rm{L_{bol,AGN}<10^{44}~erg~s^{-1}}$ \citep[see also][]{Rosario12}. They justified this possible lack of correlation as: 1) secular SF is perhaps not associated with BH accretion, or 2) BH accretion rate and SFR could be delayed removing any correlation \citep[see also][]{Hopkins12}. They also found that AGN with low or intermediate luminosity might be associated with minor merger events.

In this work we compare \citet{Hopkins10} and \citet{Neistein14} predictions with our results. We derived nuclear and circumnuclear SFRs using the PAH $\rm{11.3~\mu m}$ feature luminosities ($\rm{L_{\rm{11.3~\mu m}}}$) and applying the relation derived in \citet{Shipley16} (using \emph{Spitzer} measurements of 105 galaxies):
\begin{eqnarray}
\rm{log~SFR (M_{\odot}~yr^{-1})} &=& \rm{(-44.14 \pm 0.08)} \nonumber \\
&+& \rm{(1.06 \pm 0.03)~log~L_{11.3~ \mu m}} (\rm{erg~s^{-1}})
\label{eq:SFR}
\end{eqnarray}
\noindent
The uncertainties in the derived SFRs using equation (\ref{eq:SFR}) are typically 0.14\,dex \citep[see][for full details]{Shipley16}. As a caveat on the use of the PAH as a tracer of SF, \citet{Jensen17} recently found that the slope of the radial profile of the PAH emission are very similar, with a strength proportional to the AGN luminosity. They argue that this might imply that a compact emission source is required to explain the common slopes. Both an AGN or a nuclear star cluster are possible sources of PAH heating/excitation. Although we obtain in general a decrease of the PAH flux with the radius, a more complex  behaviour (with a deficit at the nuclear and peaks of emission on top of general decrease) is observed in most of our sources indicating in situ PAH heating.

This is not the first time such a comparison have been done. \citet{Esquej14} used a sample of 29 nuclear spectra to explore the same relation between SFR and BH accretion rate. They compared their data with the relations obtained by \citet{Hopkins10}, and they concluded that predictions for distances (D) $\rm{< 100}$\,pc reproduce their data well. We have seven sources in common with their sample\footnote{NGC\,1808, NGC\,3227, NGC\,5135, NGC\,5643, NGC\,7130, NGC\,7172, and NGC\,7582.}. Our measurements show slightly higher SFR compared to theirs (factor 2), perhaps due to a different methodology to define the continuum around the PAH feature. \citet{Ruschel16} also analyzed the presence of circumnuclear SF in a sample of 15 AGN using MIR images (with two filters centered at the $\rm{11.3~\mu m}$ PAH features and at the adjacent continuum, respectively). They compared their data with the correlation presented by \citet{Neistein14}. They concluded that SFR is correlated with bolometric AGN luminosity ($\rm{L_{bol,AGN}}$) for objects with $\rm{L_{bol,AGN}~\geq~10^{42} ~erg~s^{-1}}$, while the low luminosity AGN has larger SFR for their $\rm{L_{bol,AGN}}$. 

\begin{table}[!b]
\caption{Comparison with models by $\rm{11.3~\mu m}$ PAH. }
\begin{center}
\begin{tabular}{cccc}
\hline \hline
Distance & Theory & \multicolumn{2}{c}{Measurement} \\ 
              &              & Mean   & $\sigma$ \\ 
\hline \hline
100\,pc & 1.00 & 1.09 & 0.60 \\  
200\,pc & 1.06 & 0.88 & 0.70 \\  
500\,pc & 1.23 & 0.91 & 0.75 \\ 
700\,pc & 1.35 & 0.95 & 0.80 \\ 
\hline \hline
\end{tabular} 
\label{tab:offset}
\end{center}
\tablecomments{ For the PAH feature, we have computed the observed shift for the relation as the average and standard deviation of the relation predicted by \citet{Neistein14}. For more information see Section \ref{subsec:Coev}. }
\end{table}

\begin{figure*}
\begin{center}
\includegraphics[width=1\textwidth]{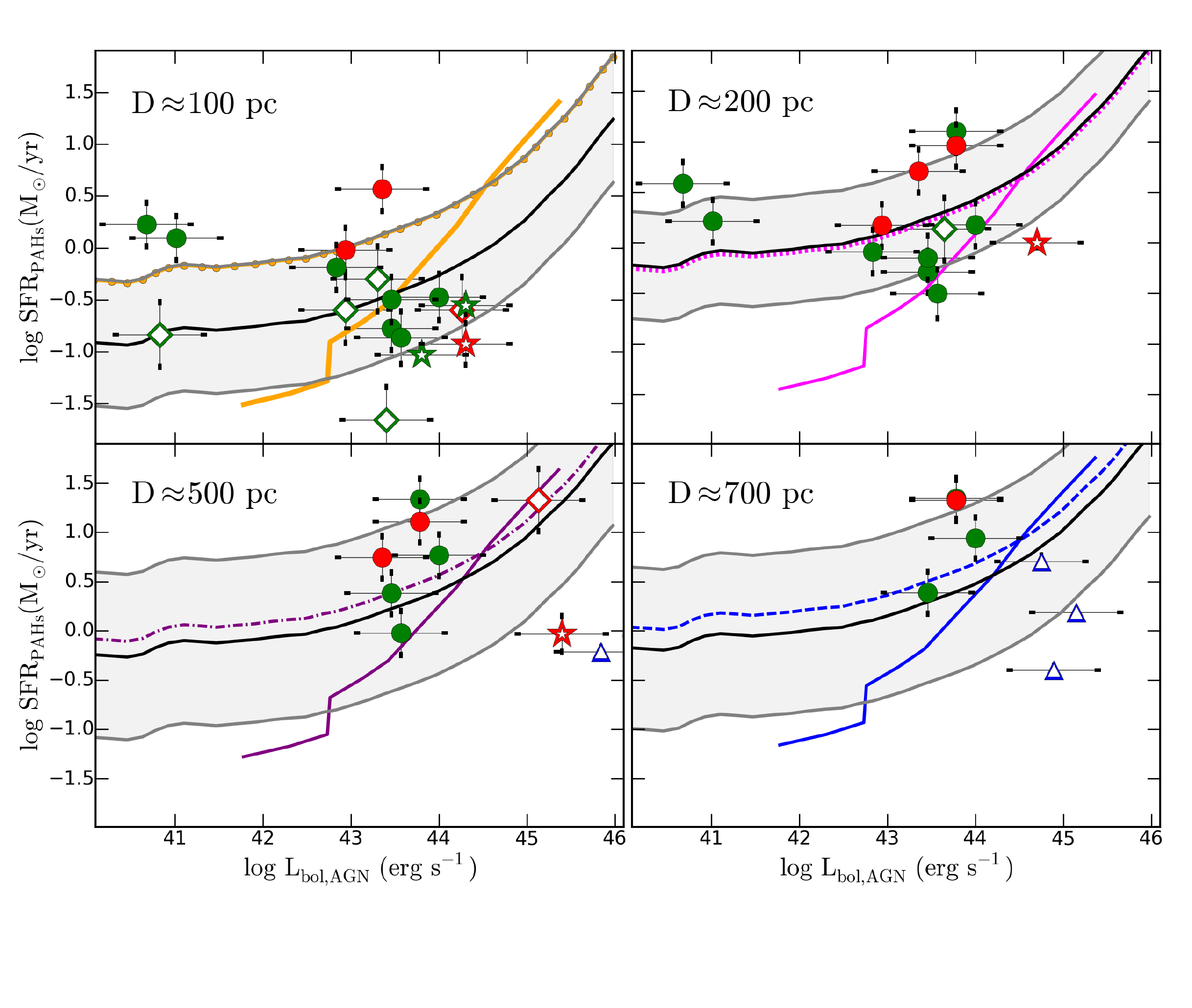}
\vspace{-2.2cm}
\caption{Star-formation rate versus bolometric AGN luminosity at different distances. Each panel corresponds to the integrated SFR$_{\rm{PAHs}}$ for the 100\,pc, 200\,pc, 500\,pc, and 700\,pc apertures, respectively. The QSOs observations from Martinez-Paredes et al. (submitted) are shown as blue triangles. The Seyfert from \citet{Esquej14}, \citet{Ruschel16}, and our are shown as starts, diamonds and dots, respectively. The Sy1 and Sy2 are shown as red and green points, respectively. The dashed and continuous lines in all panels correspond with the correlations proposed by \citet{Neistein14} as a function of radius shifted using the predictions given by \citet{Hopkins10} (see text). The dashed line is simulated SFR value for a given $\rm{L_{bol,AGN}}$, and the continuous line corresponds to the average $\rm{L_{bol,AGN}}$ for a given value of SFR according to \citet{Neistein14}. The black continuous line and the shaded area in each panel show the average and standard deviation of the best shift to the relation predicted by \citet{Neistein14}, respectively.}
\label{fig.4}
\end{center} 
\end{figure*}

Compared to previous works, our analysis has the advantage that it allows to explore the SFR at different sub-kpc scales from the nucleus. We calculate the $\rm{L_{bol,AGN}}$ from X-ray luminosities (reported in Table \ref{tab:observations}, Col.\,4) using the relation $\rm{L_{bol,AGN} = k L(2-10\,keV)}$, where the bolometric correction (k) depends on $\rm{L(2-10\,keV)}$ itself with a fourth order polynomial \citep[see][]{Marconi04}. In Figure \ref{fig.4}, we present the relation between $\rm{L_{bol,AGN}}$ and SFR$_{\rm{PAHs}}$ integrated at different distances from the nucleus. Each panel corresponds to integrated SFR$_{\rm{PAHs}}$ for the 100\,pc, 200\,pc, 500\,pc, and 700\,pc apertures, respectively. Note that this plot includes the 12 sources where we measure the $\rm{11.3~\mu m}$ PAH feature (the integrated fluxes density are reported in Table \ref{tab:measurenw}). The number of sources varies for each plot depending on the resolution and spatial scale of the extended emission for each spectrum. Furthermore, Sy1 and Sy2 are shown as red and green dots, respectively. We also include the measurements for QSOs from Martinez-Paredes et al. (submitted, triangles) as well as Seyferts from \citet[][starts]{Esquej14} and \citet[][diamonds]{Ruschel16}.

Two of our objects (NGC\,1808 and NGC\,4569) are in the range of low-luminosities (${\rm{L_{bol,AGN}}}<10^{42}$ erg s$^{-1}$). \citet{Hopkins10} predictions are not able to reproduce these low-efficiency objects. \citet{Ruschel16} suggest that the low-luminosity AGN have high circumnuclear SF. However, our objects with high luminosity have a similar or higher SFRs. \citet{Neistein14} presented two correlations: (1) the average SFR value for a given ${\rm{L_{bol,AGN}}}$ in their models, and (2) the average of ${\rm{L_{bol,AGN}}}$ for a given value of total SFR. Indeed, the first relation flattens towards low-luminosities, as seen by our two low-luminosity AGN. In Figure \ref{fig.4}, we show these relations shifted as predicted by \citet{Hopkins10} for different apertures (dashed and continuous lines with different colors in each panel):

\begin{eqnarray}
\rm{SFR}_{\rm{PAHs}} (R<100\,pc) & = & \rm{SFR}_{\rm{PAHs}}(<10\,pc) - 1.0  \\
\rm{SFR}_{\rm{PAHs}} (R < 1\,kpc) & = & \rm{SFR}_{\rm{PAHs}}(<10\,pc) - 1.52  \\
\rm{SFR}_{\rm{PAHs}} (\rm{total}) & = & \rm{SFR}_{\rm{PAHs}}(<10\,pc) - 2.52 
\label{eq:2}
\end{eqnarray}

\noindent These relations have been computed using equations 15-18 in \citet[][]{Hopkins10}. Note that scatter in this relations are significant. In general terms, these relations have the form:
\begin{equation} 
\rm{SFR_{PAHs} (R < Rs) = SFR_{PAHs} (< 10\,pc) - B(Rs)}
\label{eq:1}
\end{equation}
\noindent where B(Rs) is a constant that depends on the physical scale. We have interpolated the given values to obtain the expected shifts on the physical scales derived from our analysis (reported in Col. 2 of Table \ref{tab:offset}).

In order to compare predictions with models, we have computed the observed shift to this relation as the average and standard deviation of the relation predicted by \citet{Neistein14} and our data points. These shifts are reported in Cols. 3, and 4 of Table \ref{tab:offset}. This correlation and standard deviations are shown as black continuum line and shaded area in each panel, respectively.
  
Note that the results shown in Figure \ref{fig.4}, could be affected by the following errors: 1) The systematic offset due to the use of different SFR tracers. The dispersion from the correlation used to calculate the SFR from the $\rm{11.3~\mu m}$ PAH feature is similar to that obtained by other tracers. We have taken in account this dispersion in the error bars in Figure \ref{fig.4}; 2) Time-scale for the SF. According to \citet{Neistein14} a  necessary condition for agreement between data and model is that the correct time-scale for both SF and AGN activity is adopted. The models are constrained to calculate the SFR average using only the SF in the last 150 Myr.  We have calculated the SFR using the $\rm{11.3~\mu m}$ PAH  as a tracer. This feature is usually associated with B stars \citep[][]{Peeters04}; and 3) Calculation errors in the ${\rm{L_{bol,AGN}}}$. In the models, the ${\rm{L_{bol,AGN}}}$ depends on the accretion mass, while in our data, it depend on the X-ray luminosity, which might vary up to one order of magnitude. In Figure \ref{fig.4}, we have already included this uncertainty in the error bars.

We found a sensible agreement between the theoretical relations proposed by \citet{Neistein14} shifted according to \citet{Hopkins10} and our data, for most inner galaxy parts. This result is of interest as, in the simulated objects, major mergers with tidal events have been deemed responsible for both the star formation and black hole feeding.

\section{Summary and Conclusions}
\label{sec:SumCon}
In this paper, we present a sample of 19 local AGN observed with ground-based T-ReCS/Gemini and CanariCam/GTC spectra. We complemented these observations with available \emph{Spitzer}/IRS spectra. We have studied the surface brightness radial profile of the $\rm{11.3~\mu m}$ PAH feature and the [SIV] line emission.
According with the results of this research, we tried to answer the following three questions:
\begin{itemize}
\item[1)]  What is the origin of the [SIV] line emission in the nuclear region?

	 The contribution to the [SIV] line emission is not circumnuclear. Instead, it often peaks at distances greater than $\rm{1000\,R_{Sub}}$ from the nucleus.
	  We have not found a relation between the surface brightness radial profiles of the [SIV] line and the PAH feature at different distances from nucleus. If the PAH is a good tracer of SF, we speculate that the [SIV] line emission could be tracing SF with different ages than those traced by the PAH feature.
	  
\item[2)] How good is the $\rm{11.3~\mu m}$ PAH feature as tracer of SF in the vicinity of the AGN?

	We found a PAH flux deficit closer to the AGN as compared with larger apertures (toward the inner $\rm{\sim}$ 100\,pc). This deficit cannot be related to dilution by the AGN continuum. We have not found observational support for the destruction of PAH features due to the AGN radiation field. Intrinsic lack of SF toward the center is also a plausible explanation. 

\item[3)] What can we say about the connection between star-formation and AGN activity?
	We found a sensible agreement between the expected shift the ${\rm{L_{bol,AGN}}}$ - SFR theoretical relation proposed by \citet{Neistein14}, \citet{Hopkins10}, and our observations, for most inner galaxy parts.
\end{itemize}

\acknowledgments
The authors thank the anonymous referee for careful reading and constructive suggestion that improved the paper.
This scientific publication is based on observations made with the Gran Telescopio CANARIAS (GTC), installed at the Spanish Observatorio del Roque de los Muchachos of the Instituto de Astrof\'isica de Canarias on the island of La Palma. This work is based in part on observations made with the \emph{Spitzer} Space Telescope, which is operated by the Jet Propulsion Laboratory, California Institute of Technology under a contract with NASA. D. E.-A. and O.G.-M. acknowledge support from grant IA100516 and IA103118 PAPIIT DGAP UNAM. D.D. and D. E.-A. acknowledges support through grants IN108716 from PAPIIT, UNAM and 221398 from CONACyT. D. E.-A. acknowledges support from a CONACYT scholarship.
A.A.-H. acknowledges support from the Spanish Ministry of Economy and Competitiveness through the Plan Nacional de Astronom\'ia y Astrof\'isica under grant AYA2015-64346- C2-1-P, which is party funded by the FEDER program.
C.R.-A. acknowledges the Ram\'on y Cajal Program of the Spanish Ministry of Economy and Competitiveness through project RYC-2014-15779 and the Spanish Plan Nacional de Astronom\' ia y Astrofis\' ica under grant AYA2016-76682-C3-2-P.
T.D.-S. acknowledges support from ALMA-CONICYT project 31130005 and FONDECYT regular project 1151239.

\appendix
\section{Catalog of spectra and reported nuclear star forming regions}
\label{sec:Cat}
\textbf{NGC\,931} (Mrk\,1040) is a barred galaxy (Sbc) with a Sy1 nucleus. \citet{Ward78} found that this galaxy interacts with a satellite galaxy located a 10\,kpc from NGC\,931. We did not find records of SF in other works at the scales tracer with our observations.
\begin{figure}[ht]
\begin{center}
\includegraphics[width=0.9\columnwidth]{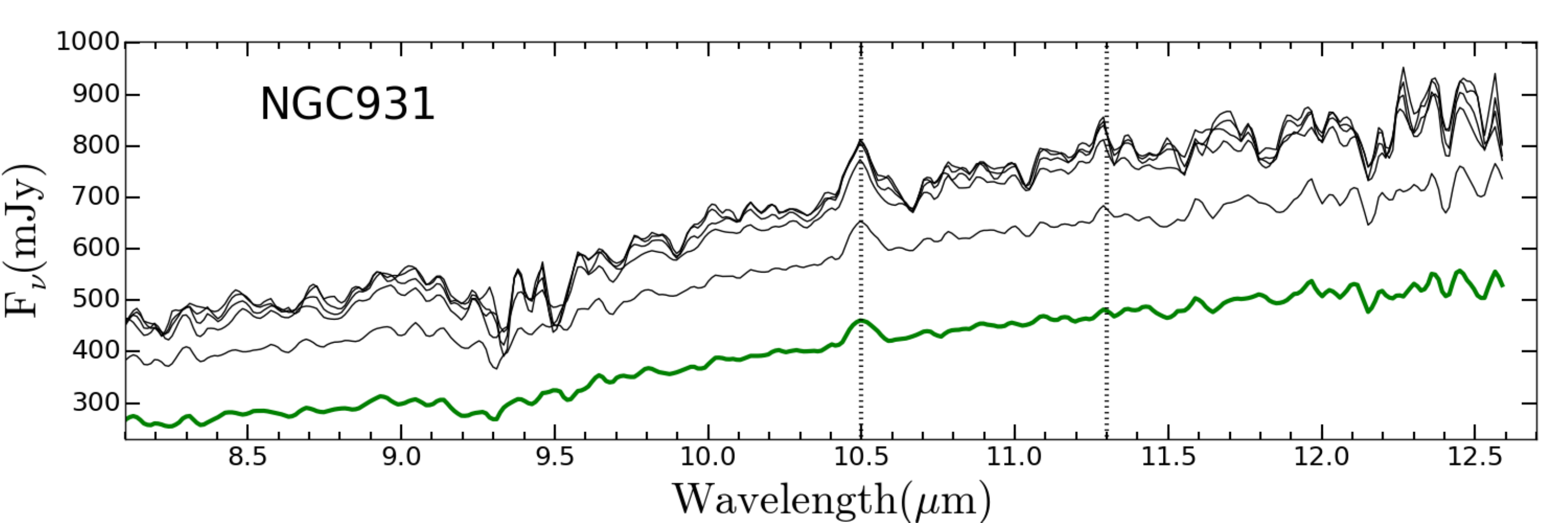}\\
\includegraphics[width=0.95\columnwidth]{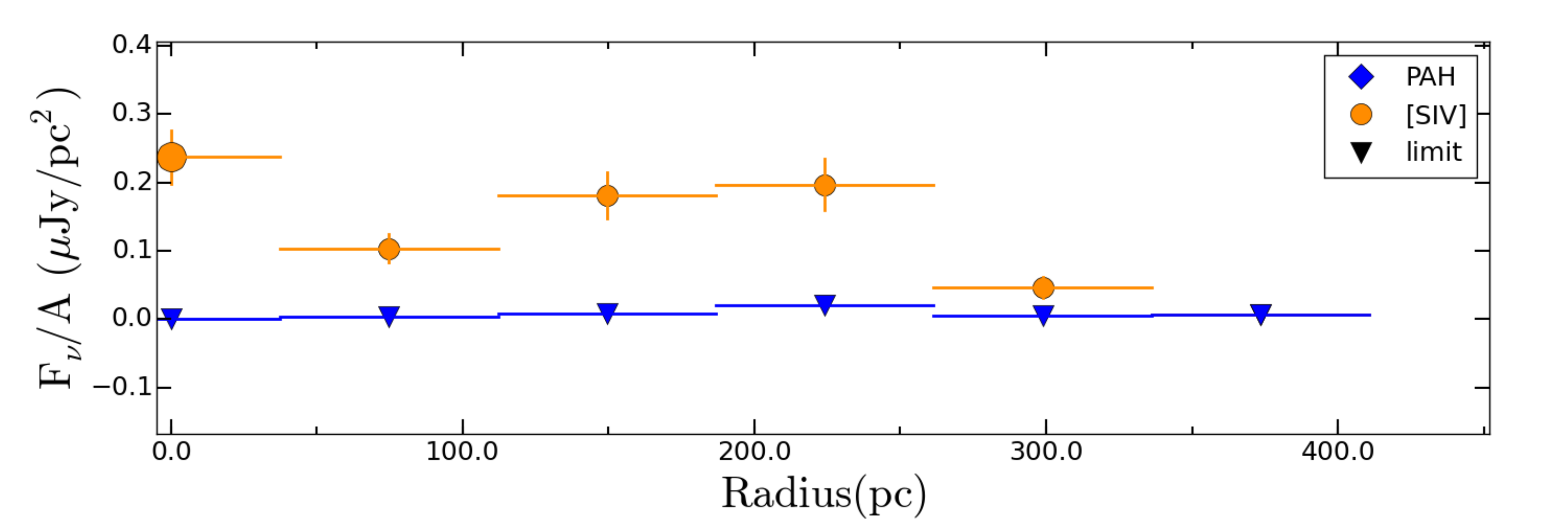}\\
\includegraphics[width=0.95\columnwidth]{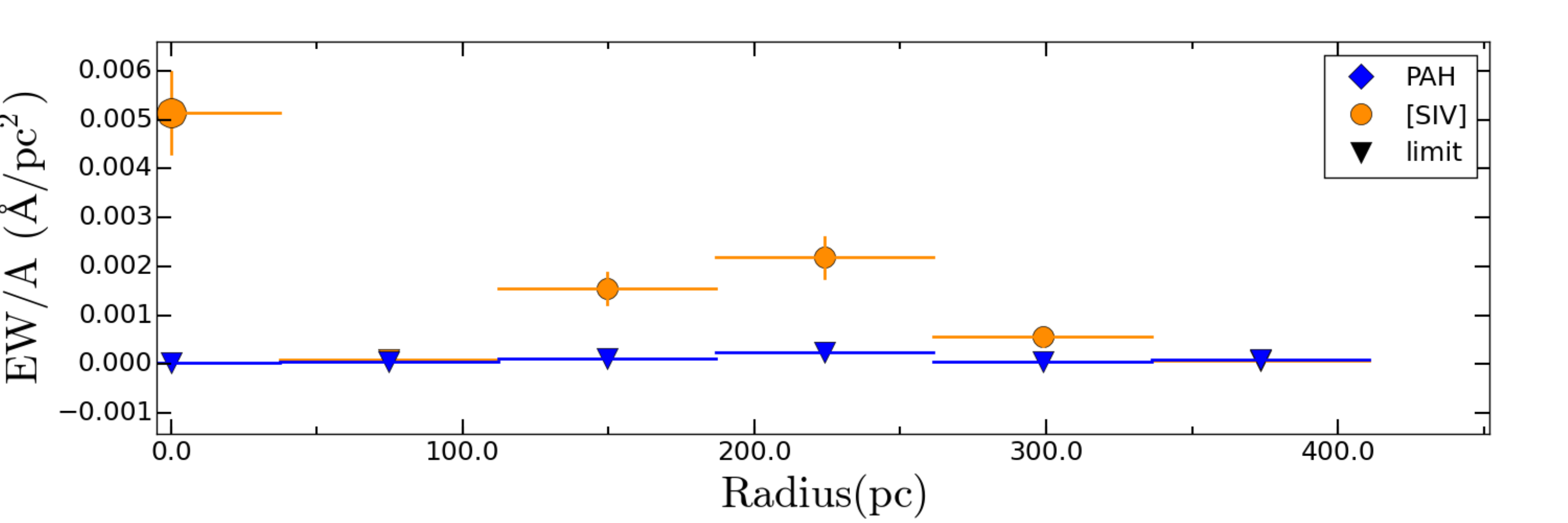}
\caption{(Top panel): Extracted spectra at different scales for NGC\,931. 
The green and black lines correspond to the nuclear spectrum and extended apertures spectra, respectively. The dotted lines show the PAH feature and [SIV] line emissions. (Middle and Bottom panels): Surface brightness and EW radial profiles, respectively. The radial profiles for $\rm{11.3~\mu m}$ PAH emission are presented with blue diamonds, while radial profiles for [SIV] line emission are shown with orange circles (the triangles are limits values). The larger circle corresponds to [SIV] emission in the nuclear spectrum. The vertical line marks the 200\,pc distance from the center.}
\label{ap:fig1}
\end{center}
\end{figure}
\clearpage

\textbf{Mrk\,1066} 
is a starburst galaxy with a Sy2 nucleus. \citet{Ramos-Almeida14} found star-forming knots at $\sim 400$\,pc of the galaxy center, after subtracting the AGN component. \citet{Alonso-Herrero14} suggest that close to the center ($\sim 125$\,pc) the near-IR lines are dominated by the AGN processes.  
\begin{figure}
\begin{center}
\includegraphics[width=0.9\columnwidth]{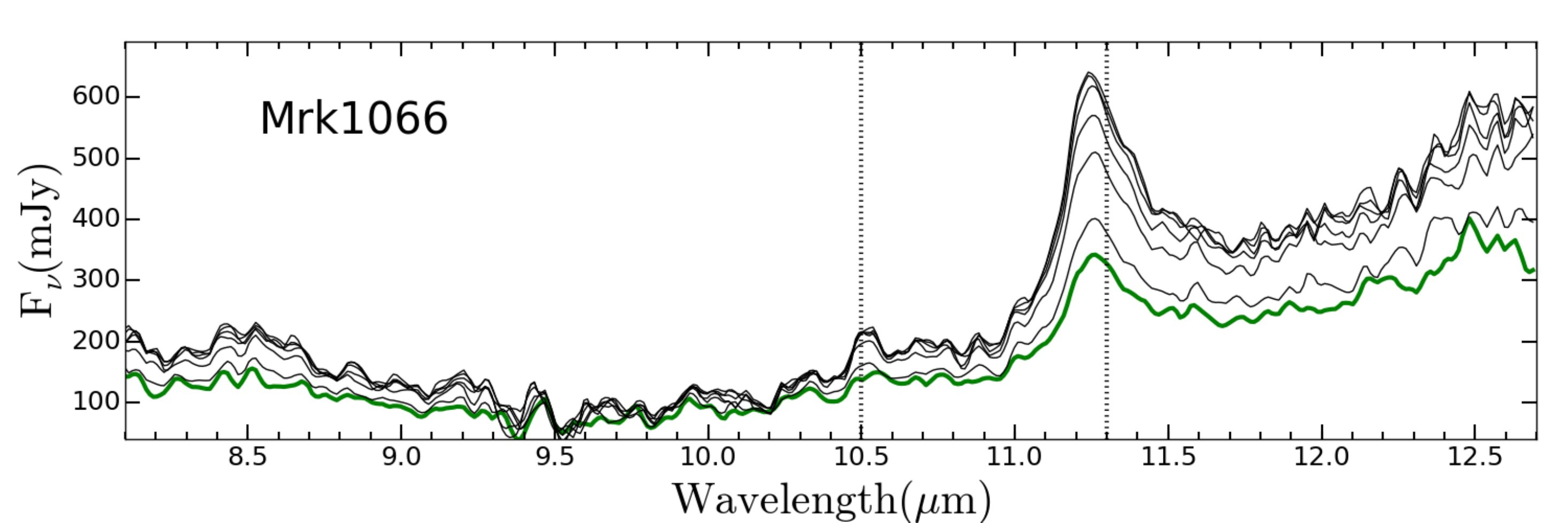}\\
\includegraphics[width=0.95\columnwidth]{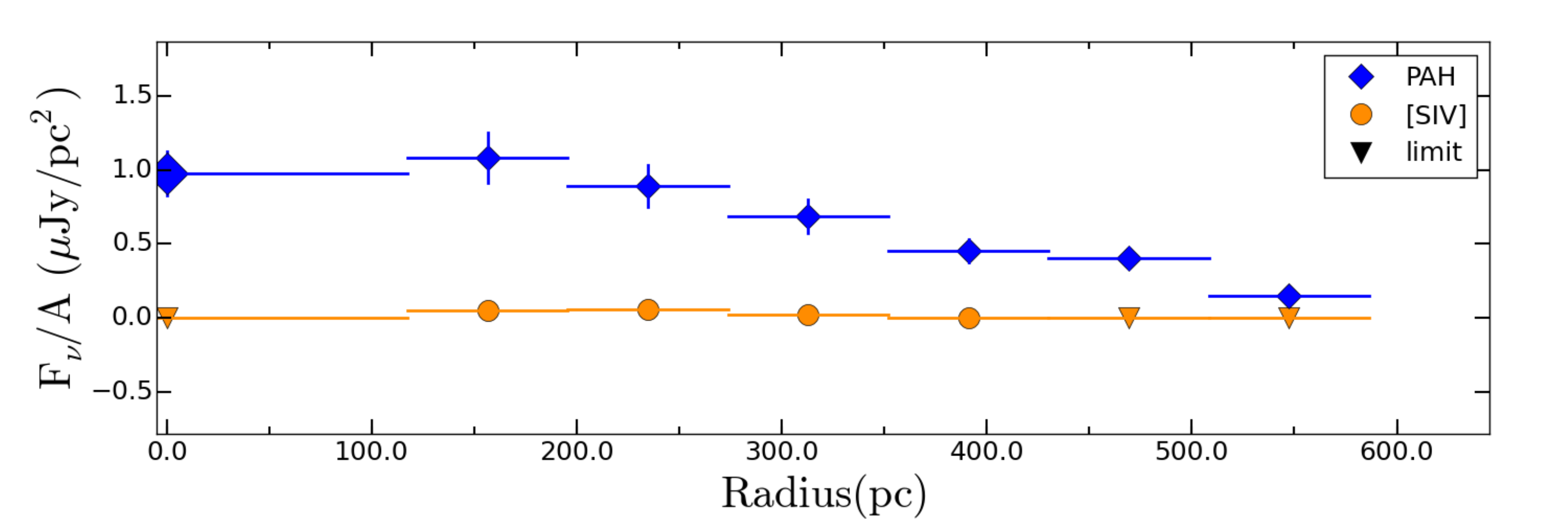}\\
\includegraphics[width=0.95\columnwidth]{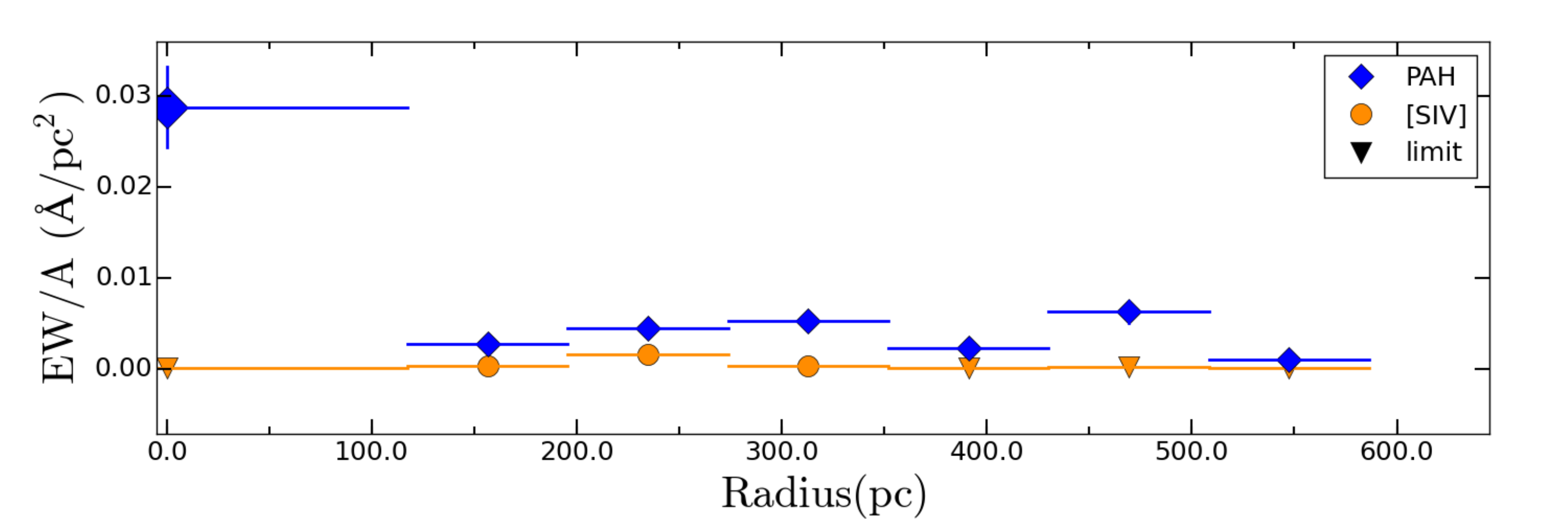}
\caption{Extracted spectra and radial profiles for Mrk\,1066. The same description that in Fig. \ref{ap:fig1}.}
\label{ap:fig2}
\end{center}
\end{figure}
\clearpage

\textbf{NGC\,1320} 
is a edge-on galaxy with a Sy2 nucleus. This source is a "warm galaxy" with a relatively high IR luminosity \citep[][]{Robertis86}. We did not find records of SF in other works at the scales tracer with our observations.
\begin{figure}
\begin{center}
\includegraphics[width=0.9\columnwidth]{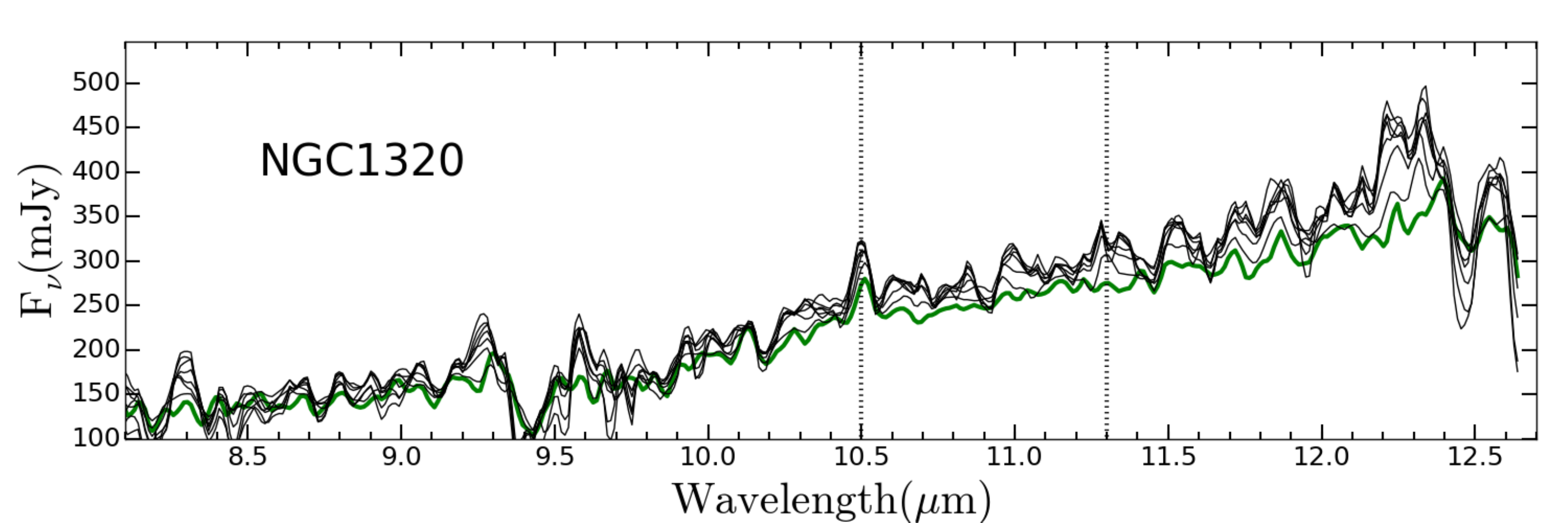}\\
\includegraphics[width=0.95\columnwidth]{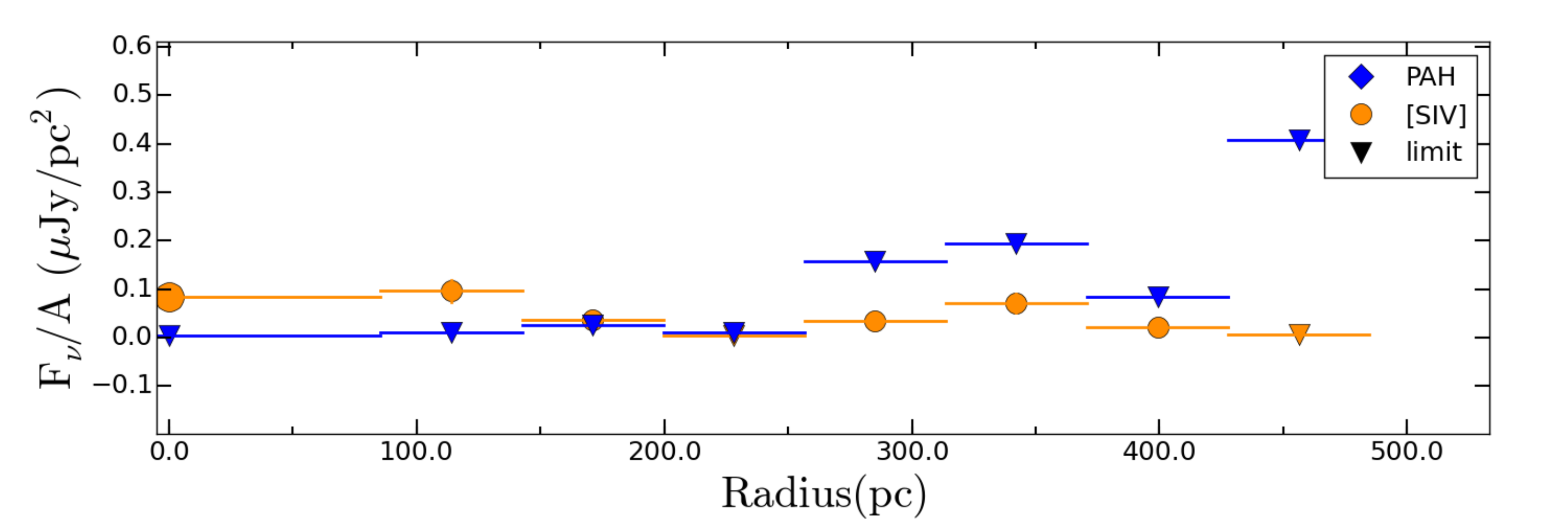}\\
\includegraphics[width=0.95\columnwidth]{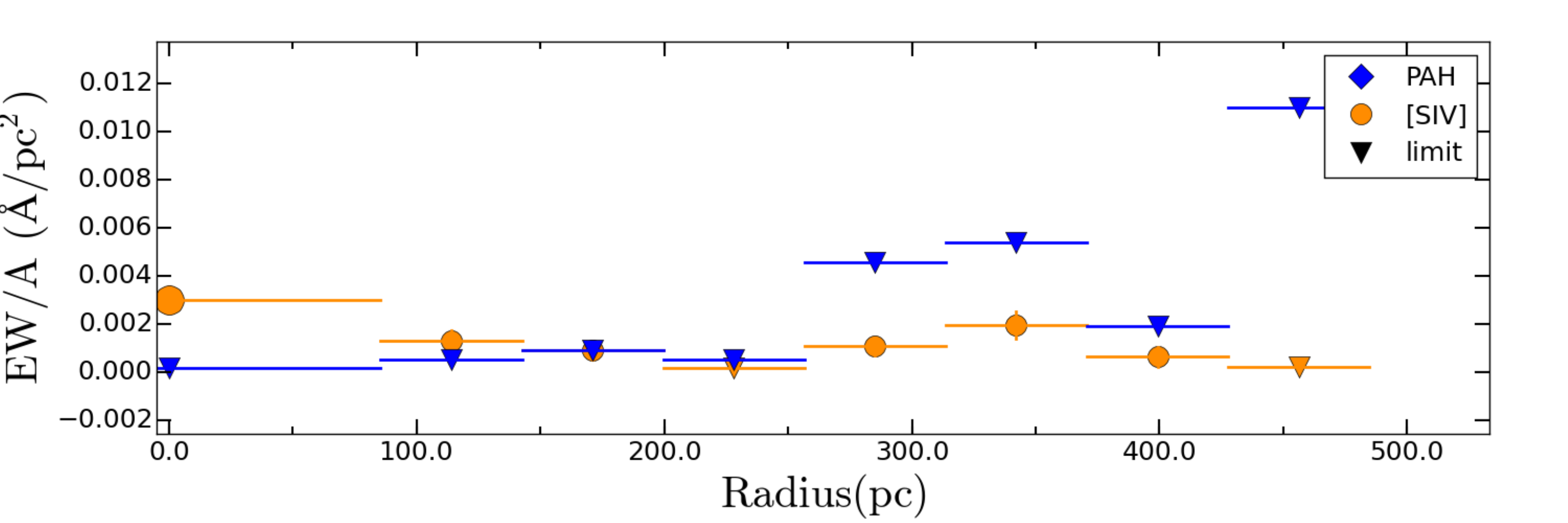}
\caption{Extracted spectra and radial profiles for NGC\,1320. The same description that in Fig. \ref{ap:fig1}.}
\end{center}
\end{figure}
\clearpage

\textbf{NGC\,1386} is an edge-on spiral galaxy with a Sy2 nucleus. \citet{Ruschel14} found that the $\rm{11.3~\mu m}$ PAH feature is more pronounced at $\rm{\sim 100\,pc}$ distance of nucleus. They also found that the [SIV] line emission is only detected in the nucleus at distances $<100$\,pc. Our observations are in agreement with these results. Optical studies show evidence of heavy obscuration \citep[][]{Weaver91,Storchi96,Rossa00}. 
\begin{figure}
\begin{center}
\includegraphics[width=0.9\columnwidth]{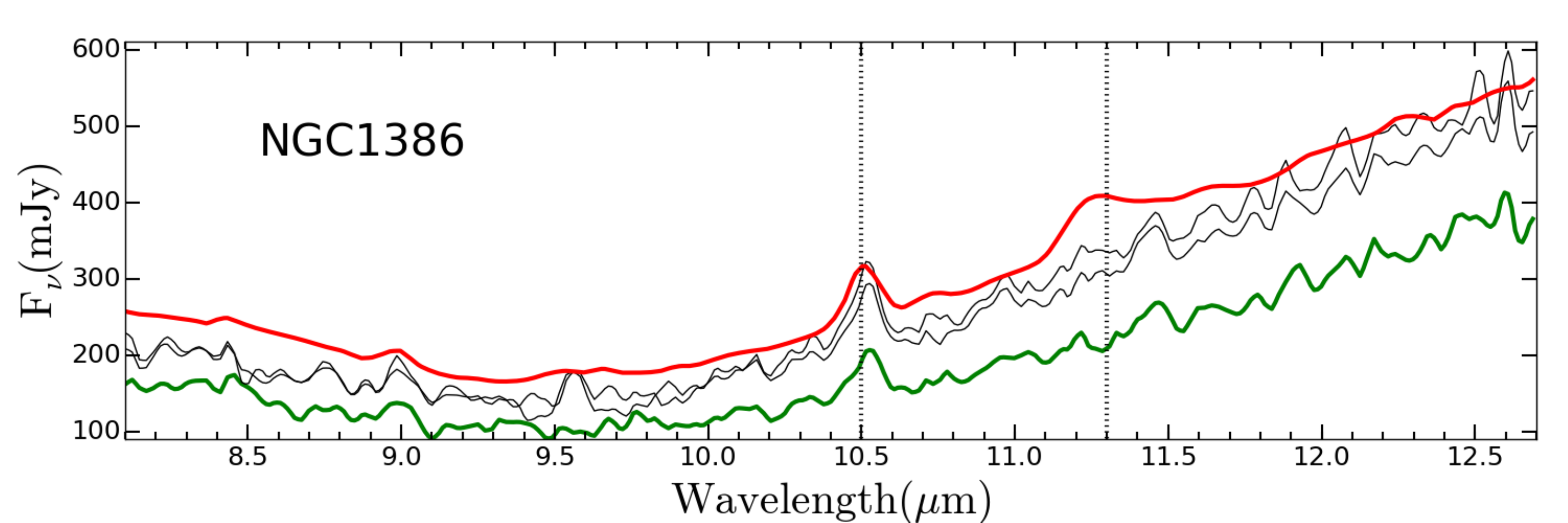}\\
\includegraphics[width=0.95\columnwidth]{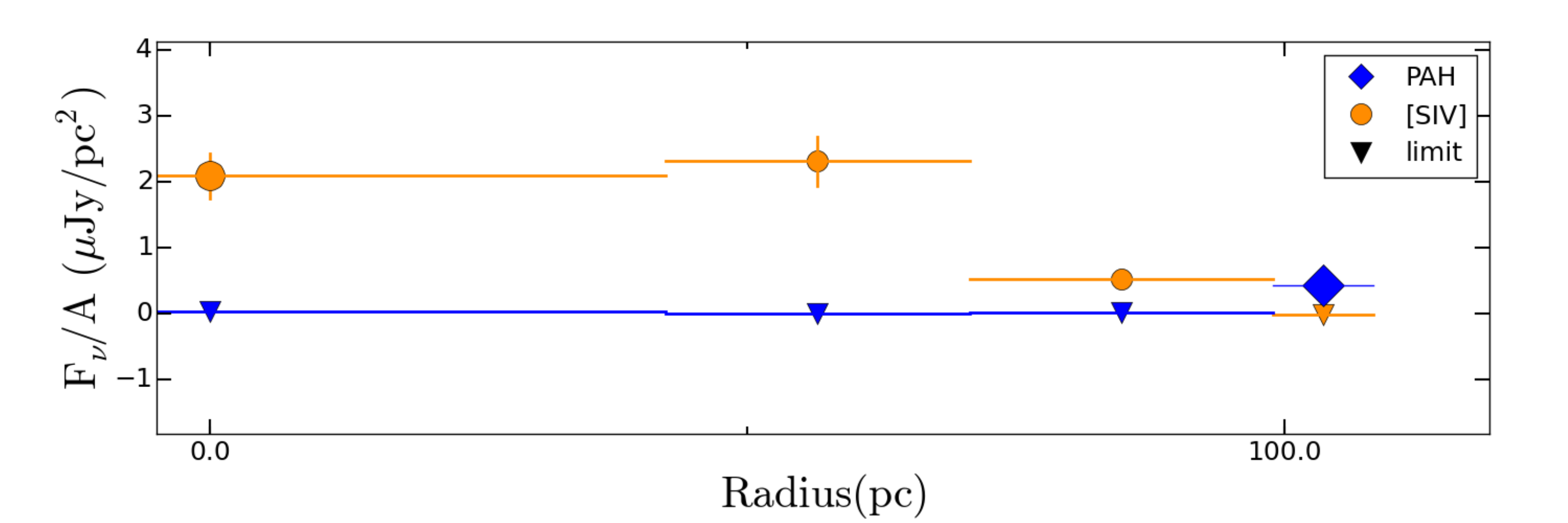}\\
\includegraphics[width=0.95\columnwidth]{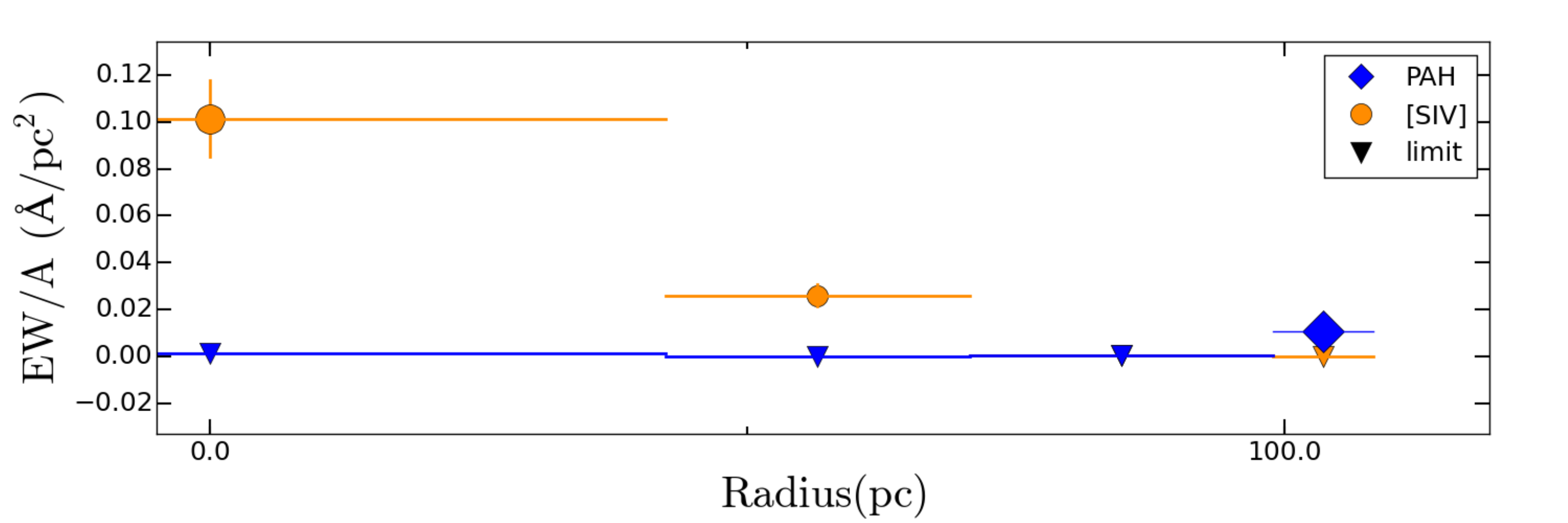}
\caption{Extracted spectra and radial profiles for NGC\,1386. The same description that in Fig. \ref{ap:fig1}.}
\end{center}
\end{figure}
\clearpage

\textbf{NGC\,1808} is an inclined spiral galaxy with a Sy2 nucleus and a prominent starburst \citep[][]{Krabbe94,Veron85}. \citet{Yuan10} considered the possibility that the nucleus is a H\,II region. \citet{Asmus14} found that star-forming regions dominate the MIR emission within $\sim 200$\,pc.
\begin{figure}
\begin{center}
\includegraphics[width=0.9\columnwidth]{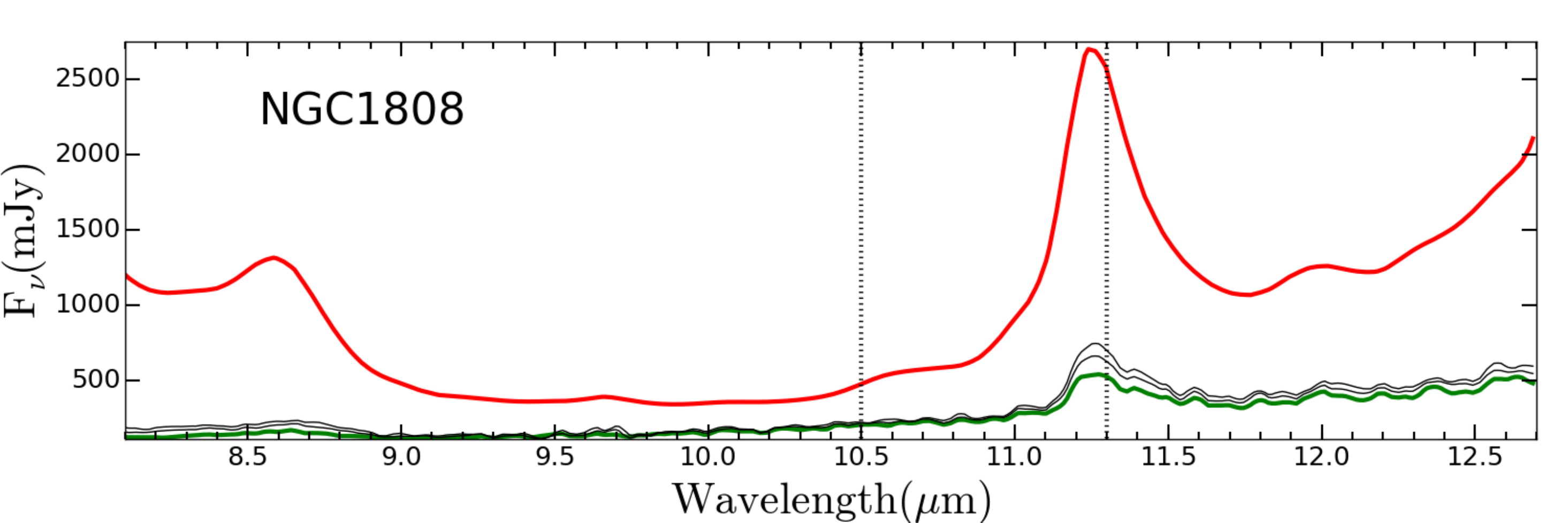}\\
\includegraphics[width=0.95\columnwidth]{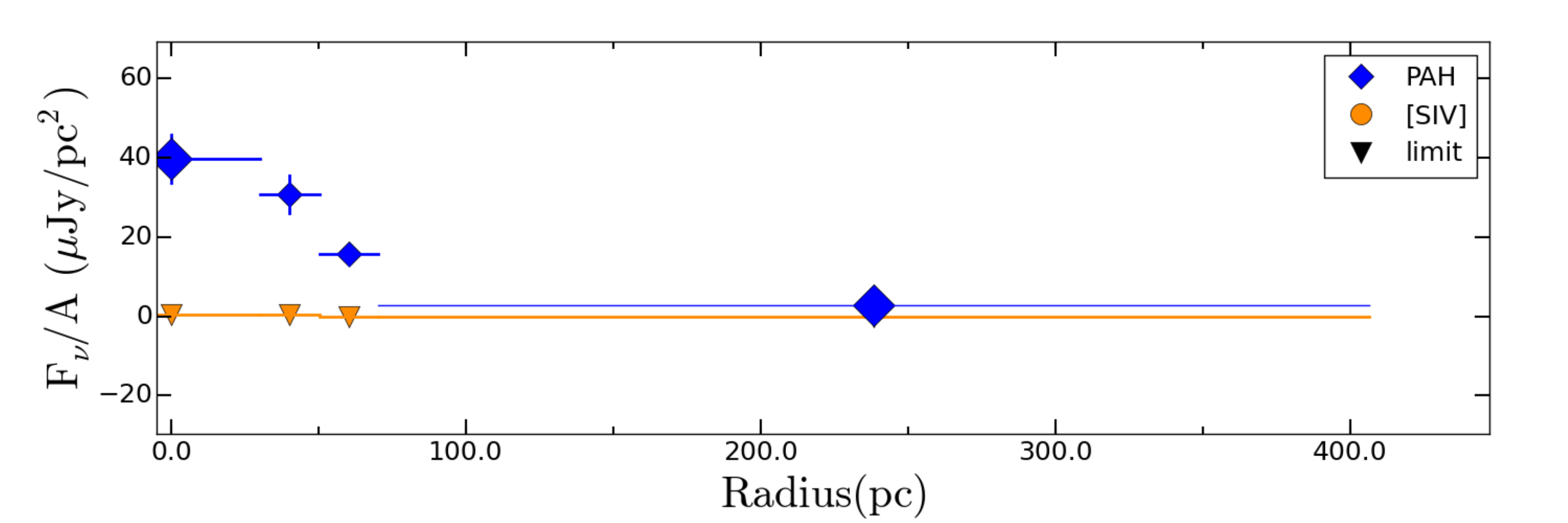}\\
\includegraphics[width=0.95\columnwidth]{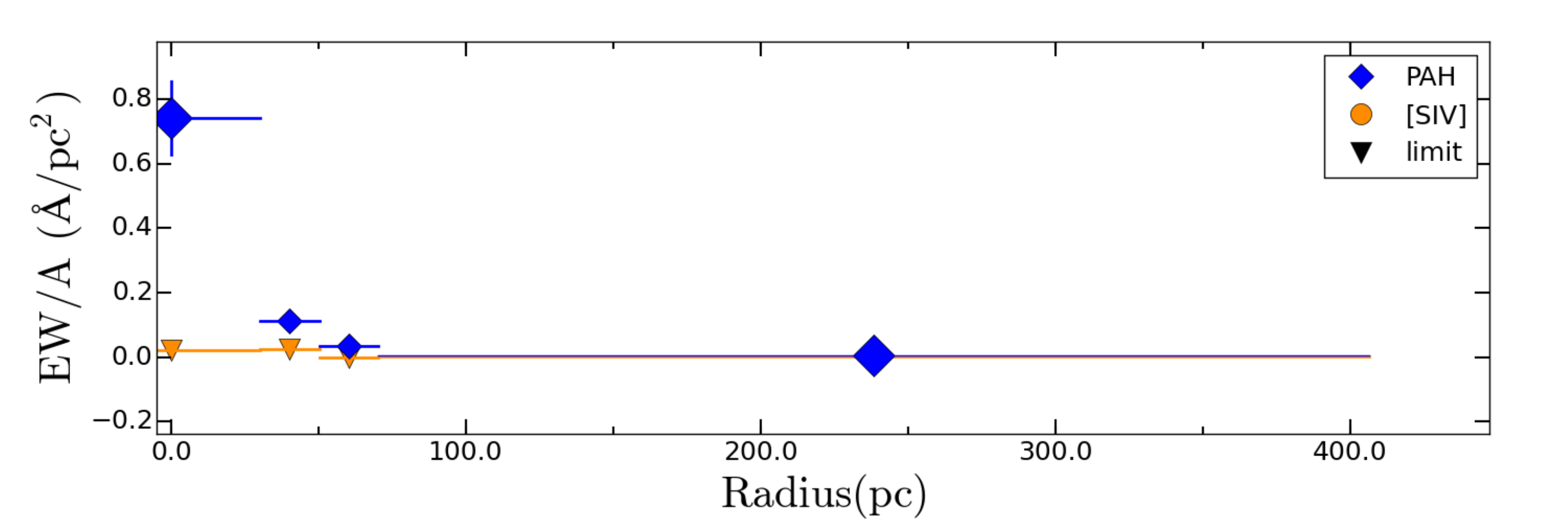}
\caption{Extracted spectra and radial profiles for NGC\,1808. The same description that in Fig. \ref{ap:fig1}.}
\end{center}
\end{figure}
\clearpage

\textbf{NGC\,2992}
is an inclined spiral galaxy \citep[][]{Vaucouleurs91} and located in the interacting system $\rm{Arp\,245}$. The nucleus of this sources is classified as a Sy1.9 in the optical. However, in other works, it is classified as Sy1.5 or Sy2 \citep[][]{Gilli00, Trippe08}.
\citet{Gilli00} suggested that the IR variations were  probably caused by a retriggered AGN. \citet{GarciaB15} found that the starburst component dominates the MIR emission, while the AGN component dominates at higher wavelengths ($\rm{\lambda > 15~ \mu m}$).
\begin{figure}
\begin{center}
\includegraphics[width=0.9\columnwidth]{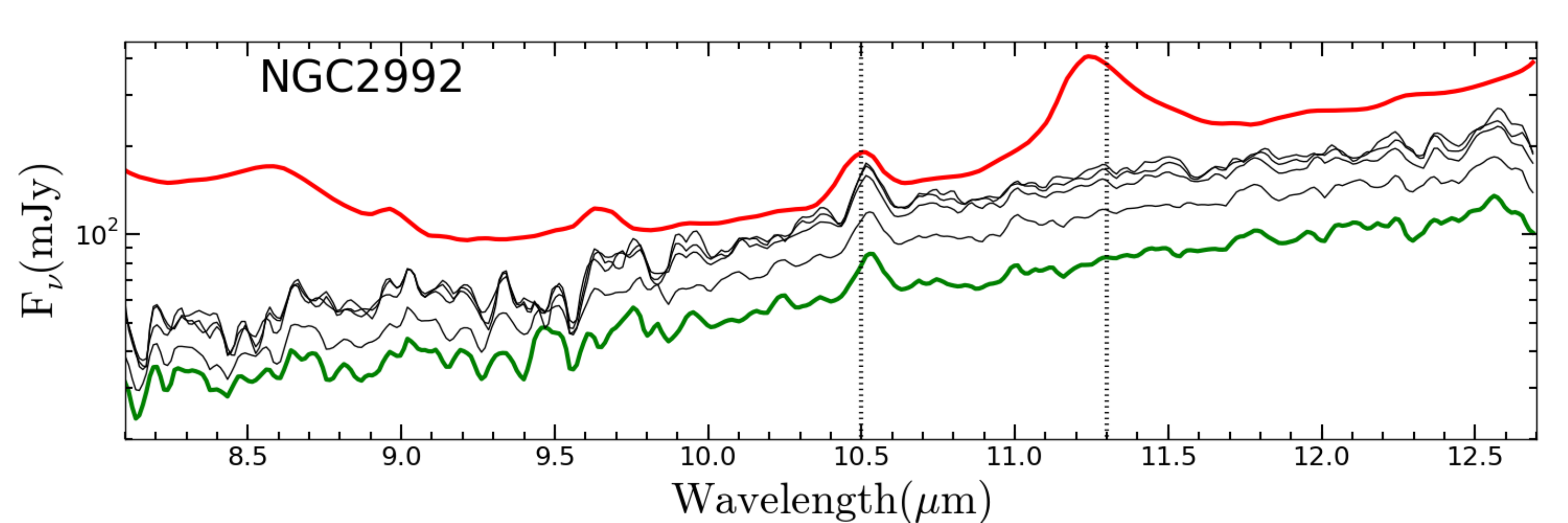}\\
\includegraphics[width=0.95\columnwidth]{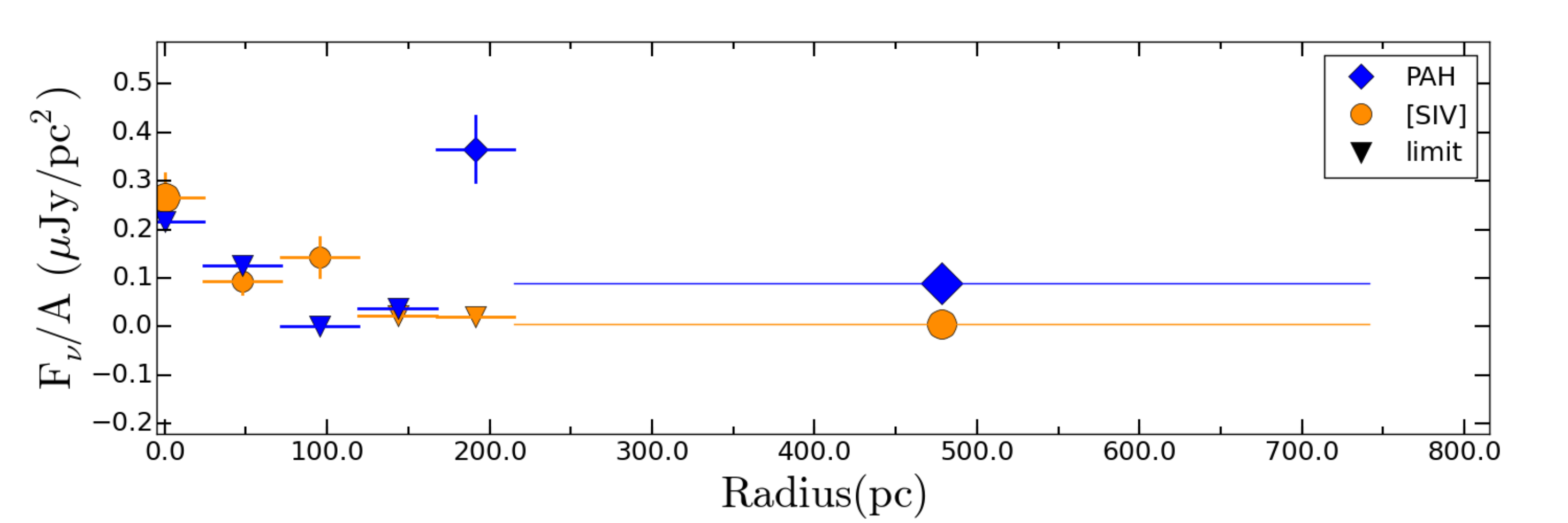}\\
\includegraphics[width=0.95\columnwidth]{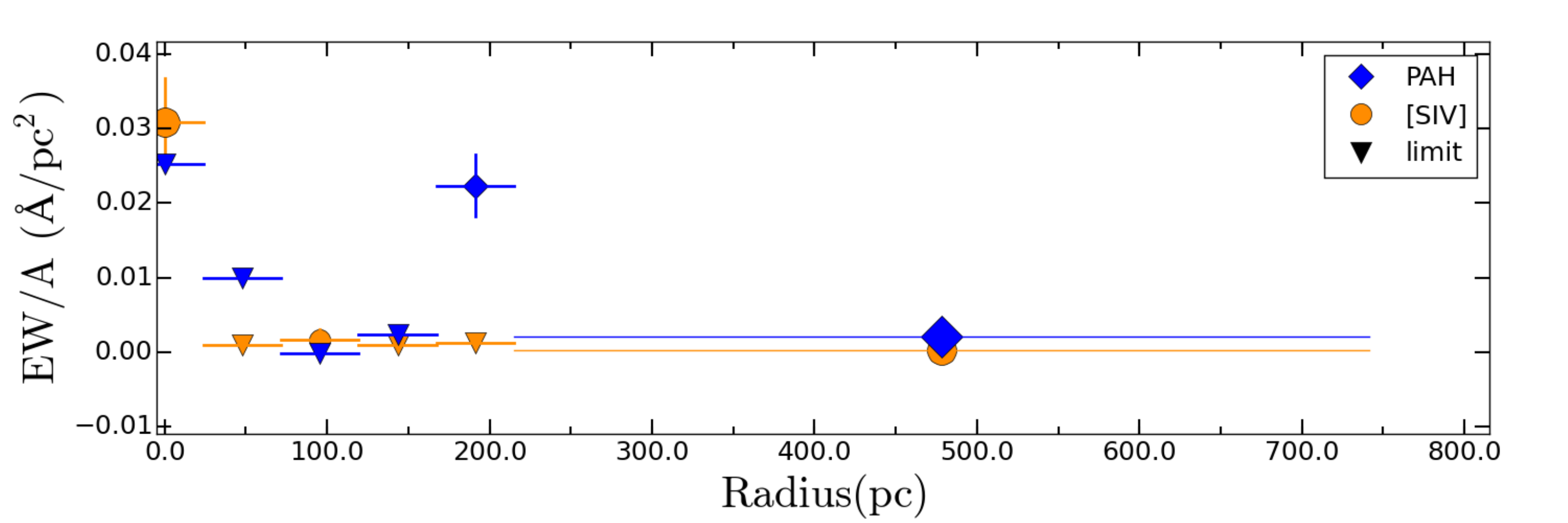}
\caption{Extracted spectra and radial profiles for NGC\,2992. The same description that in Fig. \ref{ap:fig1}.}
\end{center}
\end{figure}
\clearpage

\textbf{NGC\,3081} is a low-inclination barred spiral galaxy with a Sy2 nucleus \citep[][]{Phillips83, Asmus14}. However, \citet{Moran00} reported a type 1 optical spectrum in polarized light. \citet{Weaver10} found that the \emph{Spitzer} spectrum exhibits a weak absorption by silicate at $\rm{10~\mu m}$, a weak PAH emission, and prominent forbidden emission lines. However, \citet{Asmus14} concluded that the MIR emission mostly due.
\begin{figure}
\begin{center}
\includegraphics[width=0.9\columnwidth]{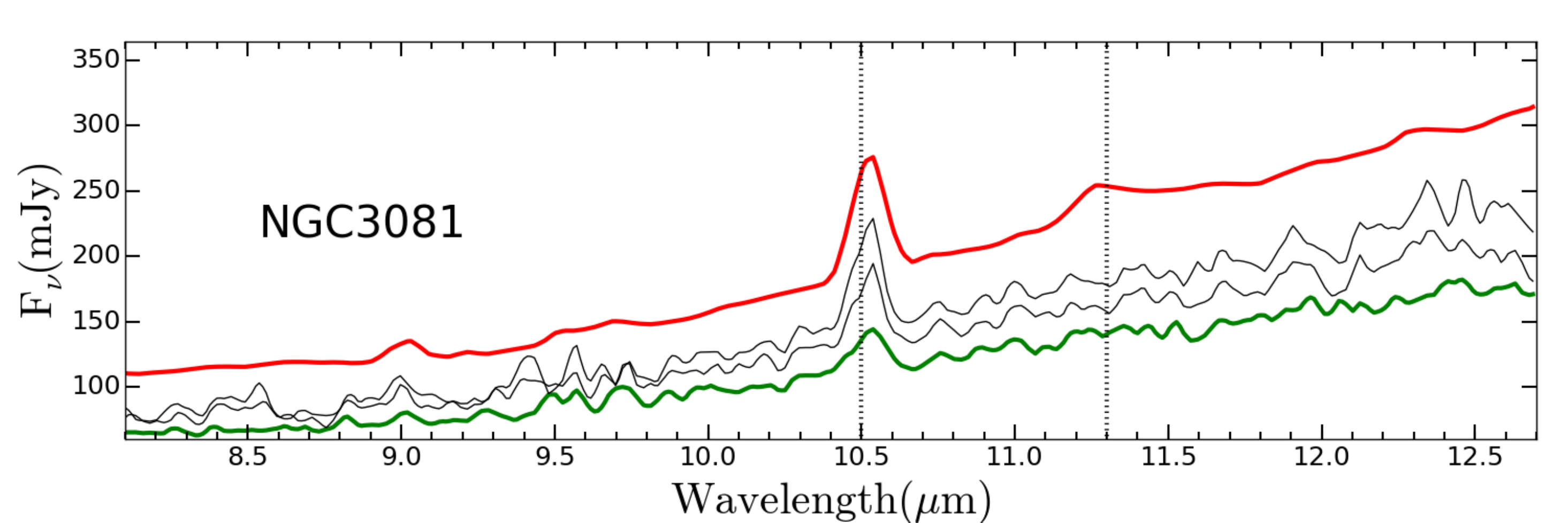}\\
\includegraphics[width=0.95\columnwidth]{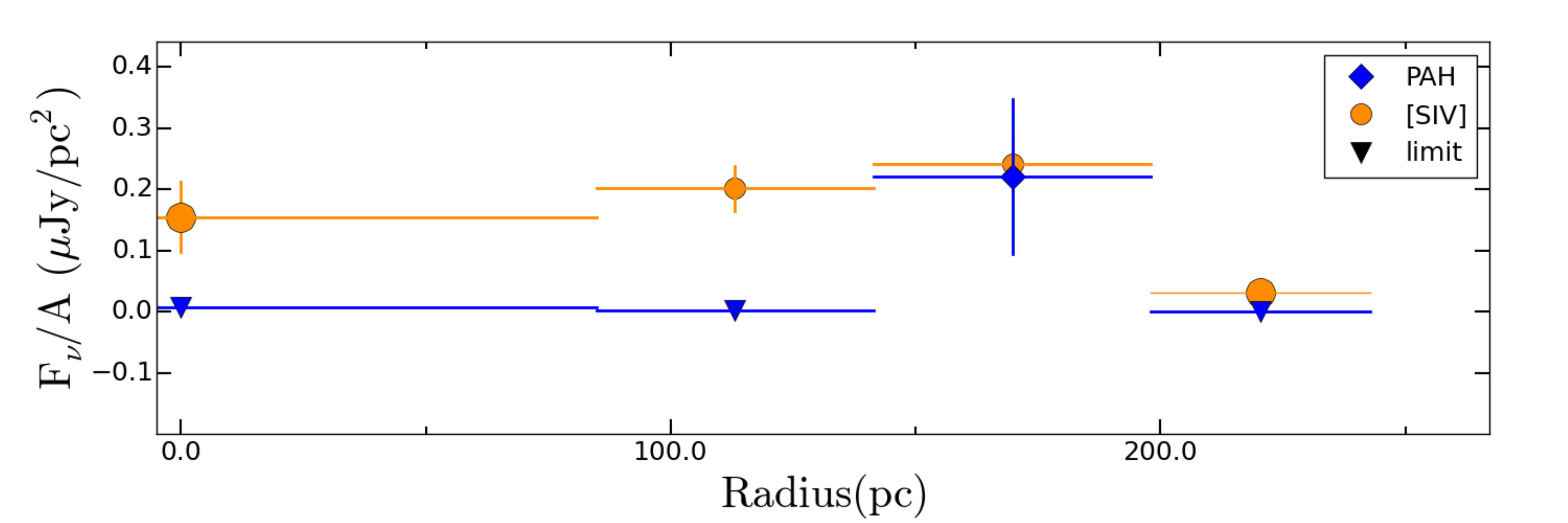}\\
\includegraphics[width=0.95\columnwidth]{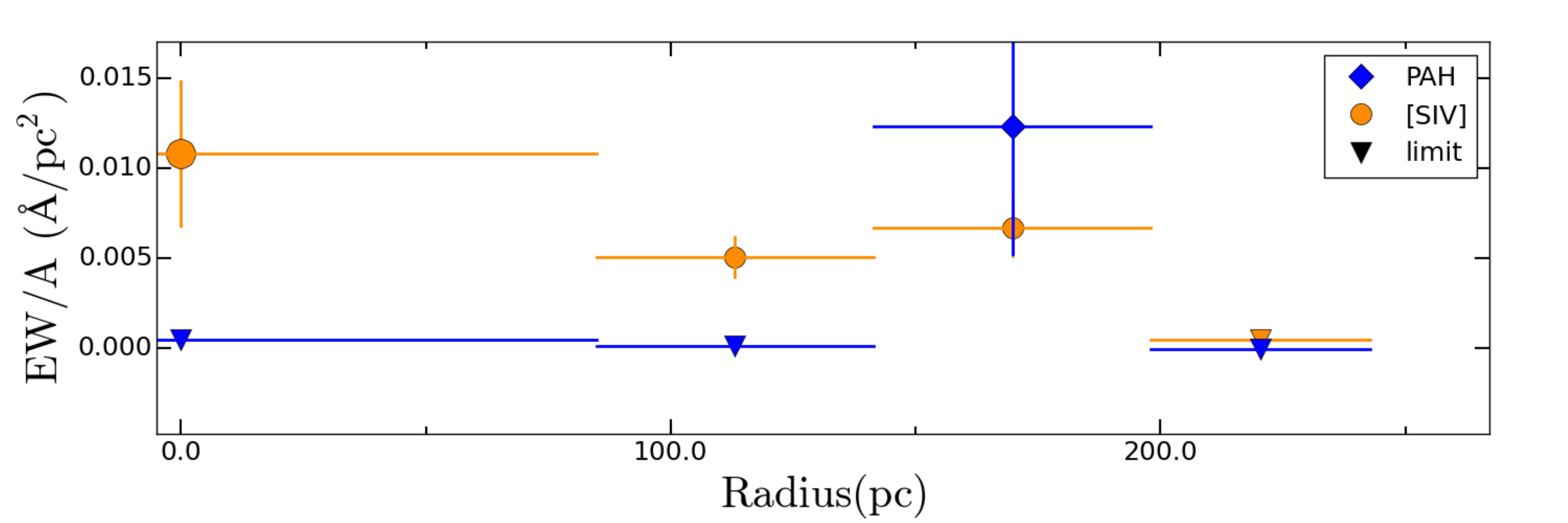}
\caption{Extracted spectra and radial profiles for NGC\,3081 with $\rm{PA = 0}$ \grad. The same description that in Fig.\ref{ap:fig1}.}
\label{ap:fig7}
\end{center}
\end{figure}
\clearpage

\textbf{NGC\,3081}: (see above).
\begin{figure}
\begin{center}
\includegraphics[width=0.9\columnwidth]{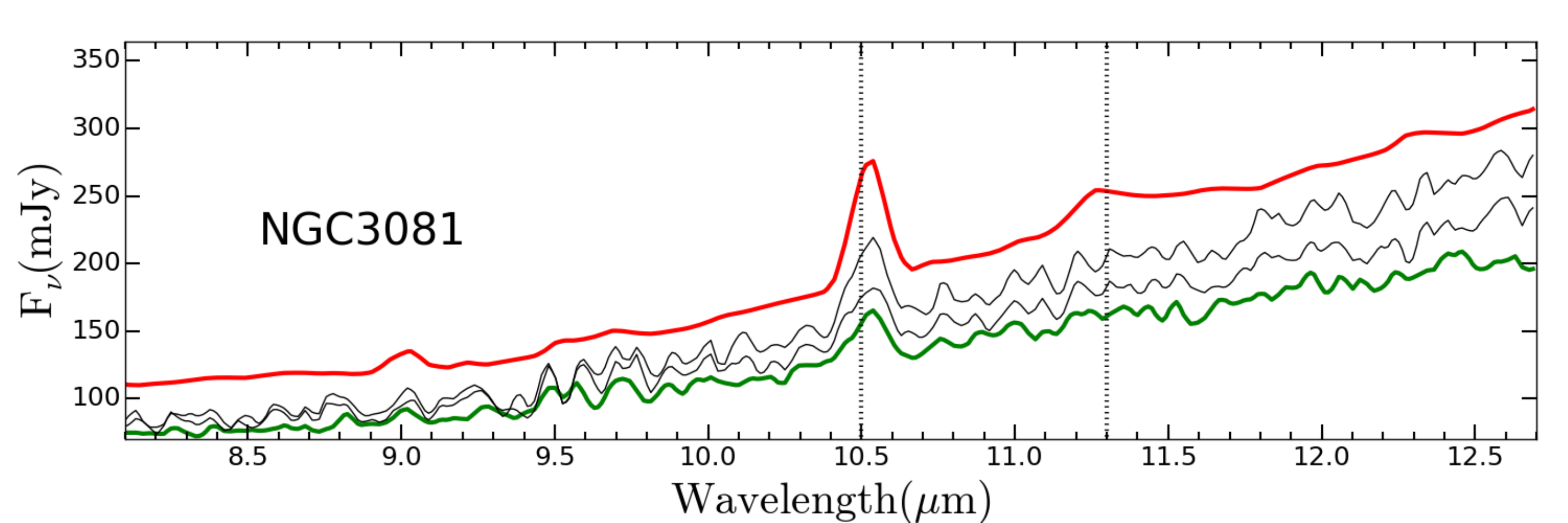}\\
\includegraphics[width=0.95\columnwidth]{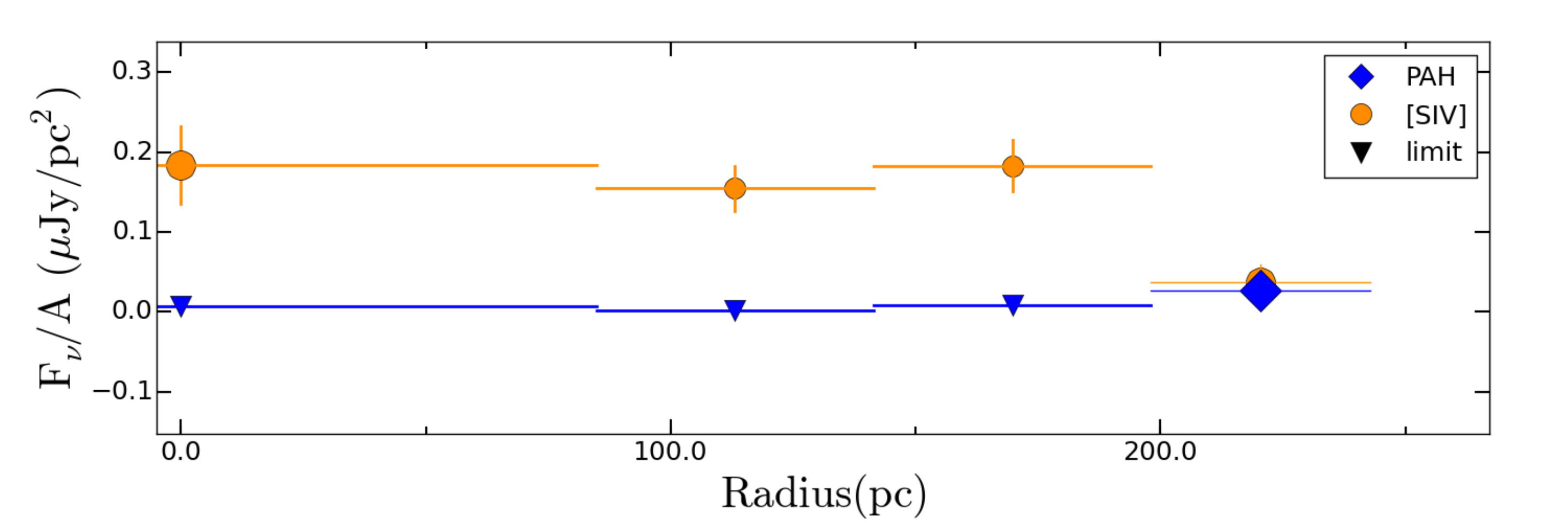}\\
\includegraphics[width=0.95\columnwidth]{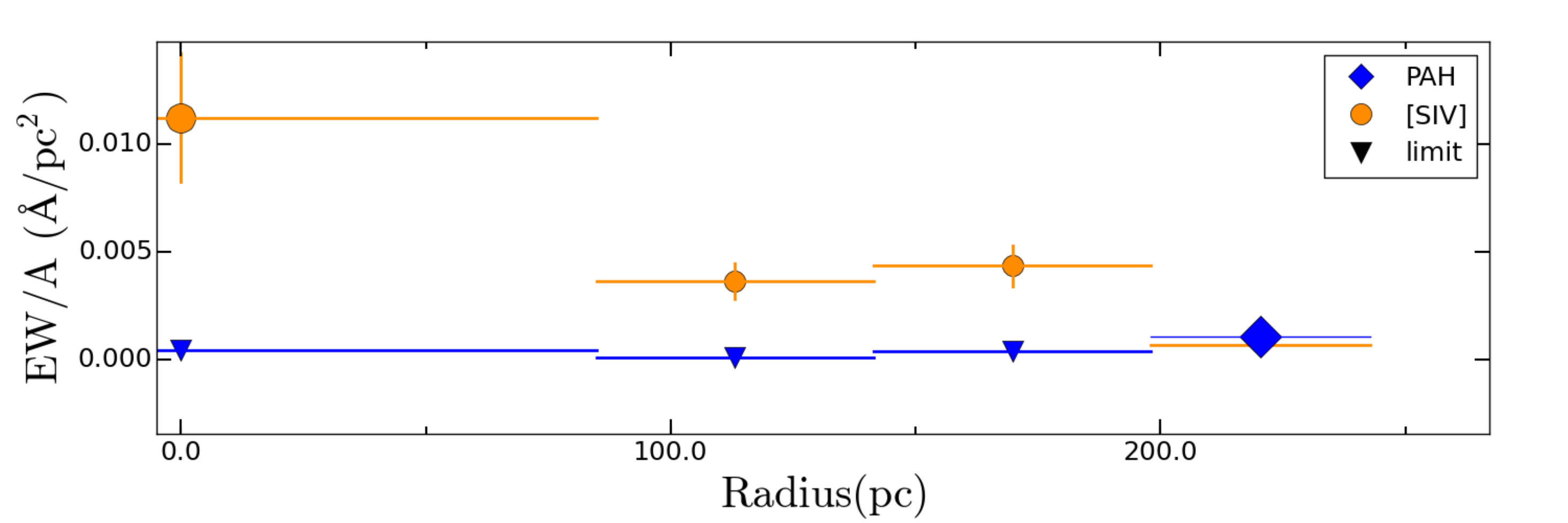}
\caption{Extracted spectra and radial profiles for NGC\,3081 with $\rm{PA = 350}$ \grad. The same description that in Fig. \ref{ap:fig1}.}
\end{center}
\end{figure}
\clearpage

\textbf{NGC\,3227} is a low-inclination barred spiral galaxy. This source is in interaction with NGC\,3226. The nucleus is classified as Sy1.5 and it is surrounded by circumnuclear starburst \citep[][]{Veron10}. \citet{Rodriguez03} and \citet{Davies06} found star-forming regions at $\sim 70$\,pc from the nucleus. \citet{Asmus14} also found that the MIR emission is dominated by star-forming regions (at arcsecond-scale).
\begin{figure}
\begin{center}
\includegraphics[width=0.9\columnwidth]{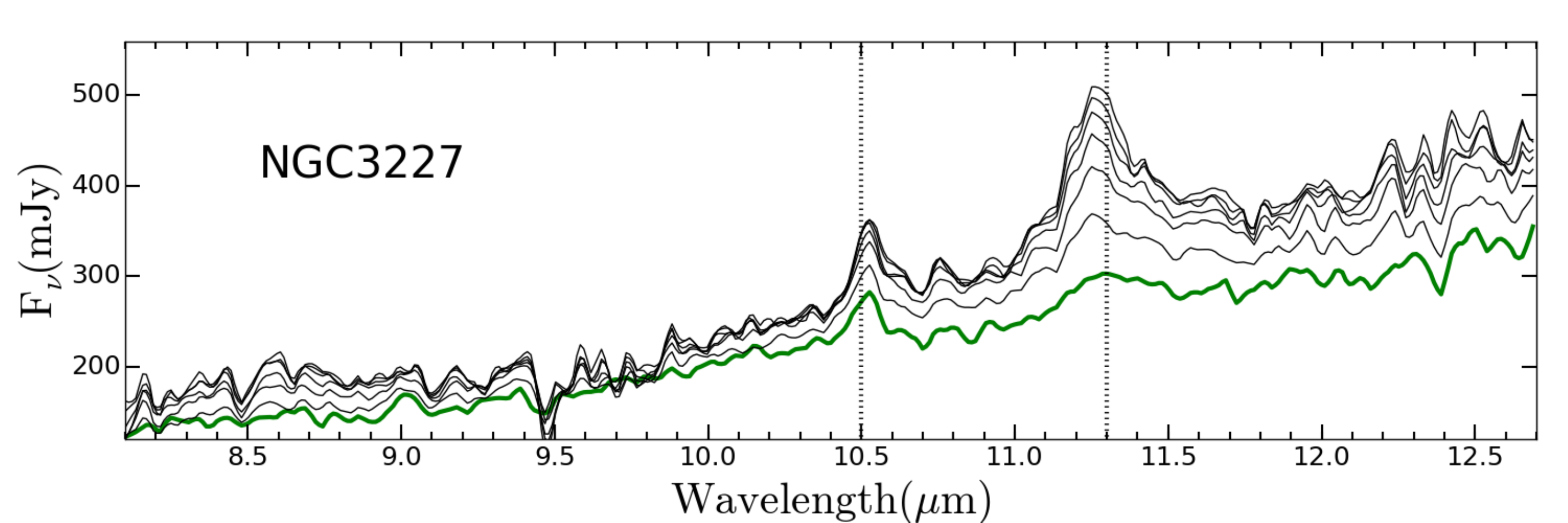}\\
\includegraphics[width=0.95\columnwidth]{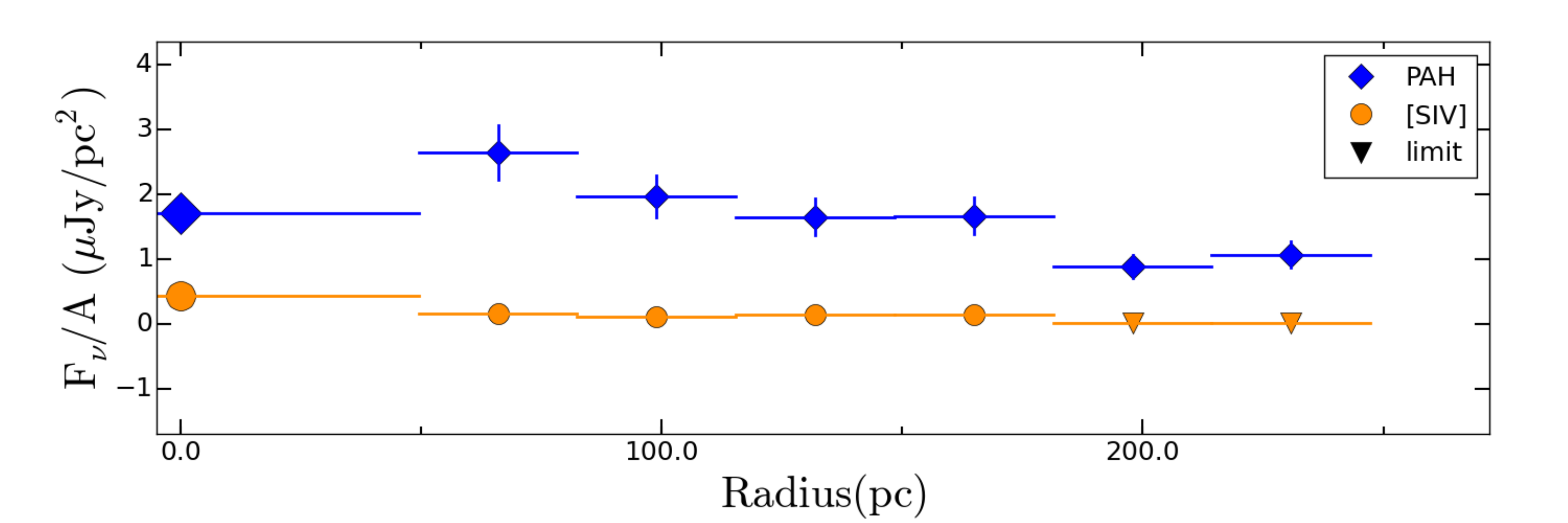}\\
\includegraphics[width=0.95\columnwidth]{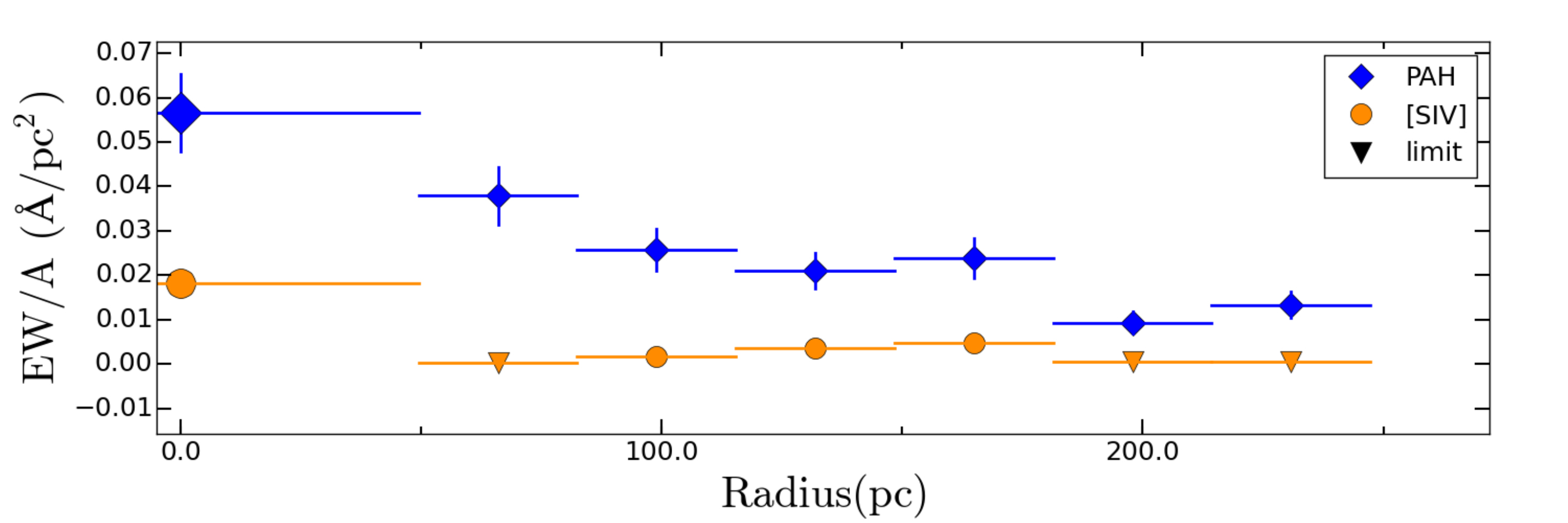}
\caption{Extracted spectra and radial profiles for NGC\,3227. The same description that in Fig. \ref{ap:fig1}.}
\end{center}
\end{figure}
\clearpage

\textbf{NGC\,3281} is an highly--inclined spiral galaxy with a Sy2 nucleus \citep[][]{Veron10}. \citet{Ramos09} and \citet{Sales11} presented observations of this source with T-ReCS with the broad N and Qa bands. They found that the spectrum of NGC\,3281 shows only a very deep silicate absorption at $\rm{9.7~\mu m}$, and some forbidden emission lines (e.g., [SIV] at $\rm{10.5~\mu m}$). They conclude that NGC\,3281 is a heavily obscured source, due to concentrated dust within a radius of 200\,pc.
\begin{figure}
\begin{center}
\includegraphics[width=0.9\columnwidth]{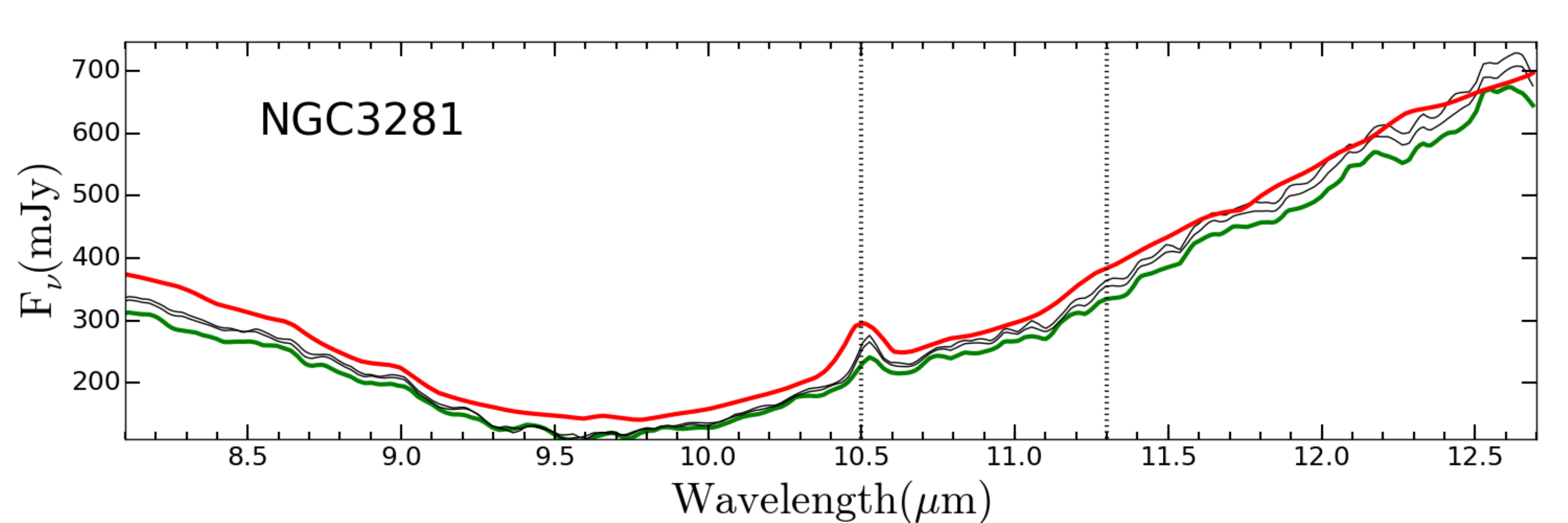}\\
\includegraphics[width=0.95\columnwidth]{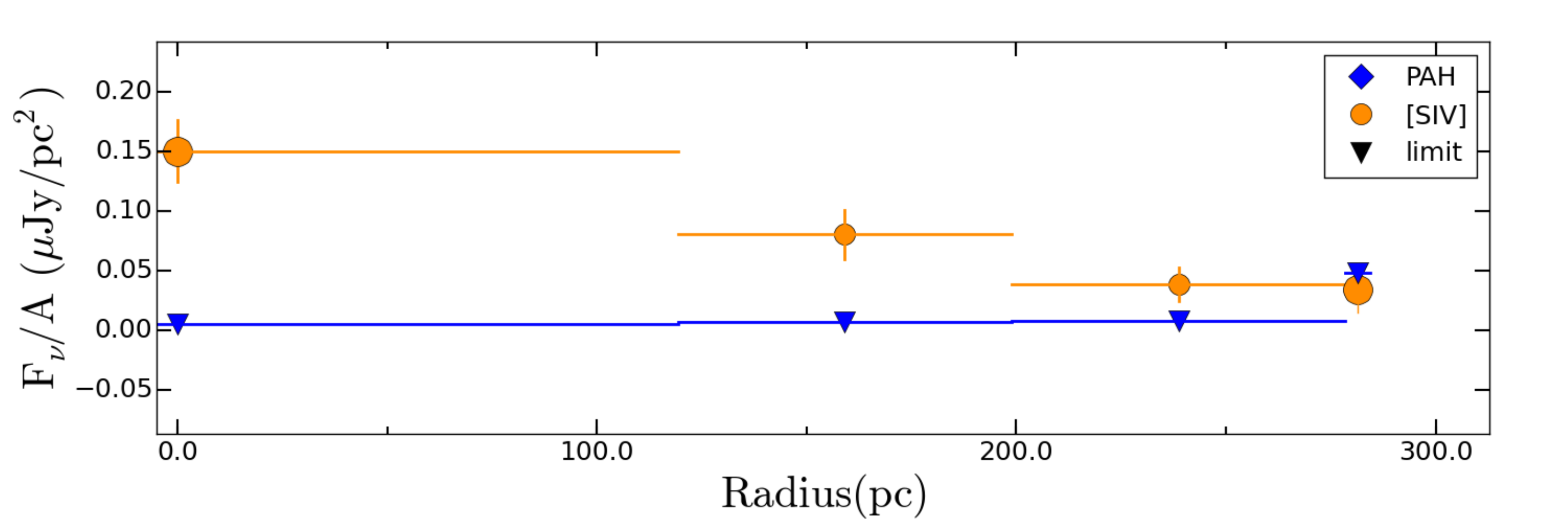}\\
\includegraphics[width=0.95\columnwidth]{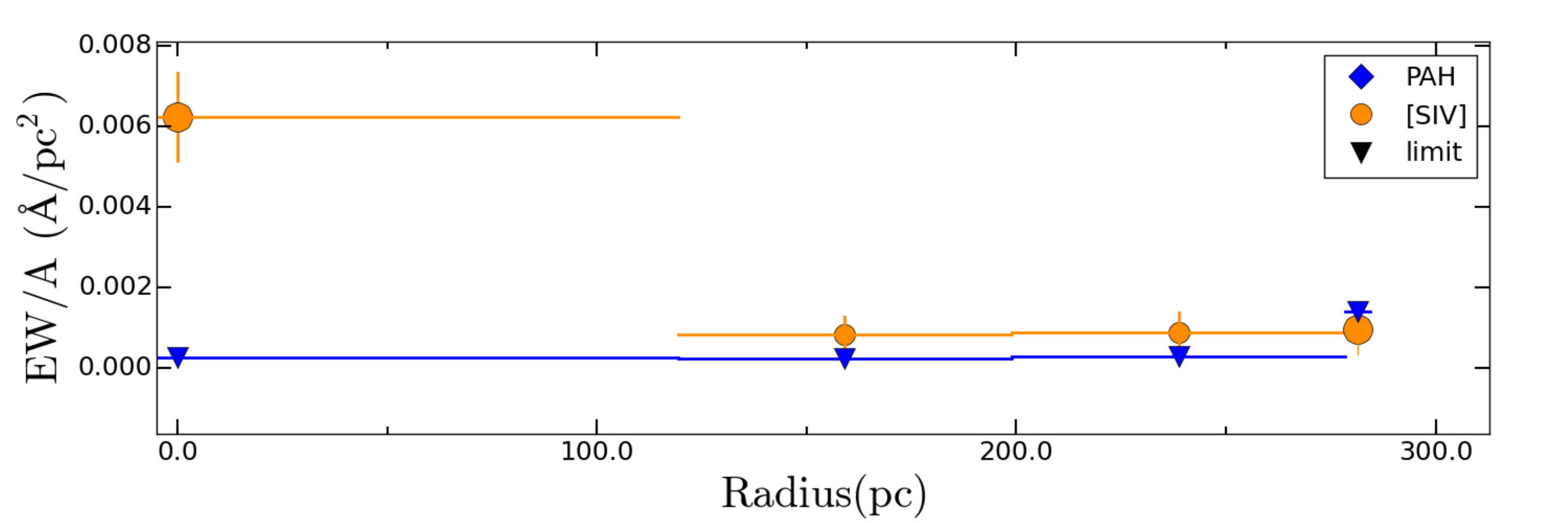}
\caption{Extracted spectra and radial profiles for NGC\,3281. The same description that in Fig. \ref{ap:fig1}.}
\end{center}
\end{figure}
\clearpage

\textbf{NGC\,4253} (Mrk\,766) is a barred spiral galaxy (SBa) with a Sy1 nucleus. The \emph{HST} images of this source show some irregular dust filaments around the nucleus \citep{Malkan98}. \citet{Rodriguez05} studied the NIR spectrum and they found permitted, forbidden, and high-ionization lines. Furthermore, \citet{Rodriguez03} found emission in $\rm{3.3~\mu m}$ PAH feature located to 150\,pc from nucleus. They considered that this emission is a signature of starburst activity.
\begin{figure}
\begin{center}
\includegraphics[width=0.9\columnwidth]{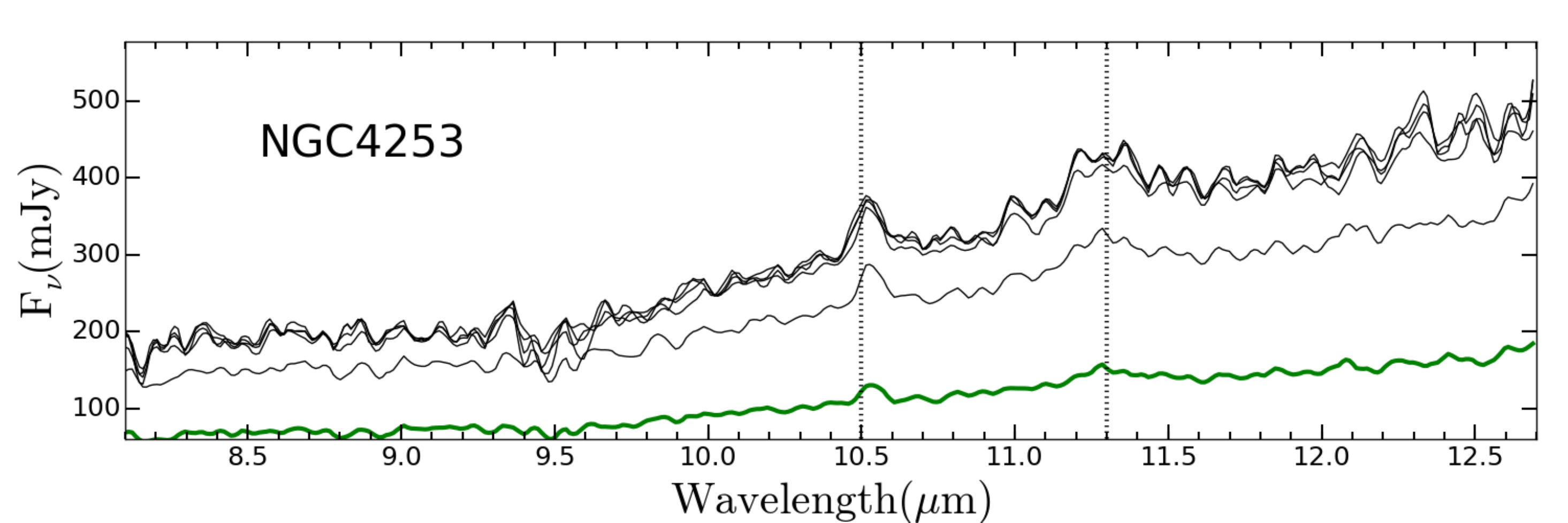}\\
\includegraphics[width=0.95\columnwidth]{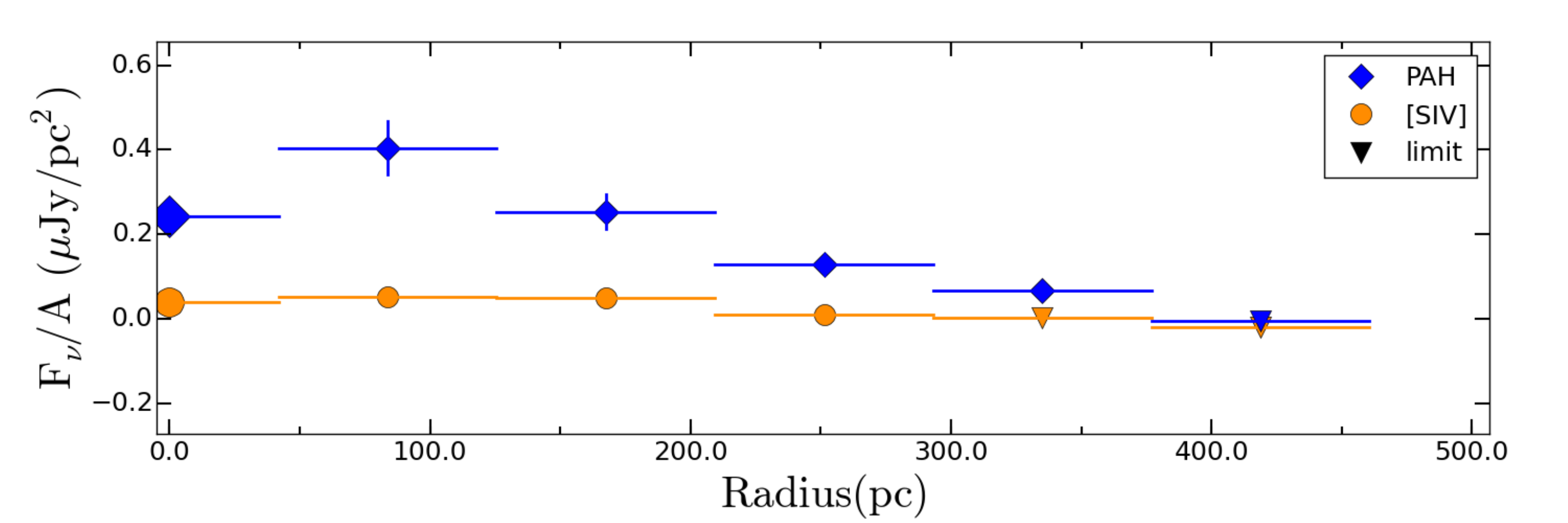}\\
\includegraphics[width=0.95\columnwidth]{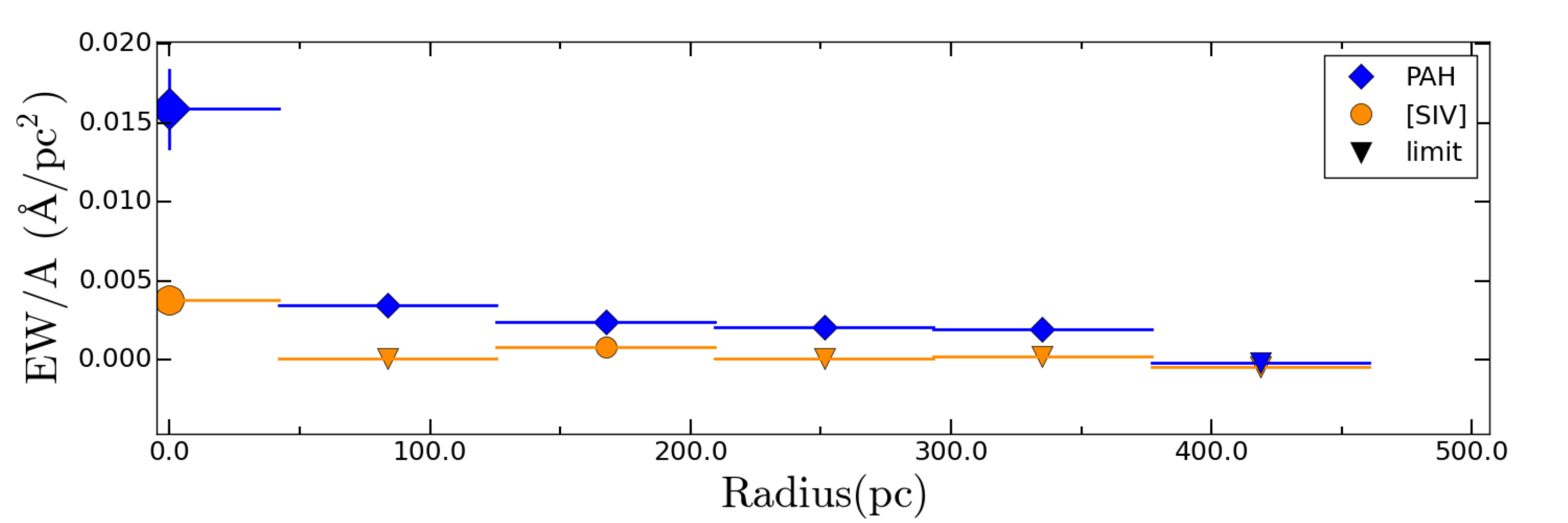}
\caption{Extracted spectra and radial profiles for NGC\,4253. The same description that in Fig. \ref{ap:fig1}.}
\end{center}
\end{figure}
\clearpage

\textbf{NGC\,4569} is the most massive, spiral, late-type, and gas-poor galaxy in the Virgo cluster \citep{Bergh76}. This source shows strong Balmer absorption lines which could be indicating SF in the last 1.5 Gyr \citep{Ho03}. \citet{Dale06} and \citet{Mason15} also suggested recent and/or ongoing SF activity based on the detection PAH emission at MIR. 
\begin{figure}
\begin{center}
\includegraphics[width=0.9\columnwidth]{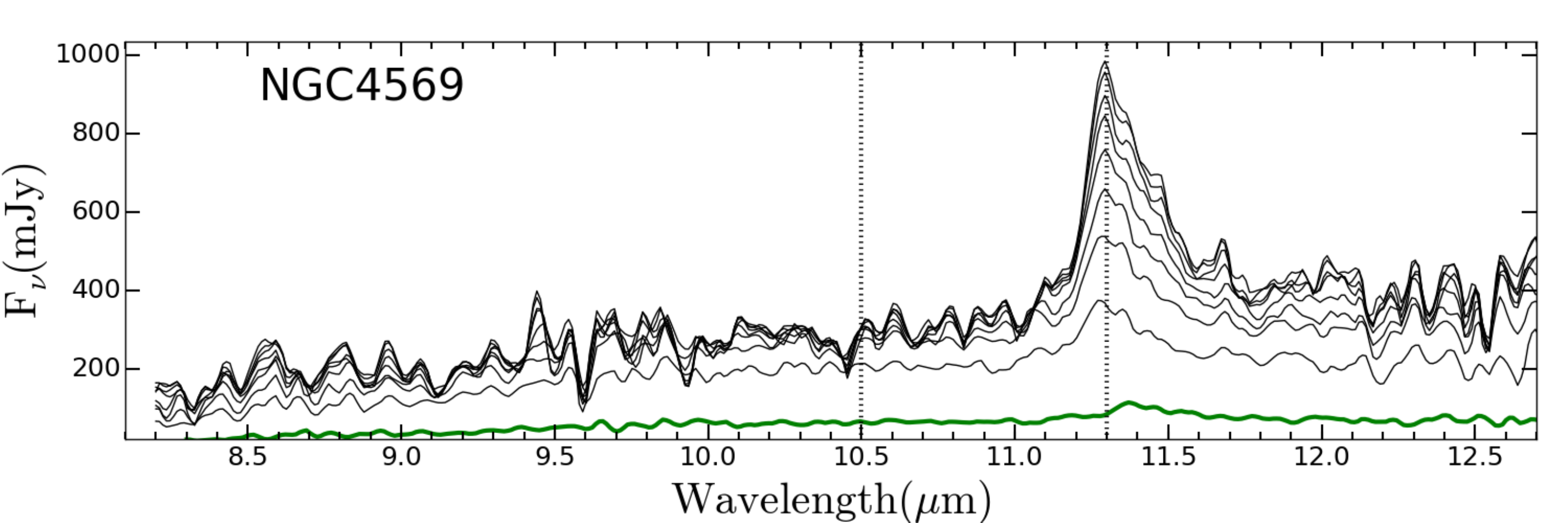}\\
\includegraphics[width=0.95\columnwidth]{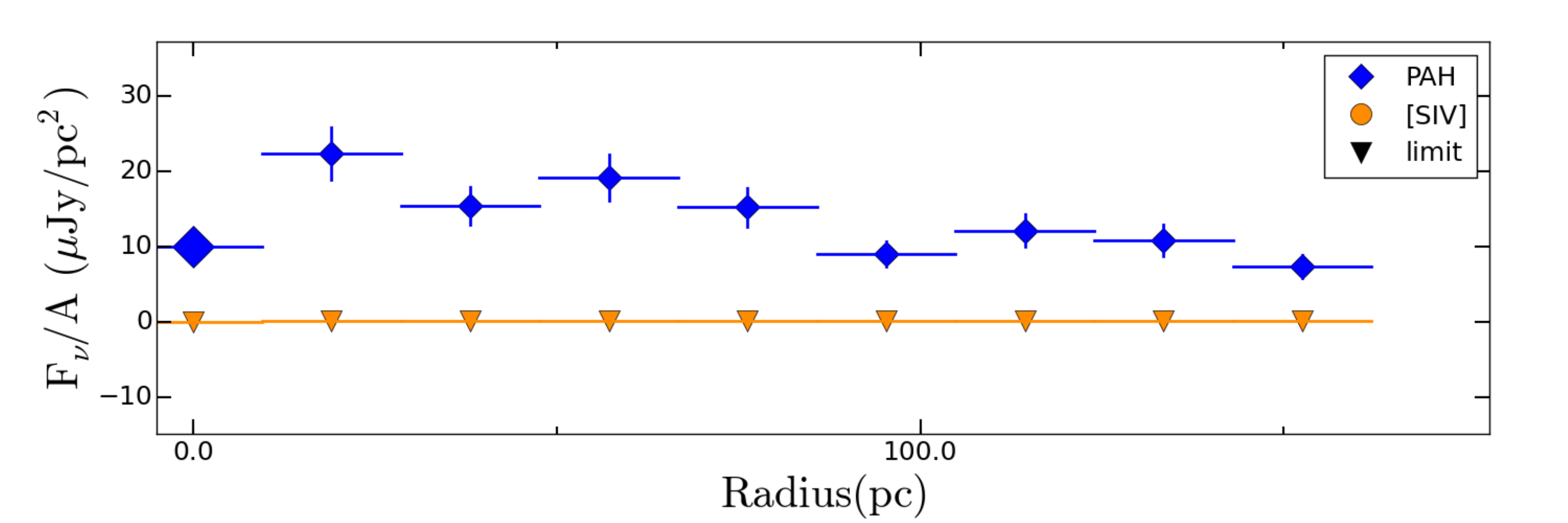}\\
\includegraphics[width=0.95\columnwidth]{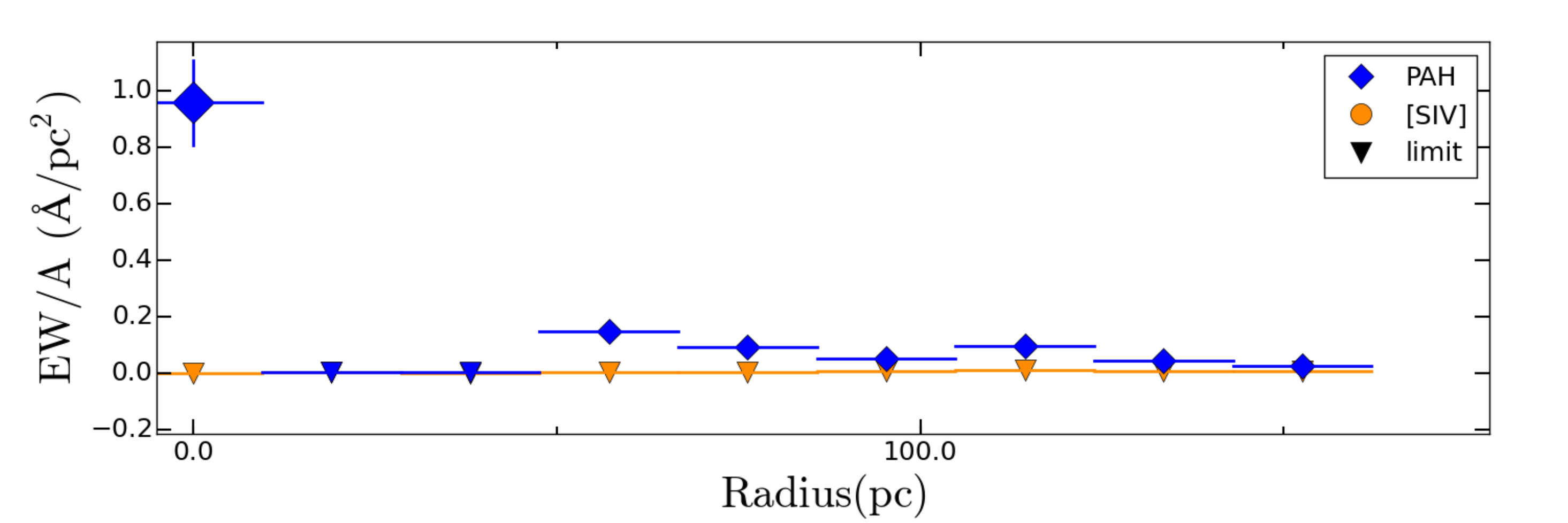}
\caption{Extracted spectra and radial profiles for NGC\,4569. The same description that in Fig. \ref{ap:fig1}.}
\end{center}
\end{figure}
\clearpage

\textbf{NGC\,5135} is an infrared--luminosity, face-on barred spiral galaxy. The nucleus is classified as a Sy2 \citep{Veron10} and it is surrounded by a circumnuclear SF in banana-shaped \citep{Gonzalez98, Bedregal09}. The inner and outer radius of the star formation emission, are located at $\sim 300$\,pc and $\sim 750$\,pc from the nucleus, respectively.
\begin{figure}
\begin{center}
\includegraphics[width=0.9\columnwidth]{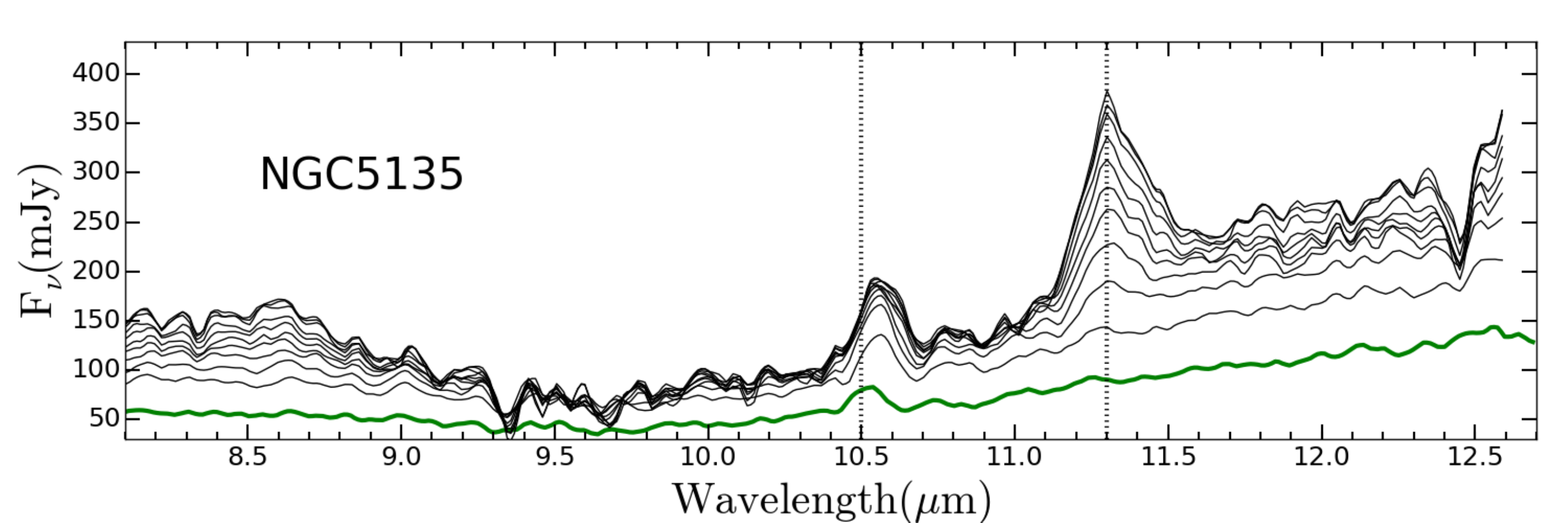}\\
\includegraphics[width=0.95\columnwidth]{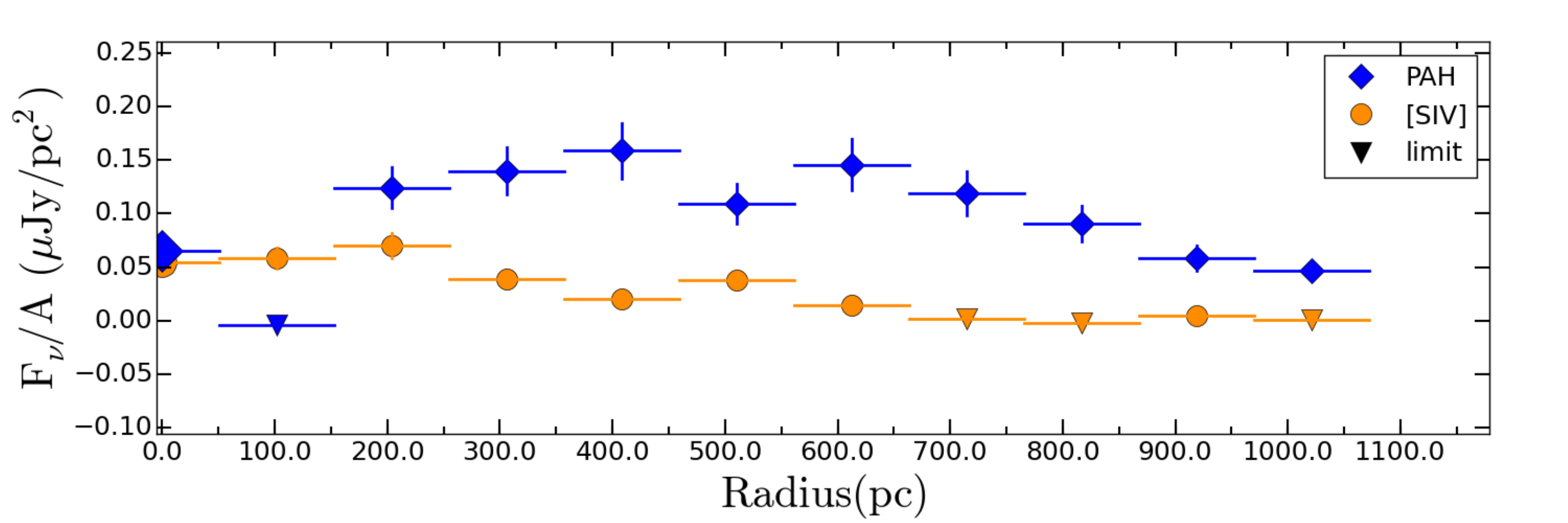}\\
\includegraphics[width=0.95\columnwidth]{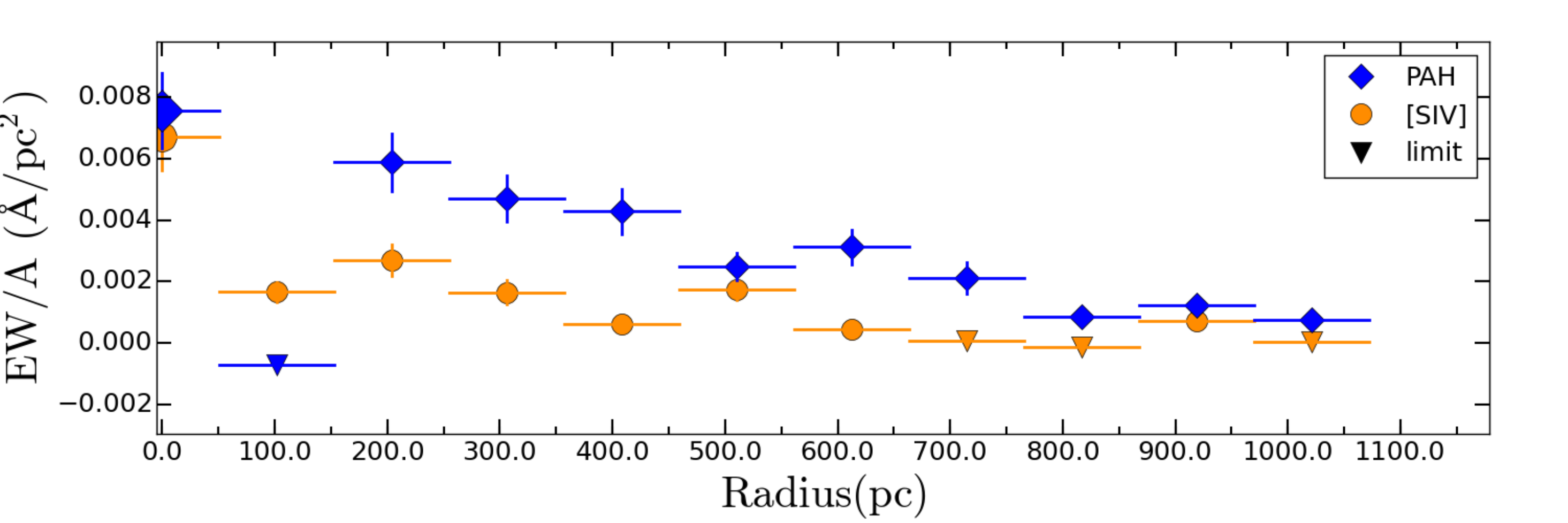}
\caption{Extracted spectra and radial profiles for NGC\,5135. The same description that in Fig. \ref{ap:fig1}.}
\end{center}
\end{figure}
\clearpage

\textbf{NGC\,5643} is a face-on barred spiral galaxy with a Sy2 nucleus \citep{Veron10}. The IRAC and MIPS images show a compact MIR nucleus embedded within the spiral-like host emission \citep{Asmus14}. Moreover, the arcsecond-scale MIR SED is significantly affected by SF \citep[e.g.][]{Shi06, Goulding09}.
\begin{figure}
\begin{center}
\includegraphics[width=0.9\columnwidth]{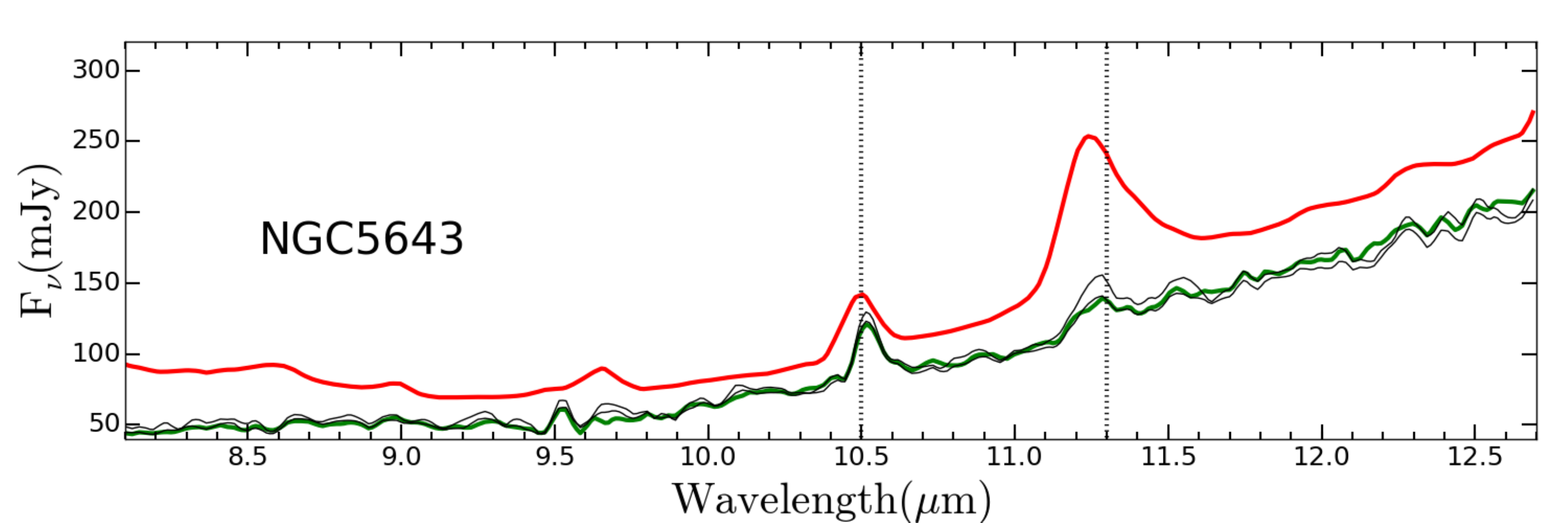}\\
\includegraphics[width=0.95\columnwidth]{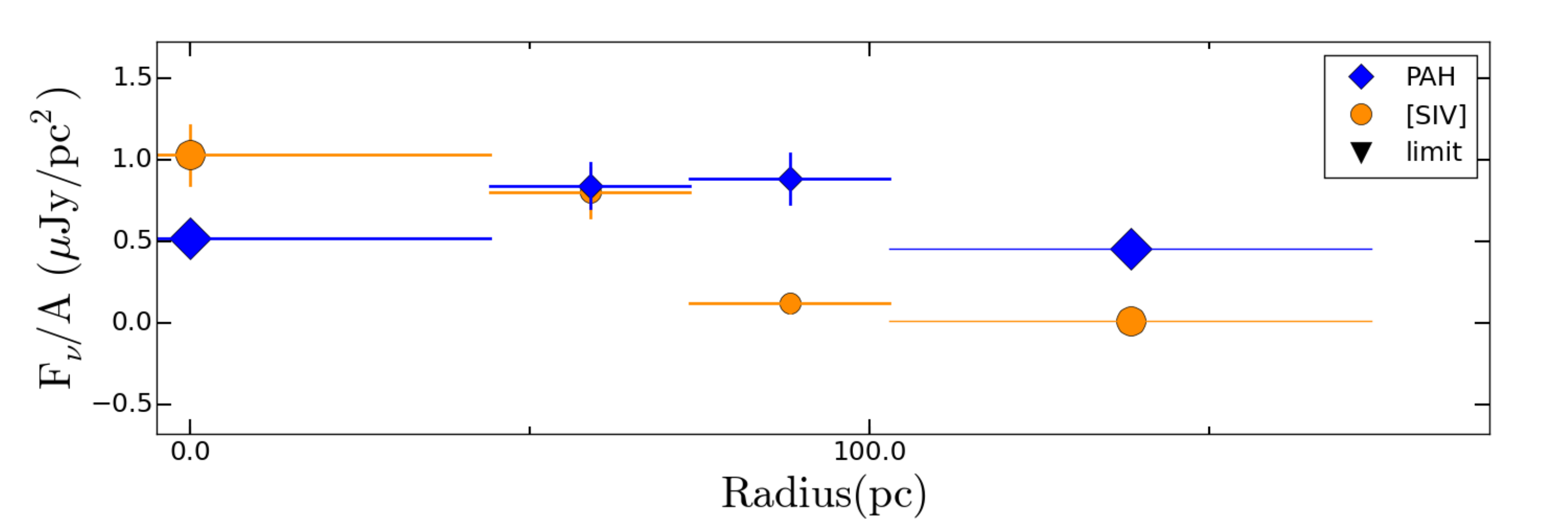}\\
\includegraphics[width=0.95\columnwidth]{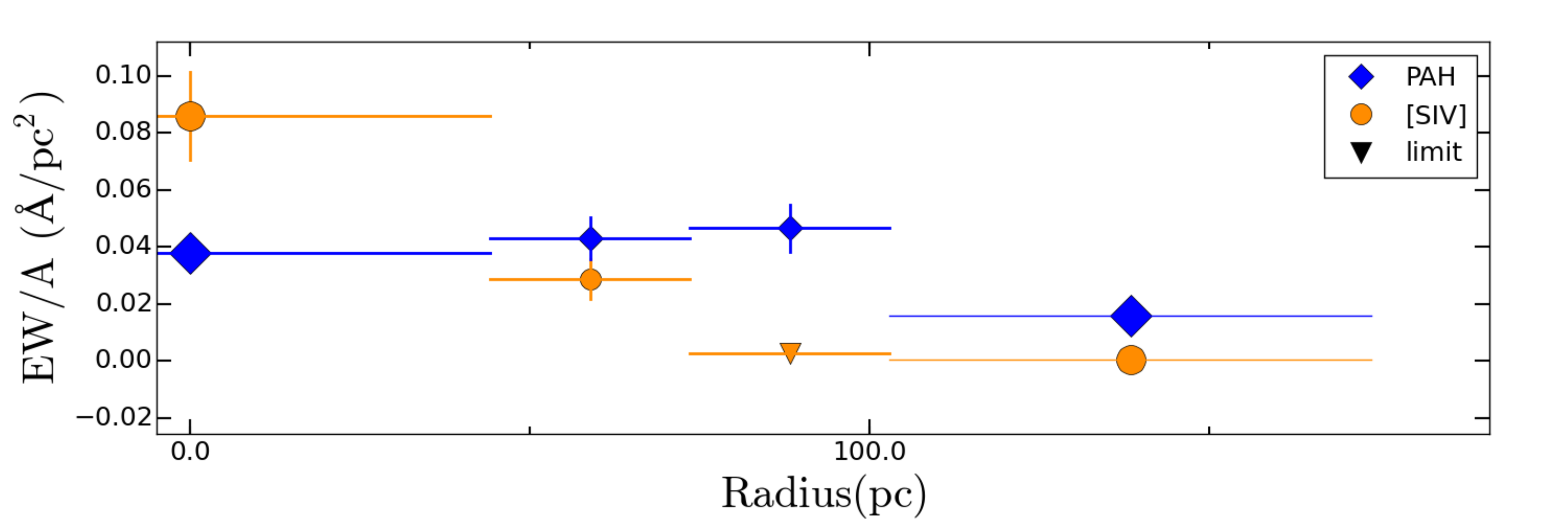}
\caption{Extracted spectra and radial profiles for NGC\,5643. The same description that in Fig. \ref{ap:fig1}.}
\end{center}
\end{figure}
\clearpage

\textbf{IC\,4518W} is a spiral galaxy with a Sy2 nucleus \citep{Veron10}. \citet{Diaz10} and \citet{Asmus14} found that the SF contribution at subarcsecond resolution is probably minor in its nucleus. \citet{Diaz10} found [SIV] line emission at $\sim 265$\,pc toward the north of the nucleus. They suggested that this emission could be related with the NLR.  
\begin{figure}
\begin{center}
\includegraphics[width=0.9\columnwidth]{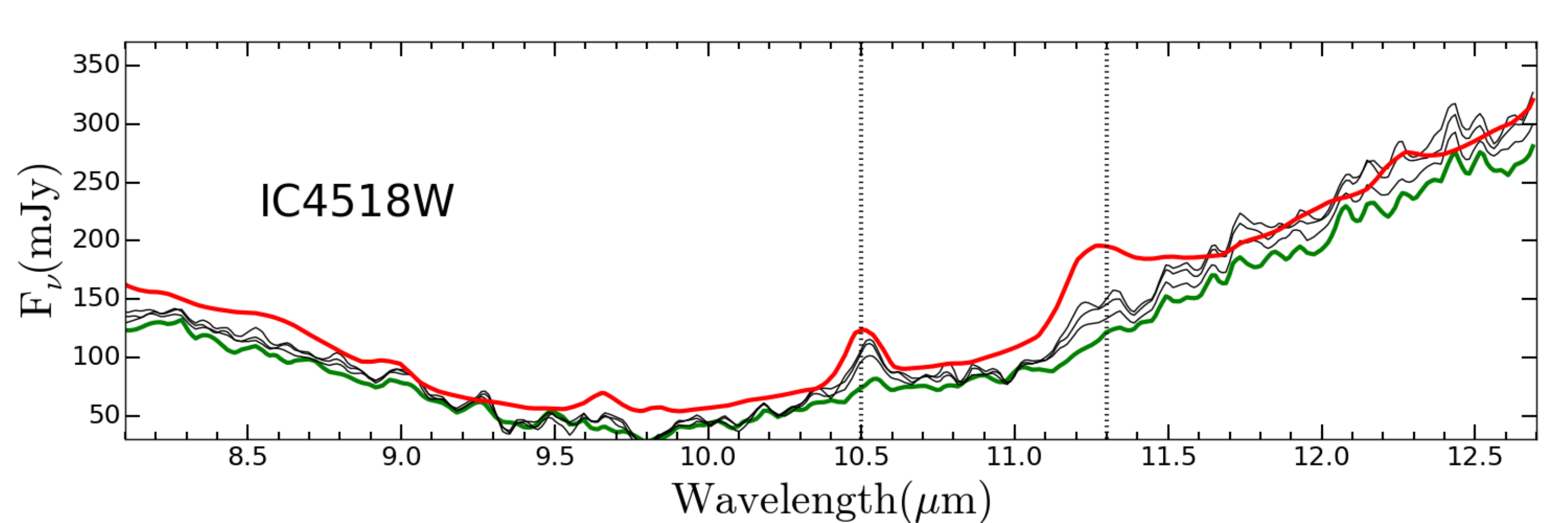}\\
\includegraphics[width=0.95\columnwidth]{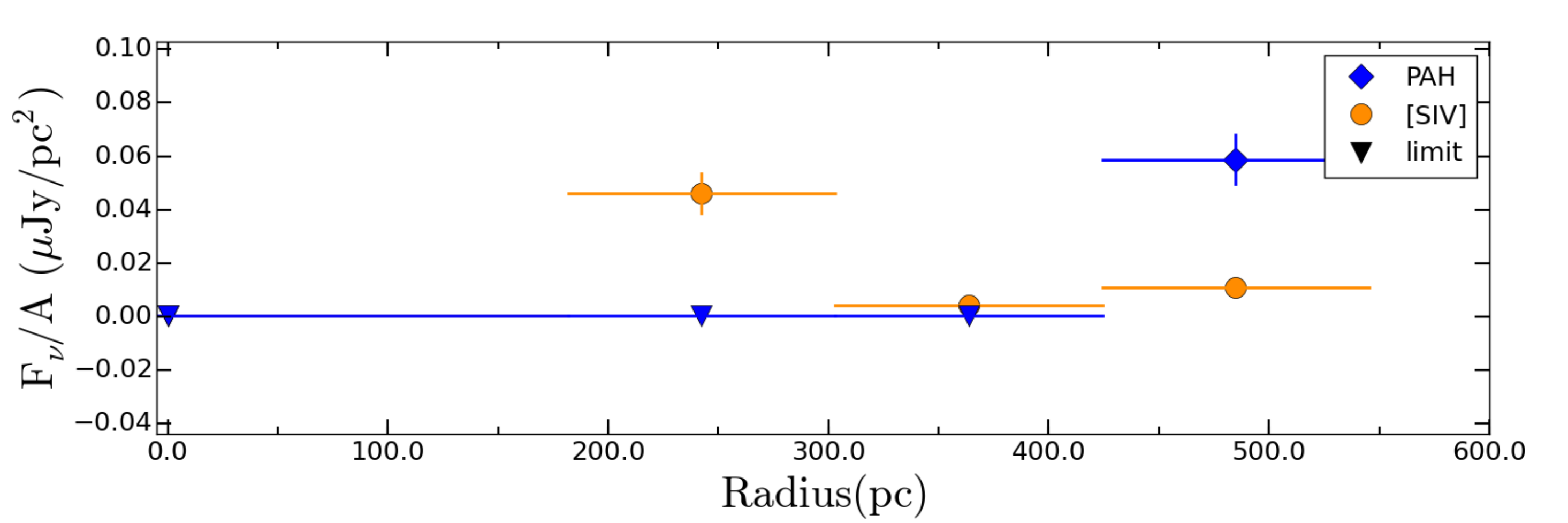}\\
\includegraphics[width=0.95\columnwidth]{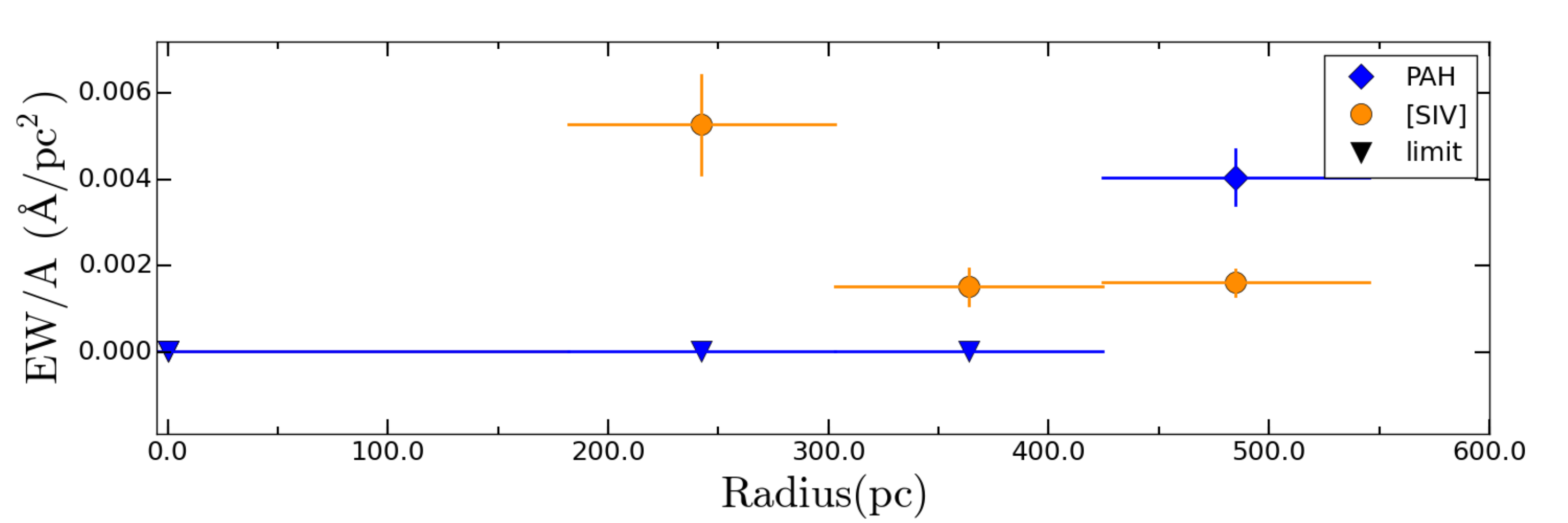}
\caption{Extracted spectra and radial profiles for IC\,4518W. The same description that in Fig. \ref{ap:fig1}.}
\end{center}
\end{figure}
\clearpage

\textbf{IC\,5063} is a peculiar galaxy with both spiral and elliptical properties with a Sy2 nucleus \citep{Kewley01}. \citet{Colina91} proposed that IC\,5063 is a remnant of a recent merger, while \citet{Martini03} speculated that the nuclear obscuration might be caused by foreground dust lanes. We did not find records of SF in other works at the scales traced by our observations.
\begin{figure}
\begin{center}
\includegraphics[width=0.9\columnwidth]{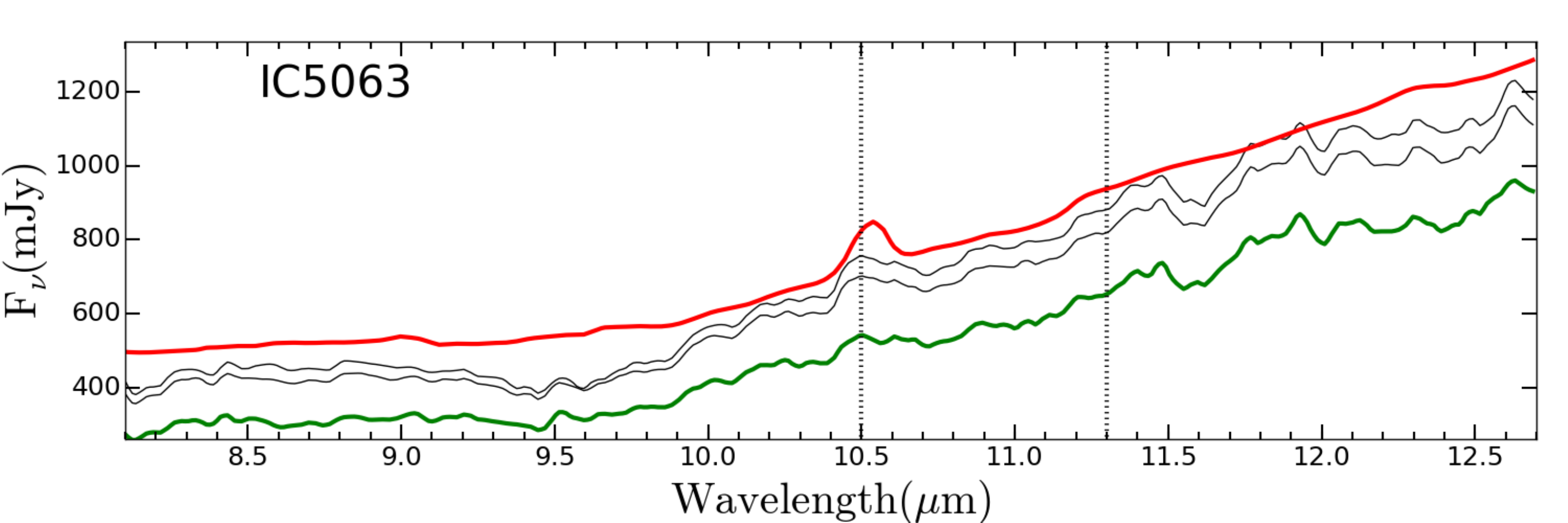}\\
\includegraphics[width=0.95\columnwidth]{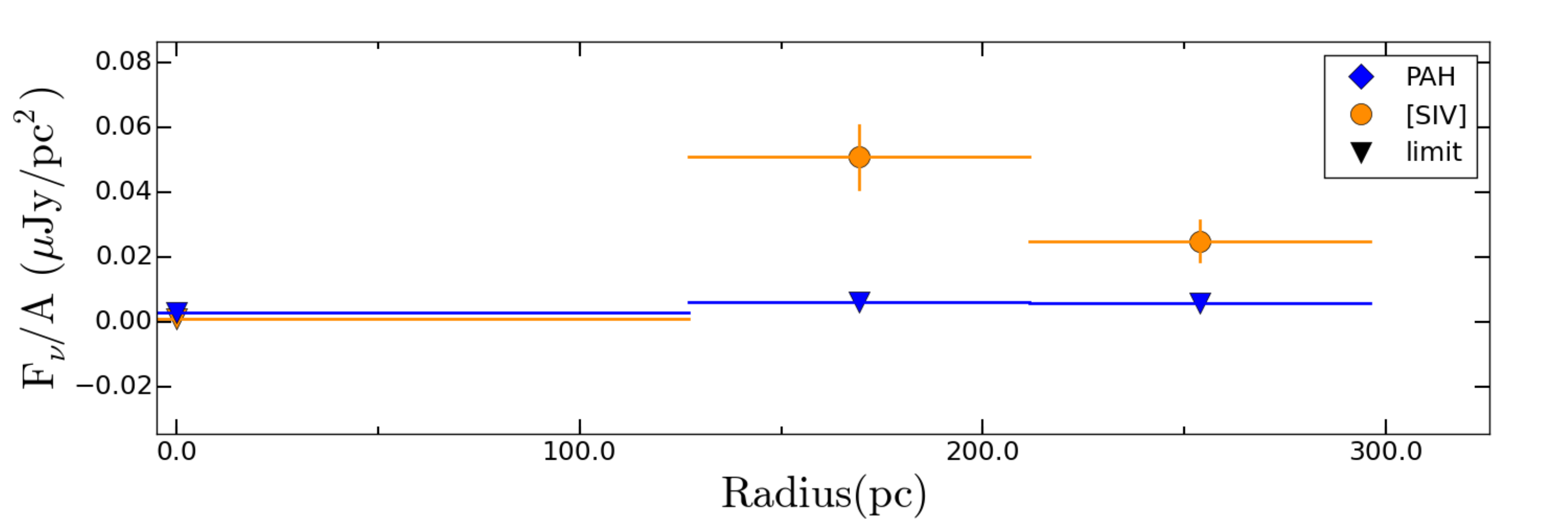}\\
\includegraphics[width=0.95\columnwidth]{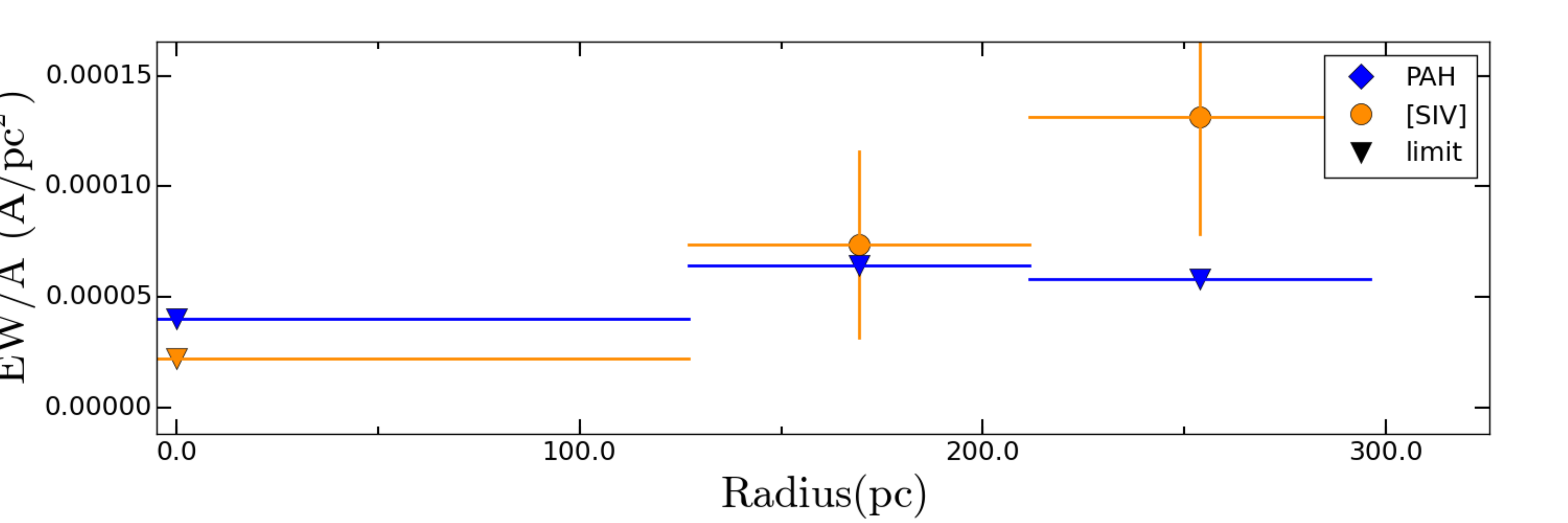}
\caption{Extracted spectra and radial profiles for IC\,5063. The same description that in Fig. \ref{ap:fig1}.}
\end{center}
\end{figure}
\clearpage

\textbf{NGC\,7130} is a peculiar low-inclination spiral galaxy with a Sy1.9 nucleus. A compact starburst is located at the center and it is extended over $\sim 300$\,pc \citep{Gonzalez98, Levenson05}. \citet{Wu09} and \citet{Alonso12} found that the arcsecond--scales MIR SED indicates obscured AGN emission with a high SF contribution. \citet{Asmus14} also concluded that the nuclear MIR SED is presumably still affected by significant SF emission.
\begin{figure}
\begin{center}
\includegraphics[width=0.9\columnwidth]{NGC7130a348_radprof.pdf}\\
\includegraphics[width=0.95\columnwidth]{NGC7130a348_flux_Comp.pdf}\\
\includegraphics[width=0.95\columnwidth]{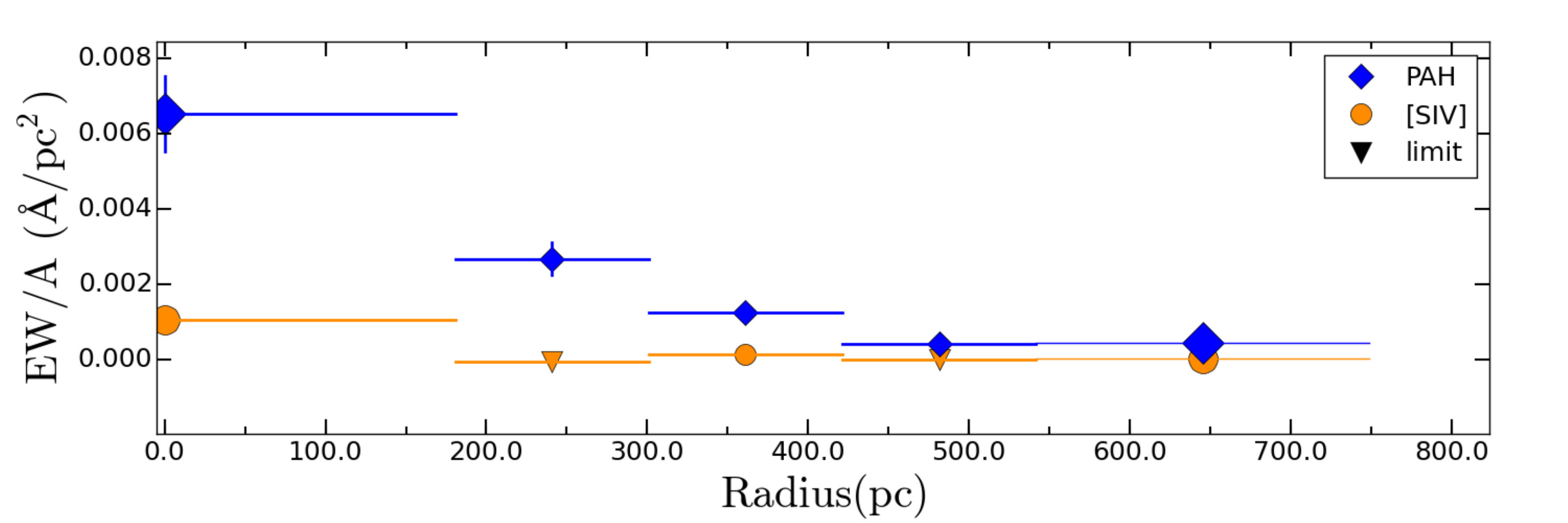}
\caption{Extracted spectra and radial profiles for NGC\,7130. The same description that in Fig. \ref{ap:fig1}.}
\end{center}
\end{figure}
\clearpage

\textbf{NGC\,7172} is an edge-on lenticular galaxy with a Sy2 nucleus \citep{Veron10}. \citet{Smajic12} found a prominent dust lane projected along the nucleus. The arcsecond--scales MIR SED might be affected by significant SF \citep{Wu09,Gallimore10}. However, \citet{Asmus14} concluded that the nuclear MIR SED is free of SF contamination.
\begin{figure}
\begin{center}
\includegraphics[width=0.9\columnwidth]{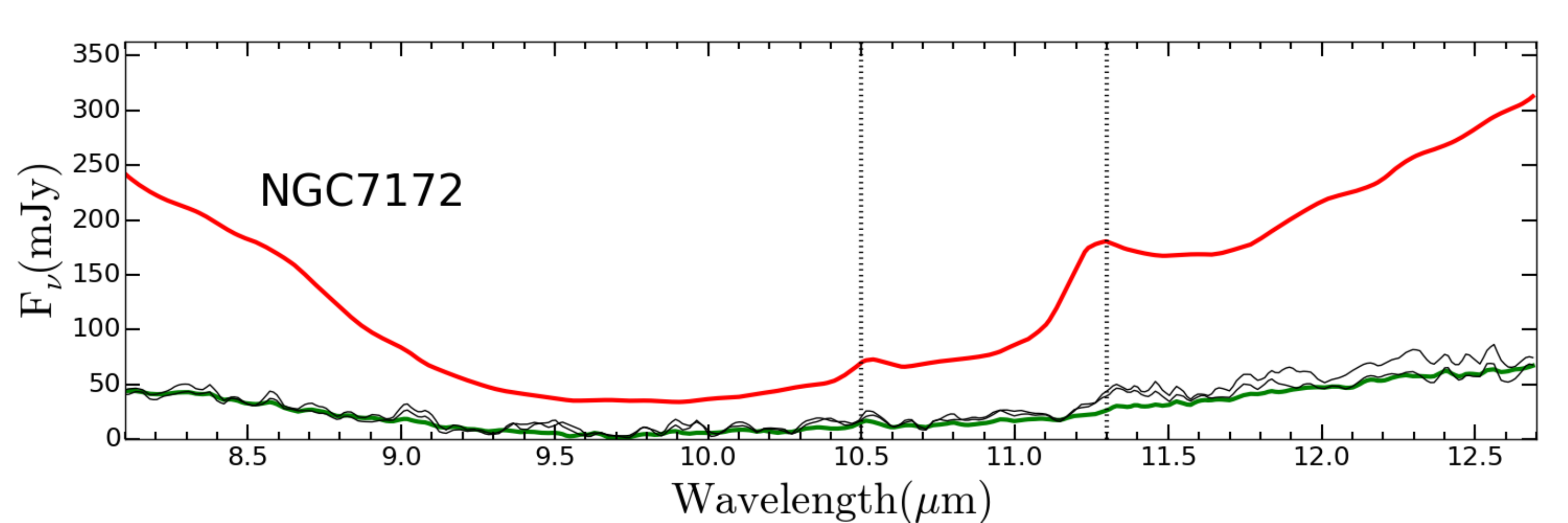}\\
\includegraphics[width=0.95\columnwidth]{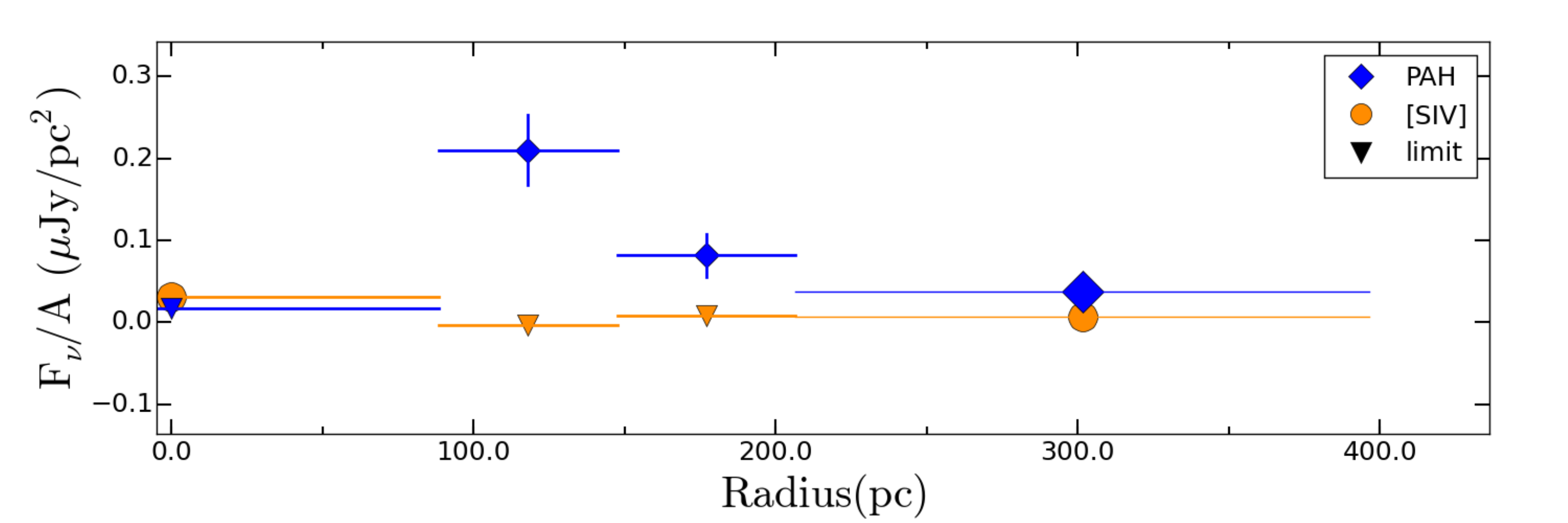}\\
\includegraphics[width=0.95\columnwidth]{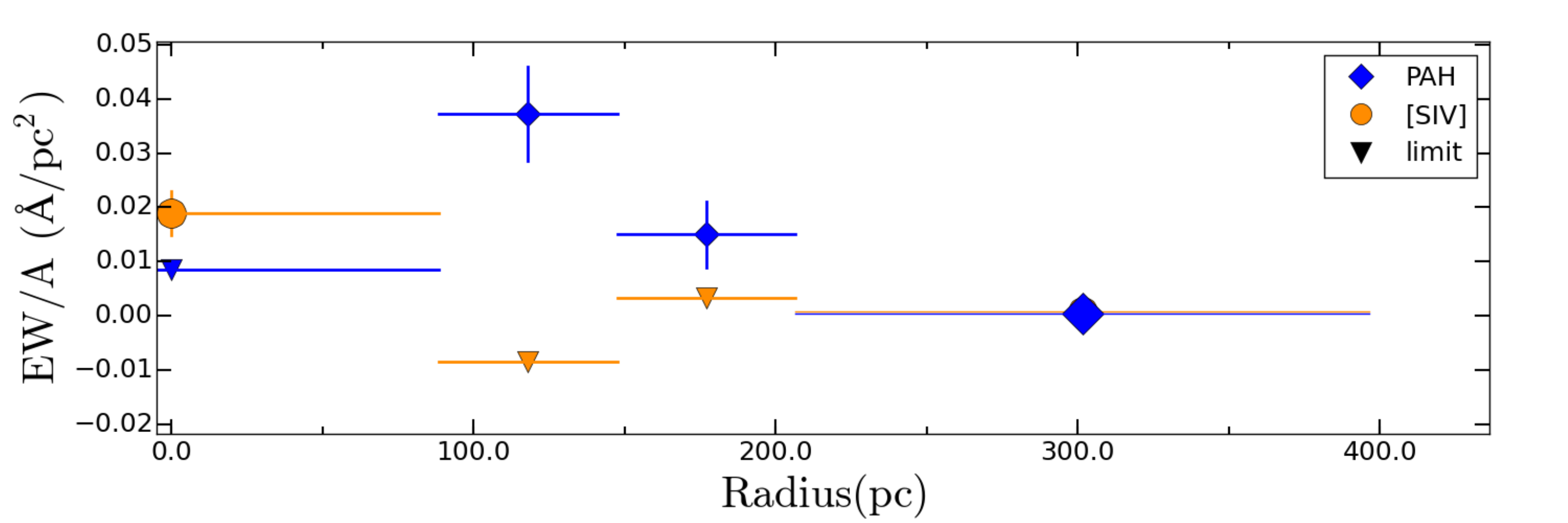}
\caption{Extracted spectra and radial profiles for NGC\,7172. The same description that in Fig. \ref{ap:fig1}.}
\end{center}
\end{figure}
\clearpage

\textbf{NGC\,7465} is a spiral galaxy with a Sy2 nucleus. This source is part of a group of nine galaxies interacting \citep{Haynes81}. The dominant stellar population in the nuclear region of NGC\,7465 corresponds to stars between K3 III and M3 III types, according to the relative absorption bands measurements \citep{Ramos09}.
\begin{figure}
\begin{center}
\includegraphics[width=0.9\columnwidth]{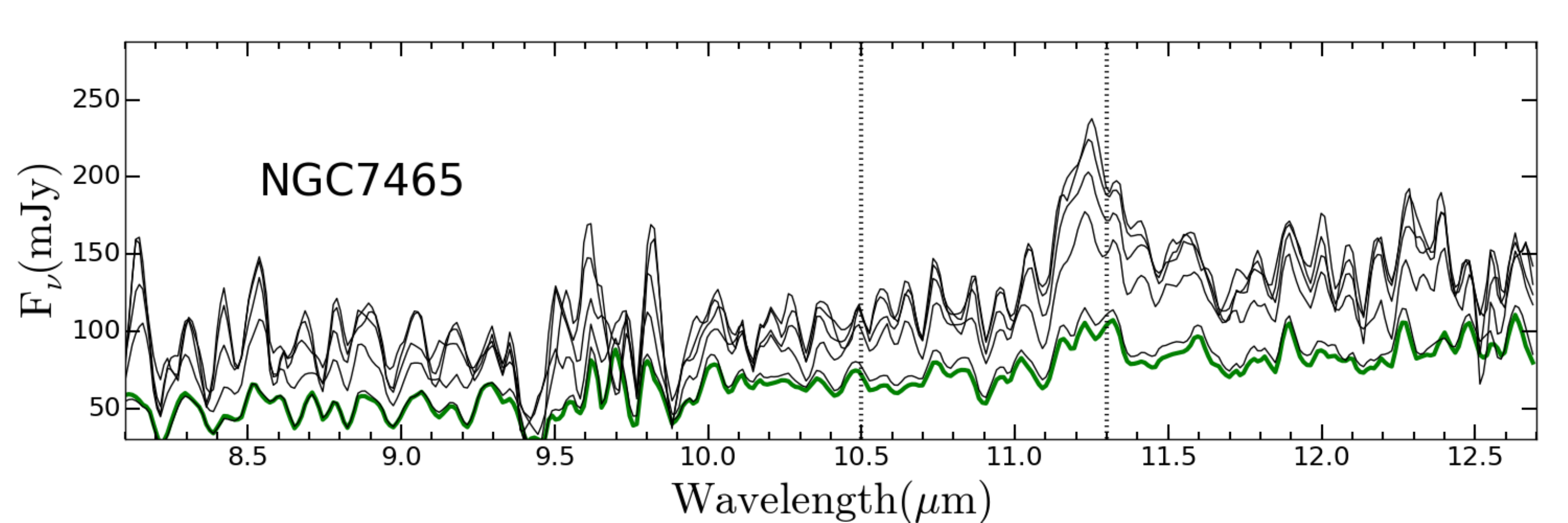}\\
\includegraphics[width=0.95\columnwidth]{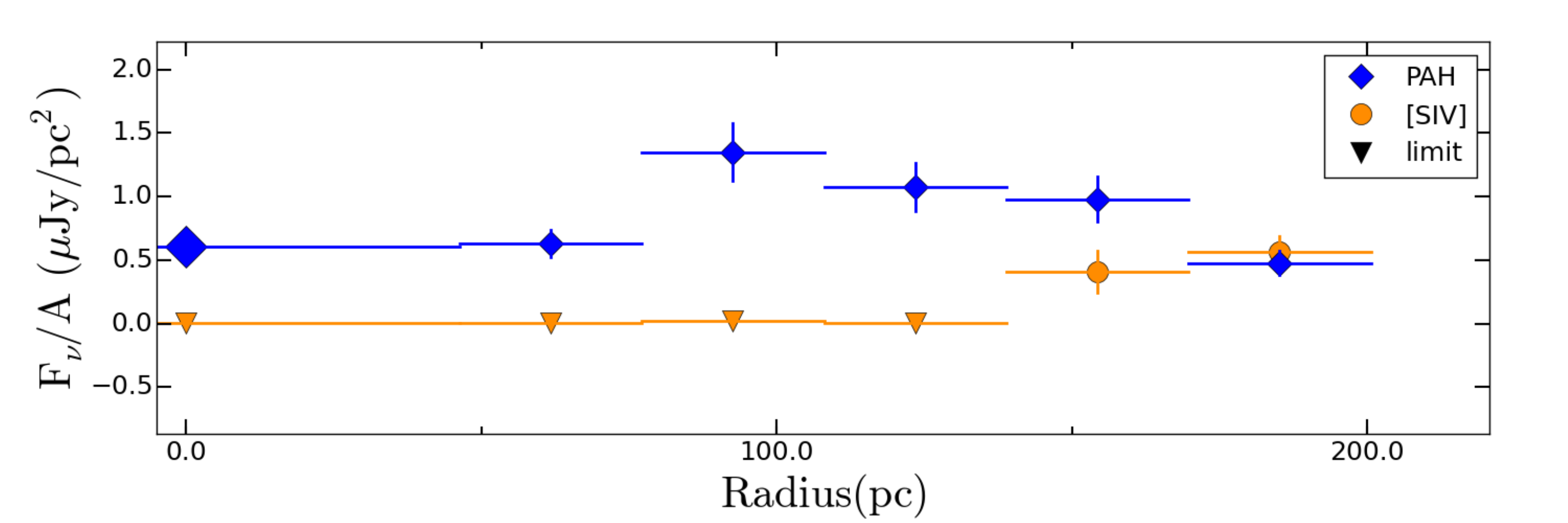}\\
\includegraphics[width=0.95\columnwidth]{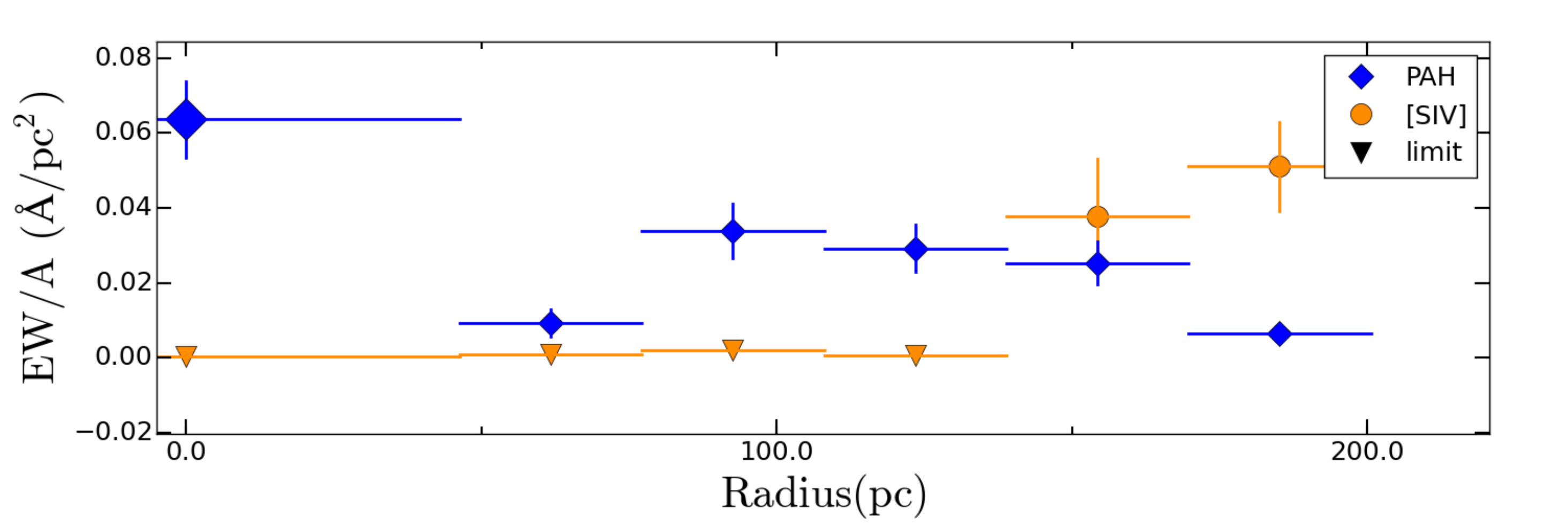}
\caption{Extracted spectra and radial profiles for NGC\,7465. The same description that in Fig. \ref{ap:fig1}.}
\end{center}
\end{figure}
\clearpage

\textbf{NGC\,7582} is a highly inclined barred spiral galaxy with an obscured nucleus. The nuclear spectrum has been studied as a composition between AGN and starburst \citep{Veron97}. The AGN is surrounded by a powerful SF disc (major axis diameter $\sim 400$\,pc) and a dust lane crossing over the nucleus \citep{Morris85, Riffel09}. \citet{Asmus14} concluded that the starburst dominates the total MIR emission.
\begin{figure}
\begin{center}
\includegraphics[width=0.9\columnwidth]{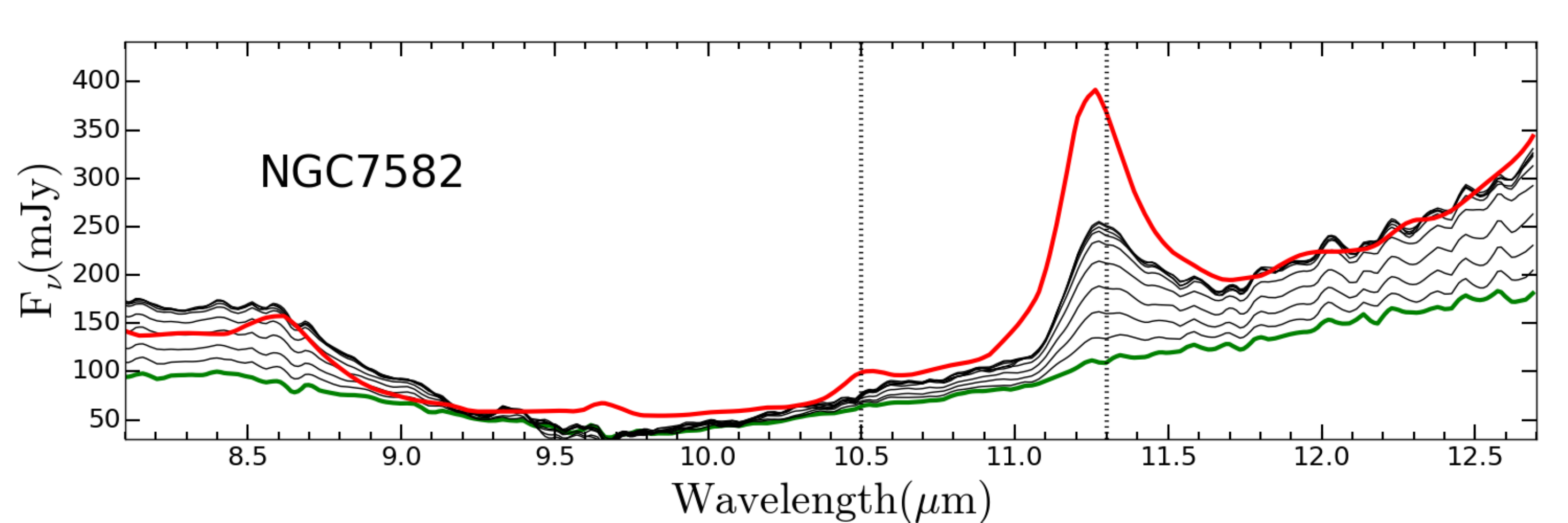}\\
\includegraphics[width=0.95\columnwidth]{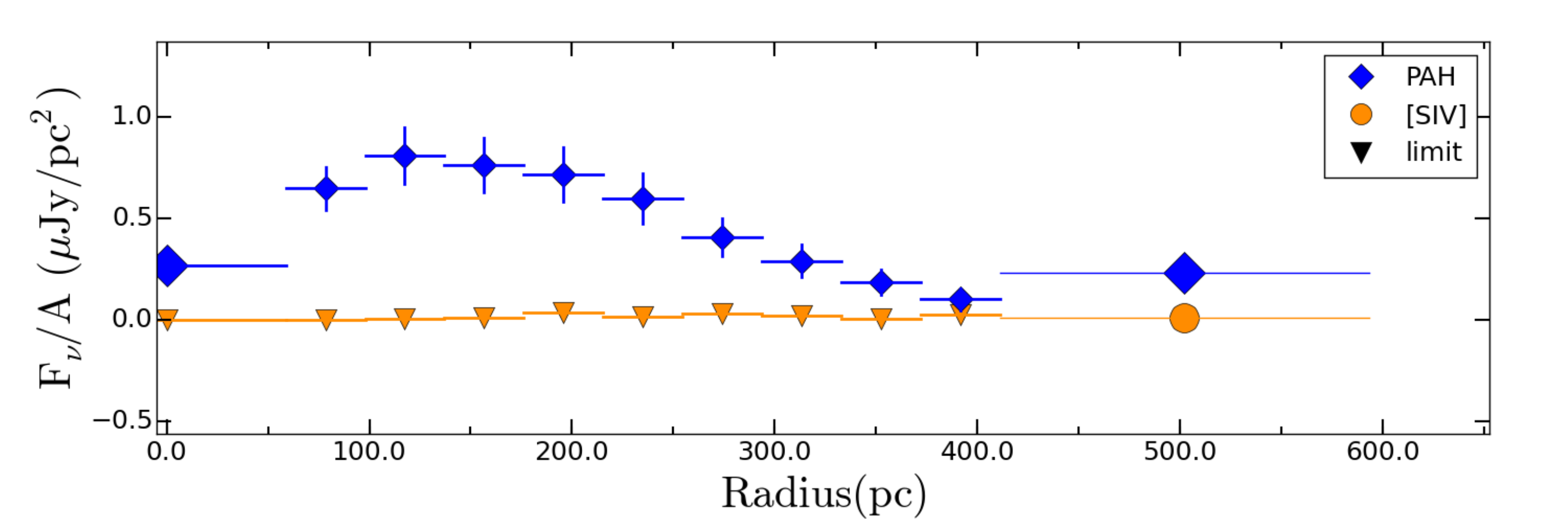}\\
\includegraphics[width=0.95\columnwidth]{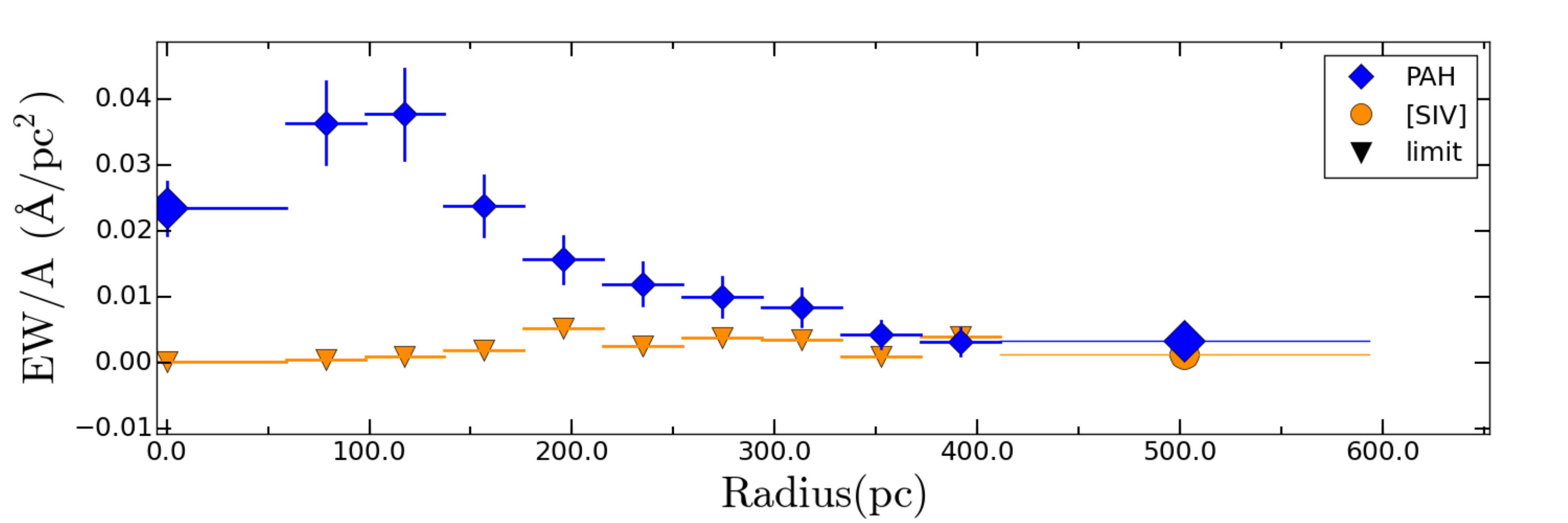}
\caption{Extracted spectra and radial profiles for NGC\,7582. The same description that in Fig. \ref{ap:fig1}.}
\end{center}
\end{figure}


\end{document}